\documentclass{ieeeaccess}
\usepackage{cite}
\usepackage{amsmath,amssymb,amsfonts}
\usepackage{algorithmic}
\usepackage{graphicx}
\usepackage{xcolor}
\usepackage{textcomp}
\usepackage{listings}
\usepackage{subfig}
\usepackage{braket}
\usepackage{hyperref}
\usepackage{cleveref}
\crefname{figure}{Figure}{Figures}
\crefname{section}{Section}{Sections}
\usepackage[htt]{hyphenat}

\def\BibTeX{{\rm B\kern-.05em{\sc i\kern-.025em b}\kern-.08em
    T\kern-.1667em\lower.7ex\hbox{E}\kern-.125emX}}
\begin{document}

\definecolor{codegreen}{rgb}{0,0.6,0}
\definecolor{codegray}{rgb}{0.5,0.5,0.5}
\definecolor{codepurple}{rgb}{0.58,0,0.82}
\definecolor{backcolour}{rgb}{0.95,0.95,0.92}

\lstdefinestyle{mystyle}{
  language=Python,
  basicstyle=\ttfamily\small,
  numbers=left,
  numberstyle=\tiny\color{codegray},
  stepnumber=1,
  numbersep=5pt,
  backgroundcolor=\color{white},
  showspaces=false,
  showstringspaces=false,
  showtabs=false,
  frame=single,
  rulecolor=\color{black},
  tabsize=2,
  captionpos=b,
  breaklines=true,
  breakatwhitespace=false,
  title=\lstname,
  keywordstyle=\color{blue},
  commentstyle=\color{codegreen},
  stringstyle=\color{red},
  escapeinside={\%*}{*)},
  morekeywords={*,...}
}

\lstset{style=mystyle}

\lstdefinelanguage{qoala}
{
    morekeywords=[1]{
        add, add, sub, addm, subm,
        jmp, bez, bnz, beq, bne, blt, bge,
        set, store, load
        meas,
        create_epr, recv_epr,
        init, meas, h, rot_x, rot_y, rot_z, cnot, cphase, cx_dir, cy_dir,
        APPID, NETQASM,
        run_request, run_routine, return_result, assign
    },
    morekeywords=[2]{
        META_START, META_END,
        SUBROUTINE,
        NETQASM_START, NETQASM_END,
        REQUEST
    },
    keywordstyle=[1]\color{blue},
    keywordstyle=[2]\color{purple},
    sensitive=true,
    morecomment=[l]{//},
    morecomment=[s]{/*}{*/},
    morecomment=[s][\color{blue}]{\#\ }{\ },
    morestring=[b]",
    escapeinside={(*@}{@*)},
}

\lstnewenvironment{qoalacode}{\lstset{style=mystyle,language=qoala}}{}

\newcommand{\revision}[1]{#1}
\newcommand{\todo}[1]{\textcolor{red}{TODO #1}}

\title{Qoala: an Application Execution Environment for Quantum Internet Nodes}

\author{
    Bart van der Vecht\authorrefmark{1,2,3},
    Atak Talay Yücel\authorrefmark{1},
    Hana Jirovská\authorrefmark{1,2,3},
    Stephanie Wehner\authorrefmark{1,2,3}
}
\address[1]{QuTech, Delft University of Technology}
\address[2]{Kavli Institute of Nanoscience, Delft University of Technology}
\address[3]{Quantum Computer Science, Electrical Engineering, Mathematics and Computer Science, Delft University of Technology}

\markboth
{Van der Vecht \headeretal: Qoala}
{Van der Vecht \headeretal: Qoala}

\corresp{Corresponding author: Bart van der Vecht (email: b.vandervecht@tudelft.nl).}

\begin{abstract}
Recently, a first-of-its-kind operating system for programmable quantum network nodes was developed, called QNodeOS.
Here, we present an extension of QNodeOS called Qoala, which introduces
(1) a unified program format for hybrid interactive classical-quantum programs, providing a well-defined target for compilers, and 
(2) a runtime representation of a program that allows joint scheduling of the hybrid classical-quantum program, multitasking, and asynchronous program execution.
Based on concrete design considerations, we put forward the architecture of Qoala, including the program structure and execution mechanism.
We implement Qoala in the form of a modular and extendible simulator that is validated against real-world quantum network hardware (available online).
However, Qoala is not meant to be purely a simulator, and implementation is planned on real hardware.
We evaluate Qoala's effectiveness and performance sensitivity to latencies and network schedules using an extensive simulation study.
Qoala provides a framework that opens the door for future computer science research into quantum network applications, including scheduling algorithms and compilation strategies that can now readily be explored using the framework and tools provided. 
\end{abstract}

\titlepgskip=-15pt

\maketitle

\section{Introduction}
\label{sec:introduction}

\begin{figure}
    \centering
    \includegraphics[scale=0.35]{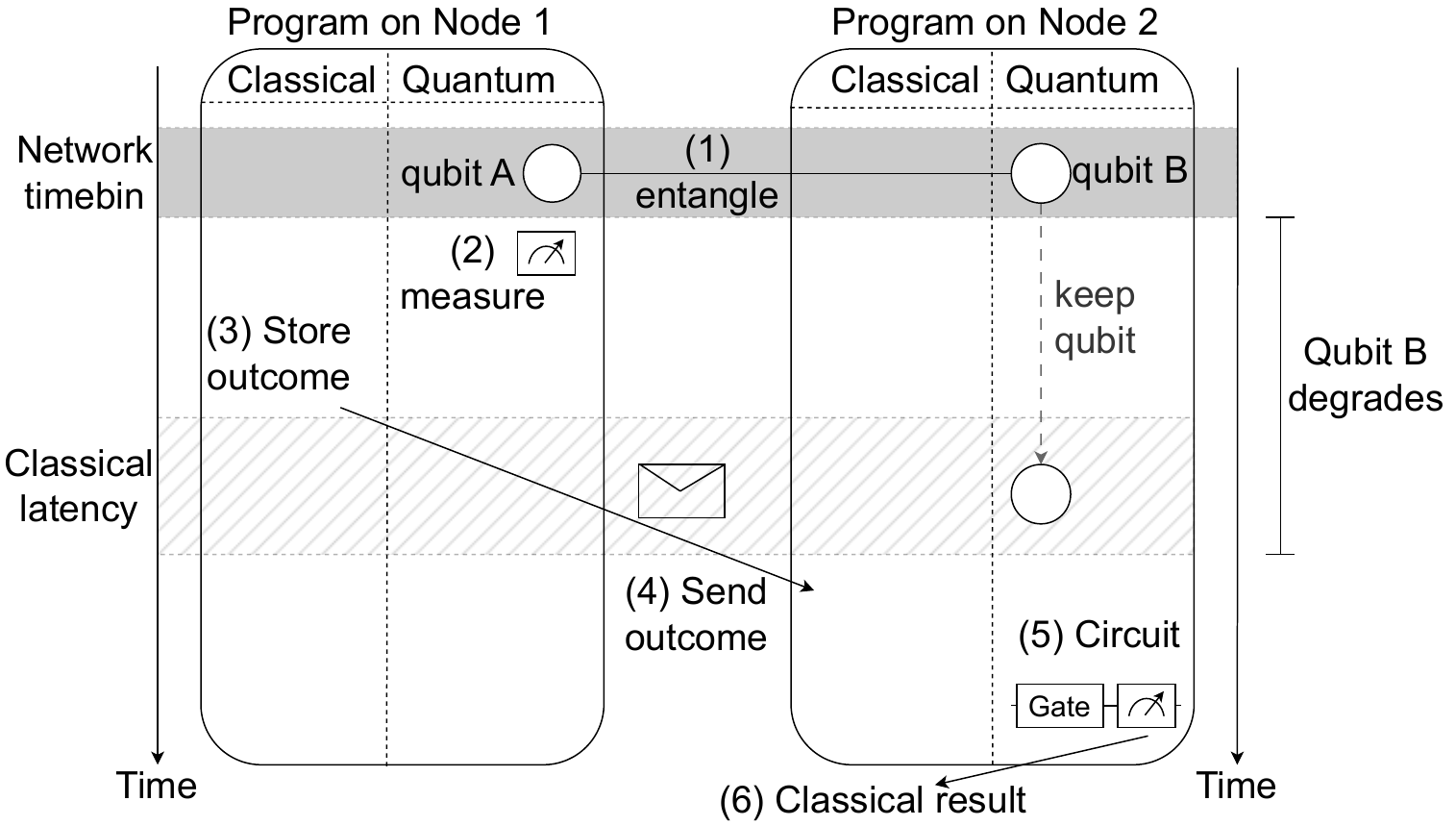}
    \caption{Example application consisting of two hybrid classical-quantum programs (on Nodes 1 and 2) including
        (1) Entanglement generation between two qubits (circles) in a synchronized time slot (defined by  network controller).
        (2) A local measurement of qubit A at Node 1 resulting in a classical outcome bit (destroying the qubit)
        (4) Communication of the classical bit from Node 1 to Node 2 (taking non-deterministic time)
        (5) Execution of a quantum circuit on qubit B at Node 2 depending on the classical bit. The quality of qubit B has degraded during the time elapsed since (1). 
        (6) Node 2 measures qubit B and outputs the classical result.
    }
    \label{fig:program_illustration}
\end{figure}

Advances in quantum computing and quantum communication technologies are paving the way for a \textit{quantum internet}~\cite{wehner2018quantum, kimble2008quantum}, where quantum applications are executed across multiple network nodes.
Examples of such applications include quantum key distribution (QKD)~\cite{bennett2014quantum, ekert1991quantum} and blind quantum computation (BQC) \cite{broadbent2009universal, arrighi2006blind} from a client to a quantum cloud server.
A multi-node quantum internet application is partitioned into separate single-node \textit{programs} (e.g. a client program and a server program in BQC) that run concurrently on different network nodes. To support security sensitive applications, each program performs local classical and quantum computations on its own private node, and programs interact with each other only via classical message passing and entanglement generation. This is in sharp contrast to distributed quantum computing (see e.g.~\cite{caleffi_distributed_2022}), where all nodes can be accessed and controlled by a single program. 

The single-node programs that constitute a quantum internet application are hybrid in nature (see \cref{fig:program_illustration}):
they can contain both classical and quantum operations, and these operations can be both local (executed fully on the node itself) or networked (interacting with another node in the network).
Quantum operations include quantum gates and measurements, e.g. to perform a server computation in BQC, (\textit{local quantum}), and entanglement generation, e.g. to produce classical bits for a secret key in QKD (\textit{networked quantum}).
Entanglement is a special property of two quantum bits (qubits) that forms a key resource for quantum internet applications. 
All quantum operations are executed on quantum processors that can store, manipulate and measure quantum information, where small networks including such processors have been realized using different quantum hardware platforms including, for example,  nitrogen-vacancy (NV) centers in diamond~\cite{pompili2021realization}, and trapped ions~\cite{krutyanskiy2023entanglement}.
Programs also need to perform classical operations, such as message passing (\textit{networked classical}, e.g. a BQC client program sending desired measurement bases to the BQC server), and local classical processing (\textit{local classical}, e.g. post-processing measurement outcomes in QKD).

Realizing the execution of quantum internet applications presents unique challenges (see \cref{sec:design_considerations}): 
First, a program for a quantum internet application is not merely a hybrid of classical and quantum code segments; these segments are also highly \textit{interactive}: classical and quantum code may run concurrently, communicating and influencing each other.
E.g., a quantum circuit (a series of local quantum gates) may ``pause'' halfway, keeping quantum states in memory, and wait for a value from a classical segment (e.g. a classical message from a remote node) before continuing.
This interactivity makes arbitrary quantum network applications more complex than simple prepare-and-measure quantum network protocols that do not require this interactivity, such as QKD.
Quantum memories have limited lifetimes, meaning qubits are subject to decoherence, degrading their quality over time. This introduces the need 
control the joint schedule of the classical and quantum segments of the program to reach desired levels of application performance.

Second, a compiler should be able to optimize the whole program including both classical and quantum code, as well as to provide information that can be used in our architecture to align and inform scheduling decisions. 

Finally, we are faced with a mix of time scales:
on the one hand, entanglement generation requires a very precise network schedule that is agreed ahead of time between the network nodes~\cite{dahlberg2019link}. On the other hand, classical messages are exchanged asynchronously between the nodes without guaranteed message delivery times. This motivates an architecture in which different segments of the system may operate at different levels of timing precision. 

\revision{In~\cite{donne2024design}, we presented QNodeOS, the first architecture for executing arbitrary programs on quantum network nodes.
QNodeOS tackles the above challenges, but we suggested that there is room for improvements in the architecture, including enabling better support for compilation and gaining better scheduling control by putting components on the same board.
In this work, we explore these improvements.
}

\subsection{Main contributions}
\revision{We propose an extension of the QNodeOS architecture~\cite{donne2024design} for program execution on quantum network nodes, called Qoala, that addresses the above challenges.
}
Qoala is an execution environment tailored to programmable quantum internet nodes, accommodating the \textbf{hybrid, interactive, networked, and asynchronous nature} of quantum internet applications. 

\textbf{Unified program format for hybrid-classical quantum programs:}
Qoala defines a unified program format for executables, encompassing classical and quantum (networked and local) code, and defining basic blocks.
This format is suitable for arbitrary quantum network programs up to the most advanced stage~\cite{wehner2018quantum}.
This paves the way for a joint optimization of the classical and quantum code by a compiler.

\textbf{Runtime representation allowing scheduling:} Qoala separates the static unified program format from a runtime representation consisting of \textit{tasks}. 
This paves the way to design and implement algorithms for scheduling the quantum program in order to meet deadlines imposed by decoherence of the quantum memory.  
To provide advice to the scheduler on deadlines to achieve a desired program performance, programs can specify advice for timing and prioritization depending 
on the quantum hardware capabilities of the node. 
The separation of a static program from its runtime tasks also allows for the programmer to define asynchronous code segments, the execution of which is decided by the scheduler alone.
This is the first architecture that allows for effective scheduling control of hybrid interactive classical-quantum programs, thus addressing a critical issues in the successful execution of quantum internet applications.

\textbf{Integration with quantum network stack:}
\revision{Qoala integrates with the existing quantum network stack~\cite{dahlberg2019link}, also present in QNodeOS~\cite{pompili2022experimental}, for realizing entanglement generation between nodes. This opens the door for Qoala to be implemented on such networks.}

\textbf{Implementation in hardware validated simulation:}
We implement the proposed architecture as a \textbf{modular and composable simulator}, which enables the evaluation of different execution strategies and techniques.
The simulation is validated against real-world quantum hardware implementations, opening the door to understand performance tradeoffs and requirements for Qoala's implementation.
Specifically, the simulator allows configuring different hardware parameters, latencies, and software component organizations, to evaluate implementation choices of Qoala in simulation. 

Using the implementation we demonstrate the effectiveness and feasibility of our proposed architecture on different types of quantum hardware, including its ability to schedule and multitask applications using a number of existing scheduling methods (EDF, FCFS).
We continue to examine tradeoffs in the classical and quantum performance metrics of using different types of scheduling approaches. 
We examine Qoala's improvement over NetQASM~\cite{dahlberg2022netqasm} in enabling hybrid classical-quantum compilation possibilities. 
Finally, we study trends in application performance when varying the amount of concurrency, and examine the impact of a network schedule for entanglement generation on the performance of Qoala.

We stress that Qoala is not just a simulator.
Qoala is an architecture for executing quantum network programs, and is not tied to specific implementations.
The simulator implementation of the architecture validates the design and opens possibilities for further research.
However, Qoala is also planned to be implemented as (part of) an operating system running on real (quantum) hardware.

We highlight the role of Qoala in opening the door for computer science research.
We make our simulator available as open source~\cite{qoala2023simulator}, paving the way for computer scientists to conduct further research, e.g., into the design of compilers, or schedulers that can readily be tested using the simulator. 

The remainder of this paper is structured as follows.
\cref{sec:related_work} compares our work to related studies.
In \cref{sec:design_considerations} we explain important context and terminology, followed by considerations that we used to design our architecture (\cref{sec:architecture}).
\cref{sec:implementation} discusses our implementation and \cref{sec:evaluation} provides evaluation results using this implementation.
We conclude and give suggestions on future work (\cref{sec:conclusion}).

\section{Related work}
\label{sec:related_work}

\begin{figure}[t]
    \centering
    \includegraphics[width=\columnwidth]{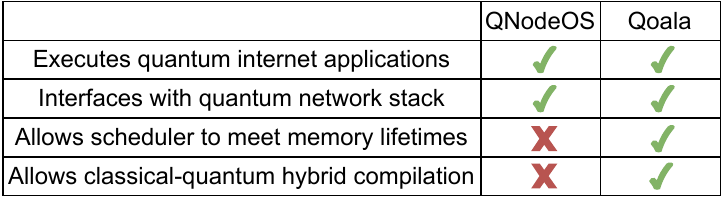}
    \caption{
        QNodeOS~\cite{donne2024design} vs. Qoala capabilities.
    }
    \label{fig:qoala_vs_qnos}
\end{figure}

Networks of quantum processors have been realized using different quantum hardware platforms including, for example, nitrogen-vacancy (NV) centers in diamond~\cite{pompili2021realization}, and trapped ions~\cite{krutyanskiy2023entanglement}.
A first operating system QNodeOS~\cite{donne2024design} including a network stack~\cite{pompili2022experimental} has recently been designed and implemented on real quantum network nodes based on NV centers in diamond.
QNodeOS makes use of the NetQASM execution framework~\cite{dahlberg2022netqasm}, where a classical network processing unit (CNPU) dispatches NetQASM routines for execution by a quantum network processing unit (QNPU).
\revision{Our work builds on top of ideas of QNodeOS and NetQASM, but addresses challenges that were suggested in~\cite{donne2024design}, including the ability to schedule hybrid programs and to optimize over the whole program code.
For example, QNodeOS was unable to have any scheduling control over the joint classical-quantum execution, which could lead to a failure in executing programs successfully (see \cref{fig:qoala_vs_qnos} for a comparison).}
Building on the only such systems that have seen real-world implementation on quantum hardware, opens the door for a later implementation of Qoala on quantum hardware by implementing an improved low-level classical control hardware architecture (\cref{sec:architecture}).

Research has been done on related topics, such as distributed quantum computing, or hybrid (non-interactive) quantum computing. Hybrid classical-quantum programs have been extensively studied in quantum computing, e.g. in the context of \textit{variational quantum eigensolvers (VQE)}~\cite{diadamo2021distributed, liu2022layer} or \textit{quantum approximate optimization algorithms (QAOA)}~\cite{farhi2014quantum}. However, they differ in two important aspects: although they are hybrid, they are not \textit{interactive} during the quantum execution:
(1) classical and quantum segments do not run concurrently, but quantum segments are executed in their entirety before returning to classical segments, i.e. no quantum state is kept in the processor between the execution of different quantum segments.
(2) such hybrid programs lack network interoperability (entanglement generation and classical message-passing between nodes), and also do not have the same timing and flexibility requirements.

Distributed quantum computing~\cite{cacciapuoti2019quantum} shares similarities with quantum internet applications but differs in several aspects.
In the former, complete control is assumed over all participating nodes, such as an application distributed across multiple cores on a single chip~\cite{ovide2023mapping, jnane2022multicore}.
Generally, the capabilities of each core and the latencies between them are fully known, allowing for precise scheduling and orchestration of individual programs running on each core to optimize overall execution.
In contrast, programs in quantum internet applications operate independently (and may even be running on different quantum hardware); therefore they have a degree of autonomy in their own scheduling, and are not fully aware of the actions or timing of other programs.

Entanglement distribution in networks is another related topic that has been extensively studied (see e.g. surveys~\cite{wei2022towards, azuma2021tools}).
However, these works do not deal with executing network applications, and give only predictions for applications in which entanglement is immediately measured (e.g. QKD).

The concept of (soft) deadlines for program execution is of course well known from classical real-time systems that are often used in domains where deterministic and time-critical response is essential, such as automotive, aerospace and medical devices~\cite{liu1973scheduling, hambarde2014survey}, including examples of systems with mixed timing precision~\cite{burns2017survey}.
We draw inspiration from this domain, and the present architecture opens the door to explore algorithms and concepts from this domain to be applied to the execution of quantum internet applications.

\section{Design considerations}
\label{sec:design_considerations}

\subsection{Background and context}
\label{sec:background_context}

\textit{Quantum nodes}.
A quantum internet connects quantum nodes on which quantum programs may be 
executed.
In their most general form, such nodes
are \textit{processing nodes} that have a quantum memory to store quantum bits (qubits) on which quantum operations (qubit initialization, quantum gates and measurements) can be performed. Pairs of nodes can establish \textit{entanglement} between them over a quantum network. Entanglement is a special property of two qubits (an \emph{entangled pair}), where one qubit is stored in the memory of each node. Nodes can also exchange classical messages (e.g. via dedicated classical links or the internet), where no guarantees are assumed on their message delivery times. 

\textit{Programs}.
A program is a series of instructions to be executed by a node.
Instructions can be categorized into four types: local classical processing, classical message-passing, quantum local processing (quantum operations), and remote entanglement generation.
A program can keep classical variables in a classical memory, and quantum variables (qubits) in the node's quantum memory during the execution.
Multiple programs, each running on their own node, together form an \textit{application} (see \cref{fig:program_illustration}), e.g. QKD (two programs, one per node),
or secret sharing~\cite{hillery1999quantum} (a program each on many nodes).
Programs may involve asynchronous operations (e.g. a server awaiting entanglement with multiple clients).

\textit{Network schedule}.
A quantum network stack has been proposed~\cite{dahlberg2019link} and implemented~\cite{pompili2022experimental} that turns entanglement generation into a robust service independent of the quantum hardware platform.
Important for the design of an architecture for the execution of quantum internet applications is that in this stack, the nodes will establish a network schedule of time slots in which they will trigger entanglement generation (due to need to synchronize entanglement generation at the physical layer~\cite{dahlberg2019link} at high-precision (ns)).
This means that once entanglement has been requested from the network, the nodes can use only the slots in the network schedule to produce entanglement between them, imposing constraints on the ability to schedule applications. What's more, in present day systems~\cite{pompili2021realization, krutyanskiy2023entanglement} limitations in the physical devices prohibit the execution of local operations while engaging in network operations (entanglement generation), creating further dependencies between the local quantum execution and entanglement generation. 
As the specifics of network scheduling~\cite{network-scheduling, skrzypczyk2021architecture} are not within scope of this paper,
we assume the existence of a \textit{network controller} that takes application demand for entanglement and issues a network schedule to the nodes. 
A schedule consists of sequential time slots, each with a start time and duration, when the node will trigger entanglement generation.
Nodes are not forced to attempt entanglement in corresponding time slots, and can instead choose to do local processing instead.

\textit{Performance metrics and noise}. Quantum internet applications have classical outcomes that are typically probabilistic in nature:
(1) applications may intentionally do measurements on quantum states that have fundamentally probabilistic outcomes (e.g. quantum cryptography),
(2) in practice, quantum hardware is imperfect (or \textit{noisy}). That is, undesired errors occur
when performing operations (such as gates, measurements, or entanglement generation) or when keeping quantum states in memory for too long.

In many quantum internet applications (e.g. BQC), a single execution of the application can result in failure or success (e.g. a BQC client receives correct measurement results from the server program~\cite{leichtle2021verifying}). Applications are often executed many times, where outcome statistics are computed in order to validate successful execution (e.g. by majority of outcomes).
We consider two metrics:
a \emph{quantum metric} --- the \textit{success probability} of executing a single instance of the application (on average), and a \emph{classical metric} --- the \textit{makespan}, i.e. the average execution time of an application instance.

\subsection{Considerations}
Considerations can be categorized into three main groups: fundamental, technological, and enabling.

\noindent\textit{\textbf{Fundamental Considerations}}
(1) \textit{Hybrid nature of applications (FC1)}: Quantum internet applications inherently consist of both classical and quantum segments, as well as local and networking operations. The execution environment must account for this hybrid nature, and the program structure should accommodate all types of operations.
(2) \textit{Interactive nature of applications (FC2)}: Quantum internet applications require classical communication between nodes. This communication may take place in between classical and quantum segments of a single program. 
This implies the need for application-level interfaces between programs on different nodes, and for interfaces between classical and quantum code segments on a single node.
(3) \textit{Multitasking (FC3)}: Programs may spend a significant amount of time waiting for messages from a remote node (ms), motivating multitasking to make optimal use of the classical and quantum computing resources at each node.  This requires scheduling of time and resources.

\noindent\textit{\textbf{Technological Considerations}}
(1) \textit{Limited qubit lifetime (TC1)}:
Quantum memory quality degrades over time, presenting a significant challenge for the execution environment,
especially in near-term hardware (sub-millisecond to multiple seconds memory lifetimes~\cite{ruf2021quantum, pompili2021realization, krutyanskiy2023telecom}).
As such, there are natural deadlines to application execution after which a desired performance (success probability) can no longer be reached.
We thus desire that a program specification allows indication of memory quality constraints (deadlines), which the runtime environment can act upon (e.g. by appropriate scheduling or restarting).
(2) \textit{Integration of processing and networking (TC2)}: We assume that near-term nodes only have a single quantum processor, which needs to perform both local quantum gates as well as remote entanglement generation.
That is, while performing local operations the processor is blocked from networking operations and vice versa, as is the case for all current implementations~\cite{pompili2021realization, krutyanskiy2023entanglement} but may be mitigated partially using future proposals~\cite{vardoyan2022quantum}. 
The node must hence allocate time for local computation while at the same time adhering to the network schedule which constrains timing of the entanglement operations.

\noindent\textit{\textbf{Enabling Considerations}}
(1) \textit{Different compilation strategies and programming languages (EC1)}: The execution environment should support various compilation strategies and accessible programming languages. 
In order to enable compilation, we furthermore want a representation of the program that can be integrated with existing compiler frameworks.
(2) \textit{Different scheduling strategies (EC2)}: Since we expect that scheduling plays a vital role in optimizing application performance, the execution environment should enable scheduling, and support different scheduling algorithms and policies, 
allowing for their comparison and evaluation.
(3) \textit{Different (control) hardware implementations (EC3)}: The architecture should make minimal assumptions on the classical control hardware, and be independent on the choice of quantum hardware platform~\cite{donne2024design},
allowing for integration with multiple (future) technologies such as NV centers~\cite{pompili2022experimental} or trapped ions~\cite{drmota2023robust}.

\section{Architecture}
\label{sec:architecture}

\begin{figure}
    \centering
    \includegraphics[width=\columnwidth]{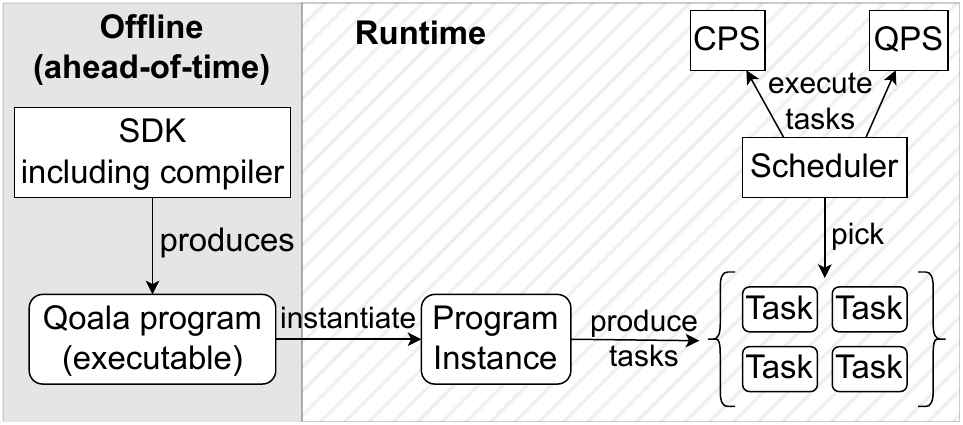}
    \caption{High-level overview of Qoala: An SDK allows program code in a high-level language (e.g. Python).
    A compiler translates this code into a Qoala program (specific compiler not in scope of this work).
    To run, a program is instantiated with concrete values for program arguments.
    Tasks are created for the program instance, which are scheduled and executed by the scheduler.
    Multiple program instances may exist at the same time (both multiple instances of the same or different programs).
    All tasks from all instances are added to a single task graph (\cref{sec:tasks}) used by the scheduler.
    }
    \label{fig:runtime_overview}
\end{figure}

Based on these design considerations, we propose Qoala (see \cref{fig:runtime_overview}), an execution environment for programmable nodes in a quantum internet. 
Provided minimal hardware assumptions are met (\cref{sec:minimal_hardware_assumptions}), 
each node implements its own Qoala execution environment, supporting a specific \textit{program structure} (\cref{sec:program_structure})
and implementing a specific \textit{runtime environment} (\cref{sec:runtime_environment}) that is able to \textit{schedule tasks} (\cref{sec:tasks,sec:program_instantiation,sec:scheduling}).
Details in \cref{app:program_structure,app:runtime_environment,app:scheduling_execution}.

\subsection{Minimal hardware assumptions}
\label{sec:minimal_hardware_assumptions}

\begin{figure}
    \centering
    \includegraphics[width=\columnwidth]{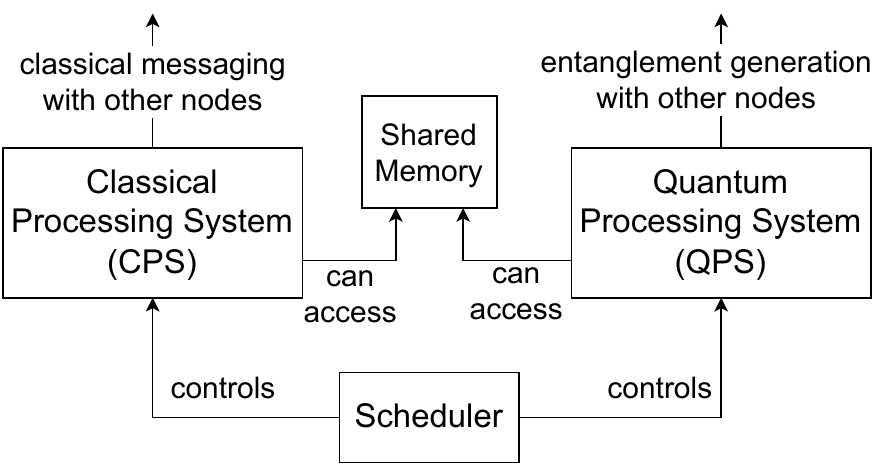}
    \caption{Minimal hardware assumptions for a single node.
    \revision{A Classical Processing System (CPS) can execute classical code and can communicate classical messages with other nodes in the network.
    A Quantum Processing System (QPS) can execute quantum code and can realize entanglement (quantum connections) with other nodes in the network.
    The CPS and QPS are controlled by a scheduler, and have access to shared memory.
    In the QNodeOS architecture~\cite{donne2024design}, the CPS is realized as the CNPU, and the QPS as the combined QNPU-QDevice system.
    For Qoala, we only focus on the classical-quantum distinction, and not the internal implementation (such as a QNPU-QDevice separation), hence the different terminology.}
    }
    \label{fig:minimal_hardware_assumptions}
\end{figure}

Qoala is based on only a few core assumptions on the processing node (consideration EC3, \cref{fig:minimal_hardware_assumptions}):

\textit{CPS-QPS distinction}. We assume the node distinguishes between a \textit{classical processing system (CPS)} managing classical computing resources (e.g. CPU, classical memory and networking), and a \textit{quantum processing system (QPS)}, responsible for executing quantum operations (gates, measurements, entanglement generation) on quantum hardware including a quantum memory as in~\cite{dahlberg2022netqasm, pompili2022experimental}.
\revision{Unlike in the implementation of QNodeOS~\cite{donne2024design},} we assume a shared classical memory is accessible to both the CPS and QPS, enabling communication between the two processing systems, addressing the interactive property of quantum internet programs.
The CPS can act as a fully-fledged classical computer, and performs application-level classical communication with other nodes as well as with a network controller who sets a network schedule.
The QPS can execute routines consisting of low-level quantum gates, basic classical control logic (branching), and entanglement generation.
This opens the door for the QPS to be based on essentially any quantum hardware platform where a specialized microcontroller is used to control the quantum hardware, and a separate microprocessor implements the CPS, where a shared memory could be realized next to the two processors on-chip.
The scheduler controls both CPS and QPS execution, and may physically be realized on either one.

\textit{Time granularity}. Both CPS and QPS are assumed to have knowledge of time, albeit operating with different timing precision ($ms$ precision for CPS mirroring node-to-node communication latencies vs. $\mu s$ and $ns$ precision needed for synchronized entanglement generation~\cite{pompili2022experimental, dahlberg2019link}.)

 \textit{Network stack}. A quantum network stack including a network layer~\cite{dahlberg2019link} is implemented on the node with which Qoala can interface. This stack can receive and fulfill requests for remote entanglement generation.

\subsection{Program structure}
\label{sec:program_structure}
Qoala defines a hybrid format for programs, mapping naturally to their hybrid nature (consideration FC1 in \cref{sec:design_considerations}).
A Qoala program is a combination of quantum and classical instructions, organized into three main sections:
    \textit{host code} (containing classical instructions),
    \textit{local routines} (containing local quantum instructions), 
and \textit{request routines} (for remote entanglement generation).
This hybrid format allows a compiler to optimize the whole program, including critical code paths with dependencies between classical and quantum segments.
Local routines and request routines can be triggered from within host code as function calls, addressing the interactivity between them (consideration FC2).

A Qoala program is an \textit{executable} and output of a compiler.
The format is separate from any high-level language in which a programmer might write code; hence Qoala in theory allows for compatibility with any such language (consideration EC1).
Entry and exit points of a program are in host code.
\Cref{fig:example_program} shows an example program in text format.
We contrast Qoala's program format with that of~\cite{dahlberg2022netqasm}, in which there was no way to compile across classical and quantum code segments.
A Qoala program has \textit{program arguments} that are filled in during program instantiation (\cref{sec:program_instantiation}).

\textit{Host code}.
Host code, executed on the CPS, encompasses local computation, control-flow, inter-node messaging, and can initiate local and request routines.
For example, in a program that is part of a QKD application, classical post-processing (including sending bases, local error correction, and privacy amplification \cite{vidick2023introduction}) would be represented in host code.
Host code is structured as a sequence of \textit{blocks}, each holding a list of instructions.
Blocks dictate control-flow by ending with a (conditional) jump instruction (default: next block in the sequence).
This block division not only facilitates task creation and scheduling (see \cref{sec:tasks}) but also streamlines compiler integration (which may use blocks in its intermediate representation).
Blocks can contain metadata about their expected \textit{duration}, (relative) deadlines, and they may be inside \textit{critical sections}, encompassing a sequence of blocks with a maximally allowed execution duration. This metadata is propagated to corresponding tasks and used by the scheduler in order to mitigate quantum decoherence due to limited qubit lifetime (consideration TC1).
Asynchronous execution is possible by `submitting' multiple routines for execution, and waiting for all of them to finish. At runtime, the scheduler can decide in which order to execute the routines.

\textit{Local routine}.
A local routine (LR) represents a series of quantum operations (like gates and measurement), to be executed by the QPS locally (no interaction with external nodes or controllers).
An LR may also contain limited classical computation and control-flow code allowing for fast feedback, which can increase quantum performance (\cref{sec:background_context}) due to less decoherence.
An updated version of NetQASM~\cite{dahlberg2022netqasm} is used to represent the instructions, which allows both hardware-specific and hardware-agnostic instructions.
Therefore, the program format is compatible with different quantum hardware.
In contrast to \cite{dahlberg2022netqasm}, Qoala's version of NetQASM does not have instructions for entanglement generation (cleanly separating local and networked quantum operations)
nor `wait' instructions. This allows routines to be treated as atomic non-preemptable blocks.

\textit{Request routine}. A request routine (RR) consists of a request for entanglement generation with another node, and represent requests to the node's quantum network stack.
It can have local routines as callbacks, allowing quick local (quantum) processing of entangled qubits on the QPS without returning to the CPS, decreasing waiting time and decoherence.

\subsection{Runtime environment}
\label{sec:runtime_environment}
The Qoala runtime environment provides various resources that programs can leverage during execution.

\textit{Exposed Hardware Interface (EHI)}.
The Exposed Hardware Interface provides information about the hardware and software capabilities and restrictions of the node and the network,
like available quantum memory and expected latencies.
Each node provides their own EHI which is used in capability negotiation (see below), and allows a choice of executable code optimized by a compiler for those capabilities ahead of time. 

\textit{Shared memory}.
To address the classical-quantum interactivity in programs, the CPS and QPS share data with each other via \textit{shared classical memory}.
Write conflicts are avoided by explicit read/write rules for shared memory regions (see \cref{app:shared_memory}).
Our conceptual model of a shared memory leaves open different implementation choices, including a physical shared memory or a message-passing protocol.
Calls in host code to local or request routines use the shared memory to communicate routine arguments and results.

\textit{Quantum memory}.
Quantum memory is organized into a \textit{virtual quantum memory space (VQMS)} for each program instance (see \cref{sec:program_instantiation} for instantiation), represented as Unit Modules~\cite{dahlberg2022netqasm}.
Qoala maps each VQMS to the physical qubits available in the QPS.
VQMS information like qubit connectivity and noise characteristics is provided by the EHI, which a compiler can use to optimize a program.
The VQMS enables multitasking since programs have their own runtime context, while a scheduler (\cref{sec:scheduling}) sees the whole physical memory space and can schedule programs accordingly.

\begin{figure}
    \centering
    \includegraphics[width=\columnwidth]{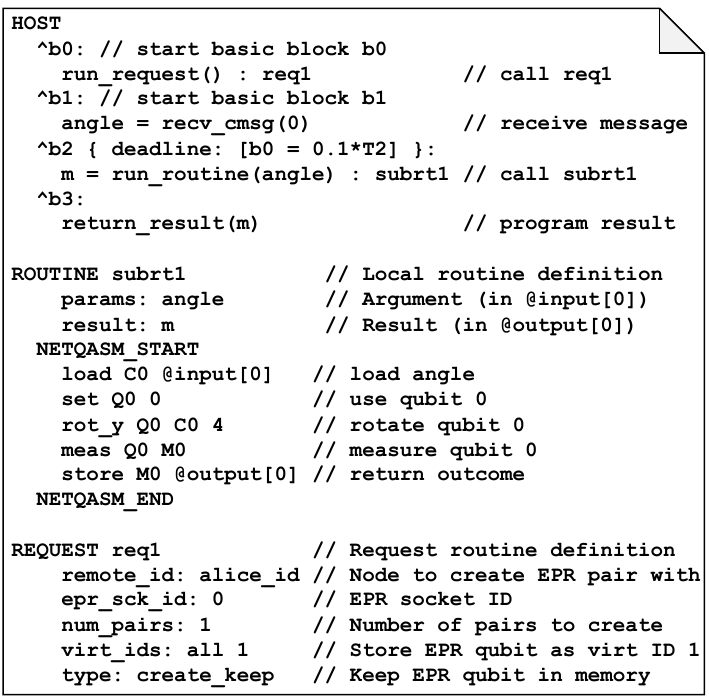}
    \caption{
        Example Qoala program containing a host section with 4 blocks, a local routine (\texttt{subrt1}),
        and a request routine (\texttt{req1}). Block \texttt{b2} has a relative deadline to \texttt{b0} of $0.1$ times qubit noise parameter $T_2$.
    }
    \label{fig:example_program}
\end{figure}

\textit{Remote interaction}. For interaction with programs on remote nodes,
the runtime provides \textit{classical sockets} and \textit{EPR sockets} based on~\cite{dahlberg2022netqasm}. Host code uses classical sockets for sending and receiving messages; EPR sockets are indicated in request routines (see e.g. \cref{fig:example_program}).

\begin{figure*}
    \newcommand{\figheight}{6.5cm}
    \centering
    \subfloat[\centering \label{fig:task_types}]{{\includegraphics[height=\figheight, keepaspectratio]{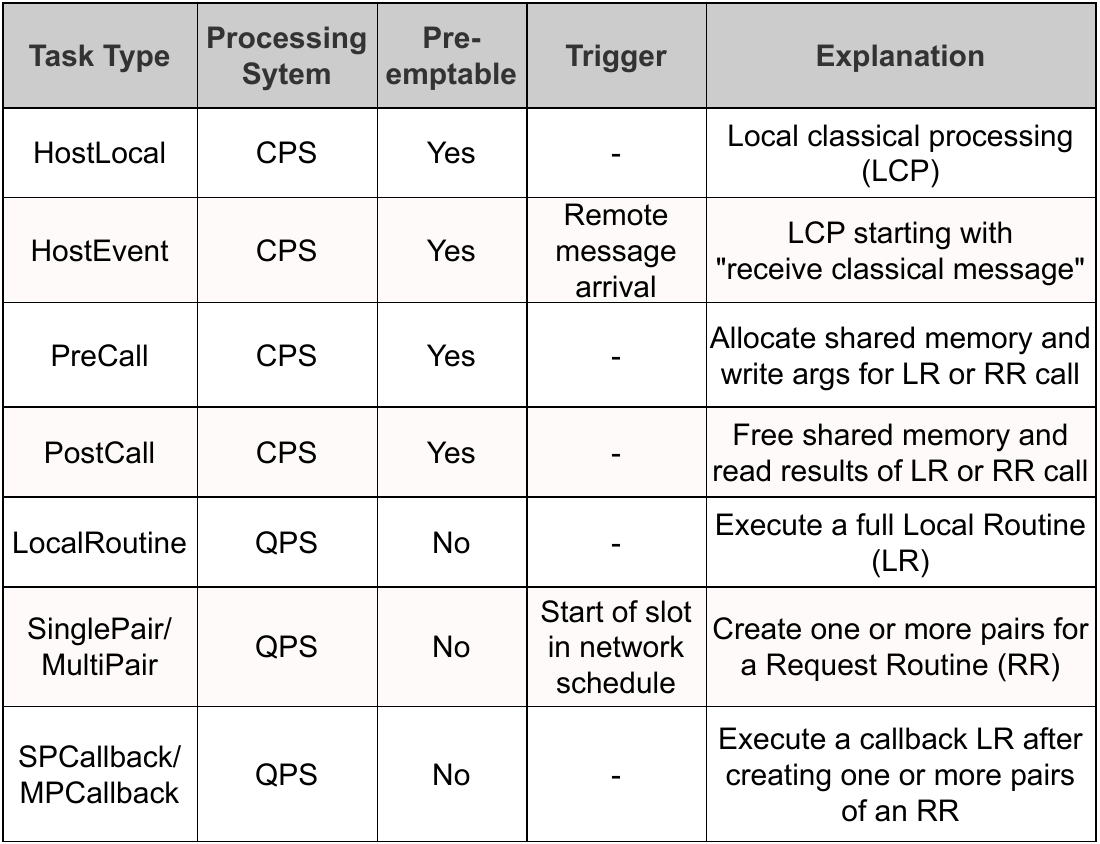}}}%
    \hfill
    \subfloat[\centering \label{fig:task_creation_cutout}]{{\includegraphics[height=\figheight, keepaspectratio]{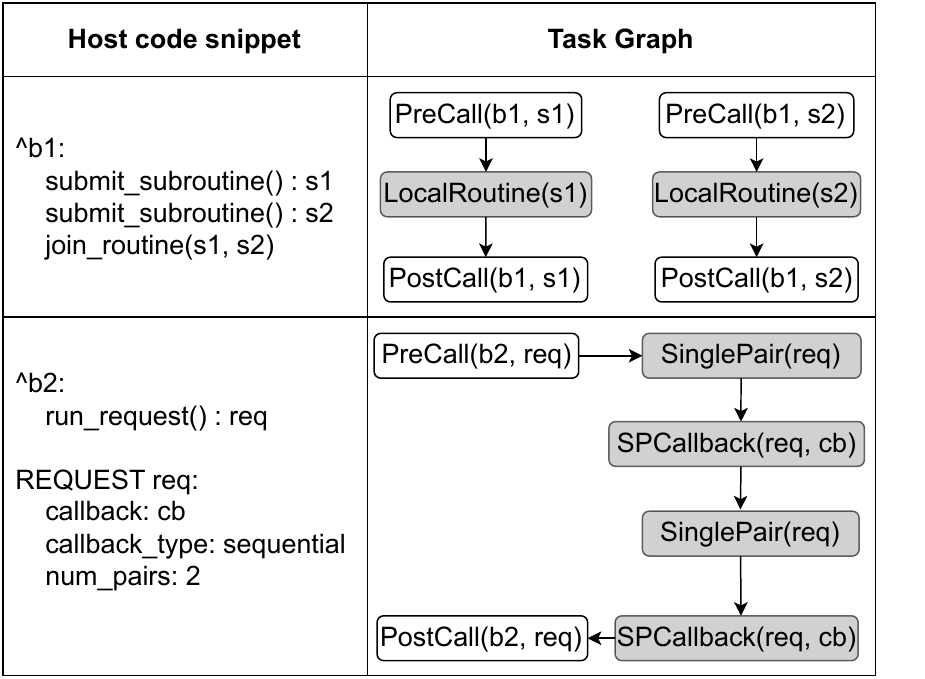}}}%
    \caption{
        (a) Overview of task types.
        (b) Examples of host code and corresponding task graphs.
        Shaded tasks are executed by the QPS, the others by CPS.
        Top: asynchronous submission of local routines \texttt{s1} and \texttt{s2}.
        The graph consists of two separate chains of tasks and the scheduler can choose in which order to execute these chains (possibly interleaved).
        Bottom: a request routine uses the callback entry to immediately add tasks for executing local routine \texttt{cb} after each entangled pair generation.
    }%
    \label{fig:tasks}
\end{figure*}

\subsection{Tasks}
\label{sec:tasks}
We introduce \textit{tasks} to enable multitasking (consideration FC3).
Each task represents a code segment of a running program with a context of runtime variables.
Tasks have different types (\cref{fig:tasks}) based on the code they represent.
By splitting a program into distinct executable tasks, 
we can utilize the parallel execution on the CPS and QPS (by assigning tasks to the corresponding system),
and we can interleave execution of multiple programs by filling waiting times of one program by execution of tasks of another.
Code segments indicated to run asynchronously (\cref{sec:program_structure}) can also be represented by tasks, the execution order of which can then be governed by a scheduler.
Further, tasks enable interleaving of local operations and quantum network (entanglement generation) operations.
A scheduler can choose when to execute entanglement tasks (with strict timing requirements from the network schedule)
and when to execute local tasks (less strict requirements), addressing consideration TC2.

\textit{Task graph}.
Tasks are organized in a \textit{task graph}, a directed acyclic graph (DAG) where each node represents a single task.
Edges can be \textit{precedence constraints} (task A must conclude before task B initiates) or \textit{relative deadlines} (task B should start within maximum duration $t$ after completion of task A).
Using a task graph introduces a well-defined and isolated scheduling problem: given a graph of tasks, which task(s) should be executed next?
Deadlines are used to assist the scheduler (see below) in mitigating the gradual quality degradation of quantum states over time (decoherence) by choosing appropriate tasks.
Some tasks are only enabled after certain events happen.
\texttt{HostEvent} tasks are enabled by an incoming classical message and \texttt{SinglePair} or \texttt{MultiPair} tasks are enabled by network schedule timestamps.
Tasks also have information about what quantum memory they use, helping the scheduler decide which tasks it can execute at a given time.

\textit{Task creation}. 
A task is created for a segment of a running program.
If a program segment is executed multiple times (e.g. because of a loop in the code), this results in multiple tasks.
A host code block is translated into a \texttt{HostLocal} task (block contains only local instructions) or a \texttt{HostEvent} task (block starts with a `receive message' instruction).
A local routine call is represented by (1) a \texttt{PreCall} task (CPS allocates shared memory and writes routine arguments), (2) a \texttt{LocalRoutine} task (QPS executes routine), and (3) a \texttt{PostCall} task (CPS reads routine results from shared memory).
Request routine calls are handled similarly (with \texttt{SinglePair} or \texttt{MultiPair}).
\texttt{MultiPair} tasks can be more time- and resource efficient since the network stack can handle multiple pair generations at once.
Callback tasks for local routines acting as entanglement generation callbacks allow quick successive execution.
For each task, its expected duration is calculated based on the metadata of the corresponding block or routine in the program, together with information from the EHI (see below).
See \cref{fig:task_types} for task types and \cref{fig:task_creation_cutout} for examples of host code and corresponding tasks (details in \cref{app:scheduling_execution}).

\subsection{Program instantiation}
\label{sec:program_instantiation}
A program is part of an application that uses entanglement generation orchestrated by a network controller (\cref{fig:program_illustration}).
Therefore, before execution, the program must align with the other programs of its application as well as with the network controller.
(1) \textit{Capability negotiation and entanglement demand registration}.
First, all collaborating nodes exchange their EHI and agree on concrete values for deadlines and task duration estimations (using advice pre-computed by the compiler).
These values are needed to do effective scheduling at runtime.
Second, the nodes together register their entanglement demands to the network controller, which then creates a \textit{network schedule} based on these.
This schedule consists of time slots, each of which is assigned to an individual \textit{application instance} (tuple of program instances, one per node).
(2) \textit{Program instantiation}. Concrete values for program arguments can be filled in such as deadlines, durations and program-specific input values.
Typically, for a given application, the involved nodes create many \textit{program instances} of the same program (to gather statistics, \cref{sec:background_context}).

\subsection{Scheduling and execution}
\label{sec:scheduling}
Tasks produced for program instances are executed by the \textit{node scheduler}.
This scheduler manages a global task graph containing all tasks that have been created for instantiated programs and that are awaiting execution.
Among the tasks that do not have any precedence constraints going into them (anymore), the scheduler continuously chooses the next task(s) to execute.
It may choose to run a task on the CPS and a task on the QPS in parallel.
If a task completed successfully, it is removed from the task graph, and precedence constraints and relative deadlines are updated accordingly.
Based on the control flow of the program that this task was for, new tasks may be created representing the next segment of the program.
These tasks are then added to the task graph.
If a task failed (for example, entanglement generation did not succeed for a \texttt{SinglePair} task), it either (a) remains in the task graph and may be scheduled again at a later time,
or (b) the whole program instance is aborted, depending on the scheduler implementation.
For \textit{predictable programs} (where control-flow and hence all corresponding tasks are known beforehand), their entire task graph may be created ahead of execution
and (no need to add new tasks at runtime).
Tasks for entanglement generation (like \texttt{SinglePair}) additionally contain information about when they are allowed to start according to the network schedule,
allowing the scheduler to make sure that the network schedule is respected.
The scheduler allows pre-emption of CPS tasks.
For instance, the arrival of a message from a remote node might activate a \texttt{HostEvent} task with high priority;
if the CPS was executing another lower priority task, it may be pre-empted and resumed at a later time.
Since quantum tasks cannot in general be rolled back or resumed (e.g. measurements are destructive and cannot be undone), Qoala does not allow the pre-emption of QPS tasks.
Although we define a scheduling problem, and a framework for designing and implementing scheduling algorithms,
we on purpose do not prescribe an explicit implementation and leave the question of an optimal scheduling approach open for further research (consideration EC2, see also \cref{sec:conclusion}).

\section{Implementation}
\label{sec:implementation}
We implement our architecture in the form of an open-source simulator~\cite{qoala2023simulator}.
Implementation on real hardware requires developing new classical control hardware which is outside the scope of this work.
The simulator is built on top of NetSquid~\cite{coopmans2021netsquid} which can simulate quantum behavior as well as asynchronous classical processes.
Specifically, NetSquid provides detailed configuration allowing for simulations of hardware with parameters that are validated in real experiments, not possible using other simulators such as QuNetSim~\cite{diadamo2021qunetsim}, QNET~\cite{fang2023quantum}, and QuISP~\cite{satoh2022quisp}.
SquidASM~\cite{squidasmrepo} simulates the software and hardware stack used in the NetQASM/QNodeOS system mentioned in \cref{sec:related_work}, and hence misses the scheduling capabilities that we introduce in Qoala.

The simulator has on purpose been made modular and composable:
components of Qoala's architecture (like CPS, QPS, scheduler, shared memory) are provided by the simulator as building blocks that can be configured and put together in different ways (details in \cref{app:simulator}).
Both classical software parameters and quantum hardware noise models can be configured.
In this way, the simulator allows one to investigate different architecture and parameter choices.
In the simulator, a network of quantum nodes implementing Qoala can be constructed, and Qoala programs can be submitted for execution to these nodes.
Static network schedules can be provided (capability negotiation and automatic network schedule creation are not simulated).
The simulator then executes the programs, providing application results and statistics.
Our implementation allows researchers to not only test Qoala, but also configure parameters and architectures to investigate scheduling algorithms and hardware implementation choices.

\subsection{Scheduler implementation}
\begin{figure}
    \centering
    \includegraphics[width=\columnwidth]{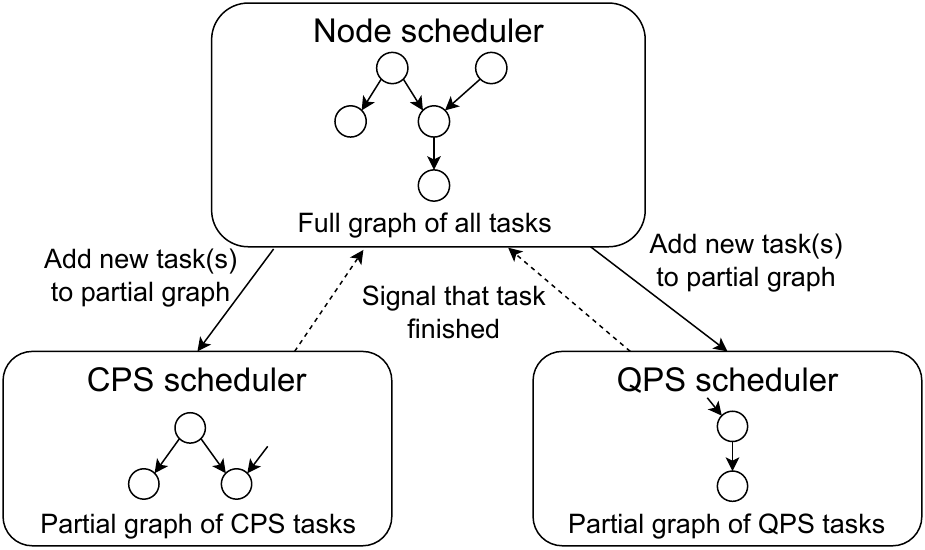}
    \caption{Overview of our hierarchical scheduler implementation.
    The node scheduler maintains a graph of all tasks. The CPS and QPS maintain partial graphs with only tasks they can execute themselves. Partial graphs are updated by the node scheduler. 
    The CPS scheduler has access to a buffer with classical messages from other nodes, activating \texttt{HostEvent} tasks. The QPS scheduler has access to the network schedule, determining allowed start times of \texttt{Pair} tasks.}
    \label{fig:scheduler_impl}
\end{figure}

In our implementation, we use a two-level hierarchical scheduler architecture,
consisting of a node-wide \textit{node scheduler} which controls two \textit{processor schedulers}, one for the CPS and one for the QPS (\cref{fig:scheduler_impl}, details in \cref{app:scheduling_execution}).
Such an approach has been used in other contexts not related to quantum networks~\cite{polychronopoulos1991hierarchical, girkar1994hierarchical}.

Each scheduler maintains their own task graph.
The node scheduler task graph contains all tasks (CPS or QPS) that are to be executed.
Each processor scheduler task graph is a partial copy of the node scheduler task graph containing only the tasks that can be executed by its own processor.
Edges in the node scheduler graph between heterogeneous tasks (i.e. between CPS and QPS tasks) are represented in the partial processor graphs by an \texttt{external-dependencies} node attribute. 
When a processor scheduler finishes a task, it is removed from the task graph and a signal is sent to the node scheduler.
The node scheduler updates its own task graph accordingly, and may then add new tasks to the task graph of the processor scheduler.
Write conflicts on the processor task graphs are avoided since tasks can only be added by the node scheduler, and tasks can only be removed by the processor scheduler.

The processor schedulers support both a first-come-first-serve (FCFS) and an earliest-deadline-first (EDF)~\cite{silberschatz2020operating} scheduling mechanism.
In our evaluation (\cref{sec:evaluation}), deadlines are used as \textit{soft deadlines}, i.e. there is no guarantee about meeting deadlines.

\section{Evaluation}
\label{sec:evaluation}
All simulations were run on a machine using 80 Intel Xeon Gold cores at 3.9 GHz and 192 GB of RAM.
The specific code and data used for the results in this section can be found at~\cite{evaluation-data}.
Each subsection describes an independent evaluation (details in \cref{app:evaluation}):

\subsection{Demonstrating the architecture's effectiveness}
\label{sec:demonstrating_architecture_effectiveness}
We first validate the functionality of our architecture by demonstrating that applications of different CPS-QPS interaction types can successfully be executed on two or more nodes.
Using our implementation (\cref{sec:implementation}), we report that we successfully simulated the following applications:
(A1) quantum key distribution (2 nodes, first QPS generating $10^3$ EPR pairs followed by only CPS actions (classical computation and messaging)),
(A2) blind quantum computation (1 client and 1 server node, first QPS generating 2 EPR pairs, then CPS performing rounds of classical messaging followed by local quantum gates by QPS),
(A3) single-qubit teleportation across two nodes (1 sender and 1 receiver node, QPS generating one EPR pair followed by QPS measurement by the sender, CPS classical messaging and QPS local quantum gates by the receiver),
(A4) a ping-pong application which repeats the single-qubit teleportation application to transfer states back and forth,
and (A5) a multi-node GHZ-state~\cite{greenberger1989going} creation application (3 nodes, QPS creating a tripartite entangled state using multiple EPR pairs, using CPS classical messaging and QPS local quantum gates).

Each program is instantiated 1000 times and all tasks are immediately added to the task graph (since the the programs are \textit{predictable} (\cref{sec:scheduling})).
Precedence constraints are added such that instances are executed sequentially for simplicity.
We use a fixed network schedule (no demand registration (\cref{sec:program_instantiation}) since network schedule generation is not handled by Qoala itself and hence not part of the evaluation).
To demonstrate the hardware independent performance of Qoala, all simulations are performed on three different hardware models: a generic quantum platform (uniform qubit connectivity and \textit{vanilla} NetQASM instruction set~\cite{dahlberg2022netqasm}),
and two models based on data validated on real hardware (NV centers~\cite{bradley2019ten, hermans2022qubit} and trapped ions~\cite{krutyanskiy2023entanglement}). 
We observe successful execution (desired deterministic outcome when setting noise parameters (\cref{sec:background_context}) to 0, and expected non-deterministic outcome distributions with realistic noise parameters) for all types of applications (details in \cref{app:evaluation}).

\begin{figure}
    \centering
    \includegraphics[width=1.0\columnwidth]{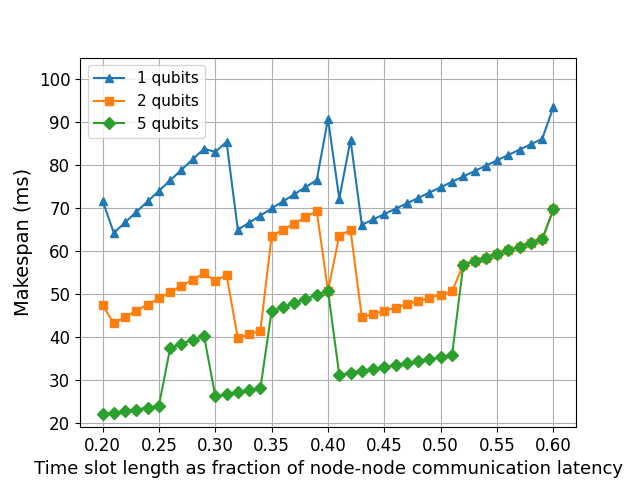}
    \caption{Self-preemption of a teleportation program.
    For certain durations of the time slot length (as fraction of node-node communication latency, x-axis), the makespan is considerably higher (spikes in the plot).
    Reason: a classical message arrives for some teleportation instance $i$,
    making the node scheduler choose to perform the local quantum gates for $i$. During this,
    the time slot for instance $j > i$ starts. Since the QPS is busy with $i$, it cannot work on entanglement
    generation for $j$. Therefore, $j$ must wait for the next repetition of the network schedule, leading to a higher overall makespan.
    }
    \label{fig:teleport_self_preemption}
\end{figure}

\subsection{Demonstrating Qoala's multitasking potential}
Next, we demonstrate that Qoala can execute multiple instances of (different) programs concurrently by interleaving. We examine (1) makespan decrease (\cref{sec:background_context}) when interleaving the instances compared to sequential execution, 
(2) whether makespan depends on the network schedule.

We first evaluate multitasking instances of the same application: teleportation (same as A3 in \cref{sec:demonstrating_architecture_effectiveness}), 100 instances, with a fixed network schedule (no time slots; entanglement generation always allowed).
Sequential scheduling of instances results in makespan $N \cdot CC$
while interleaved scheduling (tasks for all instances created at the same time; no precedence constraints between instances) results in $\left\lceil N / Q \right\rceil \cdot CC$
(number of instances $N$, classical node-node communication latency $CC$, number of available memory qubits at receiving node $Q$).
We also evaluate the effect of network schedules with time slots (repeating pattern of slots assigned to A3 instances), and find that the time slots length influences the makespan (\cref{fig:teleport_self_preemption}) in a non-trivial manner due to instances pre-empting each other.
BQC (same as A2 in \cref{sec:demonstrating_architecture_effectiveness}, 100 instances) interleaved gives a makespan decrease over sequential of ($21\%, 56\%, 65\%$) for (2, 5, 10) server qubits, respectively.
The network schedule affects the makespan decrease: doubling the time slot length results in a smaller decrease ($12\%, 48\%, 48\%$). 

We then execute instances of different applications and again examine the effect of the network schedule on the makespan decrease: 50 QKD (A1 in \cref{sec:demonstrating_architecture_effectiveness}) and 50 BQC (A2) instances give a makespan decrease of $9.5\%$ (fixed schedule which first has time slots for QKD and then for BQC) and $39\%$ (schedule with time slots alternating between QKD and BQC).
We observe that multitasking can lead to improved (lower) makespan and that the network schedule can have considerable impact on the makespan.

\begin{figure}
    \centering
    \includegraphics[width=1.0\columnwidth]{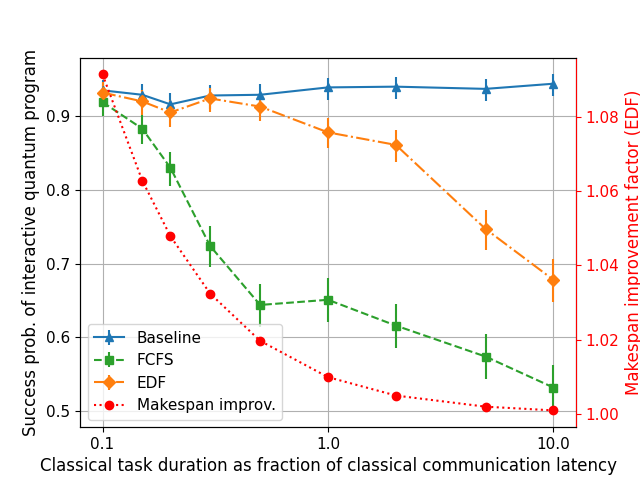}
    \caption{Execution of interactive quantum program in the presence of a `busy' CPS program (tasks with duration $f \cdot CC$ for fraction $f$ of the classical node-to-node latency $CC$, x-axis).
    \textbf{Comparison of schedulers:}
    $Baseline$ (no scheduling nor interleaving),
    $FCFS$: first-come-first-serve scheduler (interleaving possible, no deadlines to prioritize quantum tasks),
    $EDF$: earliest-deadline-first scheduler (with deadlines to prioritize quantum tasks).
    The interactive program regularly waits (duration $CC$), with quantum states in memory, for incoming classical messages. 
    Task interleaving allows busy CPS tasks to fill waiting times.
    EDF leads to higher success probability than FCFS, showcasing usefulness of deadlines.
    \textbf{Tradeoffs:}
    The baseline of sequential execution leads to the best possible success probability (quantum metric) at the expense of longest makespan. EDF allows a lowering of makespan (classical metric) at the expense of a lower succ. prob. (quantum metric). 
    }
    \label{fig:eval_tradeoffs_cq}
\end{figure}

\subsection{Improvement over NetQASM architecture}
\label{sec:improvement_over_netqasm}
We compare the Qoala architecture with the NetQASM runtime approach for executing programs on a node from~\cite{dahlberg2022netqasm}
and show that Qoala provides new compilation possibilities (optimizing across classical and quantum code) and can lead to a better application execution makespan.
We consider a remote measurement-based quantum computing program written in Python (the program format of the NetQASM runtime) which has suboptimal code logic on purpose.
Executing this program in the NetQASM runtime performs worse (success probability $66\%$) than the same program but compiled manually into a Qoala program and executed in the Qoala runtime (succ. prob. $82\%$).
We note that manual compilation allowed optimization that is not possible in the NetQASM program format, exemplifying the new compilation potential provided by Qoala.

\begin{figure*}
    \newcommand{\heatmapheight}{6.6cm}
    \centering
    \subfloat[\centering \label{fig:heatmap_teleport}]{{\includegraphics[height=\heatmapheight, keepaspectratio]{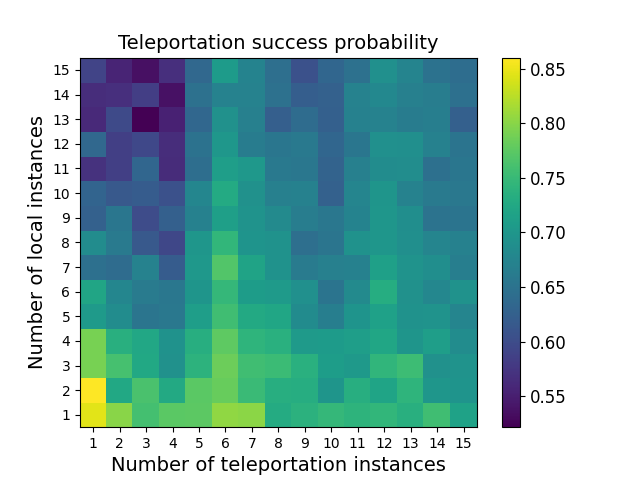}}}%
    \hfill
    \subfloat[\centering \label{fig:heatmap_local}]{{\includegraphics[height=\heatmapheight, keepaspectratio]{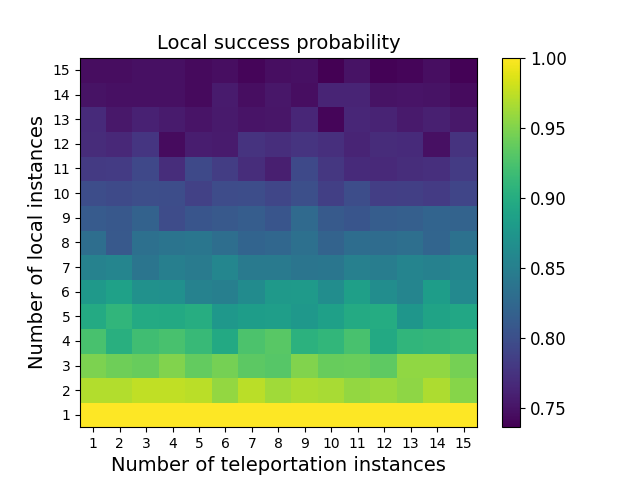}}}%
    \caption{
        Concurrent execution of teleportation (A3) and a local application (only preparing and measuring qubits).
        (a) Success probability of teleportation for different numbers of teleportation and local instances.
        More local instances lead to lower teleportation succ. prob. (effect more pronounced with few teleportation instances).
        (b) Success probability of local program. More local instances lead to lower local succ. prob., independent of the number of teleportation instances.}%
    \label{fig:quantum_multi_tasking}%
\end{figure*}

\subsection{Tradeoffs between classical and quantum performance metrics}
\label{sec:effectiveness_of_task_splitting}
We compare different scheduling modes enabled by Qoala and evaluate tradeoffs between makespan and success probability, noting that the NetQASM runtime did not allow scheduling at all (\cref{fig:eval_tradeoffs_cq}).
We expect that interleaving of tasks reduces the makespan, but may lead to lower success probability due qubits degrading in memory while tasks wait for each other.
We compare 3 scheduling modes: no scheduling (baseline), FCFS scheduling, and EDF scheduling.
We consider a simple runtime scenario with 
(1) a local quantum program which alternates between doing local quantum gates and waiting for a remote classical message before continuing and 
(2) a classical `busy program' consisting only of CPS tasks (duration defined as fraction of classical node-node latency).
We find that
(a) scheduling (FCFS or EDF) decreases success probability (EDF less than FCFS); impact larger for long task durations, but
(b) EDF provides a better makespan than no scheduling.
Note that the baseline necessarily gives the highest success probability due to no waiting, but at the expense of maximal makespan (sequential execution).

\subsection{Success probabilities with quantum multitasking}
\label{sec:quantum_multitasking}
Next, we consider a quantum multitasking scenario where we investigate trends in application success probability while varying the number of concurrent applications (\cref{fig:quantum_multi_tasking}).
In addition to a teleportation application (A3 in \cref{sec:demonstrating_architecture_effectiveness}), the receiver node also executes multiple instances of a local quantum program (only applying quantum gates).
Whenever the receiver node must wait for classical messages to come in for A3, it can work on its local quantum programs.
We find that success probability of both types of programs decreases in the presence of another program.

\subsection{Performance sensitivity}
Finally, we investigate the influence of classical message-passing latencies, internal latencies, and network schedule contents on application success probability of BQC (A2, 100 instances).
We find that the duration of sending classical messages between nodes has a large impact on the success probability:
node-node latencies [$0.01$, $0.1$, $1$] times the qubit coherence time lead to success probabilities [$0.89(2)$, $0.83(2)$, $0.54(4)$], respectively.
Internal latencies (between CPS and QPS, and between the scheduler and CPS or QPS) only have a significant impact when message-passing durations are low (0.01 times the qubit coherence time).
We also compare different network schedules (simple linear repeating schedule where each client-server pair gets a time slot consecutively; slot length is varied).
We obtain success probabilities [$0.90(2)$, $0.69(3)$, $0.48(4)$] for time slot lengths [$0.01$, $0.1$, $1$] times the qubit coherence time, respectively.

\section{Conclusion}
\label{sec:conclusion}
Qoala is the first architecture for executing quantum applications that addresses the need for scheduling and compiling hybrid classical\hyp{}quantum programs for a quantum internet.
This allows Qoala to ensure successful execution of quantum programs even in the presence of limited quantum memory lifetimes, and opens the door for compile-time optimizations of hybrid classical-quantum programs.
By building on an existing quantum network stack~\cite{dahlberg2019link, pompili2022experimental} and the implementation of QNodeOS on quantum hardware~\cite{pompili2022experimental, donne2024design} we pave the way for the real-world implementation of Qoala in a platform-independent way on diverse hardware platforms including NV centers in diamond~\cite{pompili2021realization, pompili2022experimental}, or trapped ions~\cite{krutyanskiy2023entanglement,krutyanskiy2023telecom} quantum processors. 
Such an implementation would require, however, a new classical control hardware as opposed to~\cite{pompili2022experimental, donne2024design}, e.g. by placing CPS and QPS on a single board with access to an on-chip shared memory. 

Our simulator implementation already now opens the door for further computer science research in executing quantum internet applications:

\textit{Advanced scheduling algorithms:}
More sophisticated scheduling strategies may lead to higher success probabilities and lower makespan when concurrently executing multiple program instances, where inspiration may come from~\cite{topcuoglu2002performance, baruah2011scheduling, andersson2006multiprocessor, polychronopoulos1991hierarchical}. 
In the quantum domain, missing the deadline will result in a degradation of the success probability as a function of the time by which the deadline was exceeded.
This suggest the use of time-utility functions (TUF, see e.g.~\cite{jensen1993timeliness, li2004utility}) to inform scheduling decisions, where it is an open question how such TUF could even be defined in the quantum domain.
Our work also raises the question on what fundamental tradeoffs between the classical (makespan) and quantum (success probability) performance metrics are at all possible.

\textit{Compiler design:}
Qoala's program format now allows for a compiler design that takes into account the hybrid and networked nature of programs.
It is an open question to design compilers enabling effective code optimization and translation of different types of high-level code into executables.

\textit{Capability negotiation:}
We assumed that the compiler provides advice that the nodes use in a capability negotiation and demand registration (\cref{sec:program_instantiation}).
It is an open question how to best compute such advise, and find efficient protocols for negotiating capabilities and register demand.

\textit{Network schedule:}
As expected, our evaluation shows that application performance depends on the network schedule, where we emphasize that ensuring network service is out of scope for Qoala as en environment for executing applications.
This highlights a need for understanding the quality of service a quantum network should provide, as well as to design good network scheduling algorithms to satisfy them, in order to achieve good application performance.

\section{Acknowledgements}
This research was supported by the Quantum Internet Alliance through the European Union's Horizon 2020 program under grant agreement No. 820445 
and from the Horizon Europe program grant agreement No. 101080128.
We furthermore acknowledge support from NWO (including a VICI grant).

We thank 
Przemysław Pawełczak,
Michele Amoretti,
Anabel Ovide,
Thomas Beauchamp,
Álvaro Gomez Iniesta,
Ingmar te Raa,
Diego Rivera and
Francisco Silva
for useful discussions and feedback on (early) drafts.

\section{Data availability}
The implementation of Qoala as a simulator can be found online~\cite{qoala2023simulator}.
The code and data supporting the evaluation can be found at~\cite{evaluation-data}.

\bibliographystyle{unsrt}
\bibliography{qoala-paper}

\begin{thebibliography}{10}

\bibitem{wehner2018quantum}
Stephanie Wehner, David Elkouss, and Ronald Hanson.
\newblock Quantum internet: A vision for the road ahead.
\newblock {\em Science}, 362(6412):eaam9288, 2018.

\bibitem{kimble2008quantum}
Harry~Jeffrey Kimble.
\newblock The quantum internet.
\newblock {\em Nature}, 453(7198):1023--1030, 2008.

\bibitem{bennett2014quantum}
Charles Bennett and Gilles Brassard.
\newblock Quantum cryptography: Public key distribution and coin tossing.
\newblock {\em Theoretical computer science}, 560:7--11, 2014.

\bibitem{ekert1991quantum}
Artur Ekert.
\newblock Quantum cryptography based on bell’s theorem.
\newblock {\em Physical review letters}, 67(6):661, 1991.

\bibitem{broadbent2009universal}
Anne Broadbent, Joseph Fitzsimons, and Elham Kashefi.
\newblock Universal blind quantum computation.
\newblock In {\em 2009 50th annual IEEE symposium on foundations of computer
  science}, pages 517--526. IEEE, 2009.

\bibitem{arrighi2006blind}
Pablo Arrighi and Louis Salvail.
\newblock Blind quantum computation.
\newblock {\em International Journal of Quantum Information}, 4(05):883--898,
  2006.

\bibitem{caleffi_distributed_2022}
Marcello Caleffi, Michele Amoretti, Davide Ferrari, Daniele Cuomo, Jessica
  Illiano, Antonio Manzalini, and Angela~Sara Cacciapuoti.
\newblock Distributed quantum computing: a survey.

\bibitem{pompili2021realization}
Matteo Pompili, Sophie Hermans, Simon Baier, Hans Beukers, Peter Humphreys,
  Raymond Schouten, Raymond Vermeulen, Marijn Tiggelman, Laura dos
  Santos~Martins, Bas Dirkse, Stephanie Wehner, and Ronald Hanson.
\newblock Realization of a multinode quantum network of remote solid-state
  qubits.
\newblock {\em Science}, 372(6539):259--264, 2021.

\bibitem{krutyanskiy2023entanglement}
Viktor Krutyanskiy, Maria Galli, Vojtech Krcmarsky, Simon Baier, Dario
  Fioretto, Yunfei Pu, Azadeh Mazloom, Pavel Sekatski, Marco Canteri, Markus
  Teller, Josef Schupp, James Bate, Martin Meraner, Nicolas Sangouard, Ben
  Lanyon, and Tracy Northup.
\newblock Entanglement of trapped-ion qubits separated by 230 meters.
\newblock {\em Physical Review Letters}, 130(5):050803, 2023.

\bibitem{dahlberg2019link}
Axel Dahlberg, Matthew Skrzypczyk, Tim Coopmans, Leon Wubben, Filip
  Rozp{\k{e}}dek, Matteo Pompili, Arian Stolk, Przemys{\l}aw Pawe{\l}czak,
  Robert Knegjens, Julio de~Oliveira~Filho, Ronald Hanson, and Stephanie
  Wehner.
\newblock A link layer protocol for quantum networks.
\newblock In {\em Proceedings of the ACM special interest group on data
  communication}, pages 159--173. 2019.

\bibitem{donne2024design}
Carlo~Delle Donne, Mariagrazia Iuliano, Bart van~der Vecht, Guilherme~Maciel
  Ferreira, Hana Jirovská, Thom van~der Steenhoven, Axel Dahlberg, Matt
  Skrzypczyk, Dario Fioretto, Markus Teller, Pavel Filippov, Alejandro
  Rodríguez-Pardo Montblanch, Julius Fischer, Benjamin van Ommen, Nicolas
  Demetriou, Dominik Leichtle, Luka Music, Harold Ollivier, Ingmar~te Raa,
  Wojciech Kozlowski, Tim Taminiau, Przemysław Pawełczak, Tracy Northup,
  Ronald Hanson, and Stephanie Wehner.
\newblock Design and demonstration of an operating system for executing
  applications on quantum network nodes.
\newblock {\em arXiv preprint arXiv:2407.18306}, 2024.

\bibitem{pompili2022experimental}
Matteo Pompili, Carlo Delle~Donne, Ingmar te~Raa, Bart van~der Vecht, Matthew
  Skrzypczyk, Guilherme Ferreira, Lisa de~Kluijver, Arian Stolk, Sophie
  Hermans, Przemys{\l}aw Pawe{\l}czak, Wojciech Kozlowski, Ronald Hanson, and
  Stephanie Wehner.
\newblock Experimental demonstration of entanglement delivery using a quantum
  network stack.
\newblock {\em npj Quantum Information}, 8(1):121, 2022.

\bibitem{dahlberg2022netqasm}
Axel Dahlberg, Bart van~der Vecht, Carlo Delle~Donne, Matthew Skrzypczyk,
  Ingmar te~Raa, Wojciech Kozlowski, and Stephanie Wehner.
\newblock Netqasm—a low-level instruction set architecture for hybrid
  quantum--classical programs in a quantum internet.
\newblock {\em Quantum Science and Technology}, 7(3):035023, 2022.

\bibitem{qoala2023simulator}
Qoala simulator implementation.
\newblock https://github.com/QuTech-Delft/qoala-sim.

\bibitem{diadamo2021distributed}
Stephen DiAdamo, Marco Ghibaudi, and James Cruise.
\newblock Distributed quantum computing and network control for accelerated
  vqe.
\newblock {\em IEEE Transactions on Quantum Engineering}, 2:1--21, 2021.

\bibitem{liu2022layer}
Xiaoyuan Liu, Anthony Angone, Ruslan Shaydulin, Ilya Safro, Yuri Alexeev, and
  Lukasz Cincio.
\newblock Layer vqe: A variational approach for combinatorial optimization on
  noisy quantum computers.
\newblock {\em IEEE Transactions on Quantum Engineering}, 3:1--20, 2022.

\bibitem{farhi2014quantum}
Edward Farhi, Jeffrey Goldstone, and Sam Gutmann.
\newblock A quantum approximate optimization algorithm.
\newblock {\em arXiv preprint arXiv:1411.4028}, 2014.

\bibitem{cacciapuoti2019quantum}
Angela~Sara Cacciapuoti, Marcello Caleffi, Francesco Tafuri, Francesco~Saverio
  Cataliotti, Stefano Gherardini, and Giuseppe Bianchi.
\newblock Quantum internet: networking challenges in distributed quantum
  computing.
\newblock {\em IEEE Network}, 34(1):137--143, 2019.

\bibitem{ovide2023mapping}
Anabel Ovide, Santiago Rodrigo, Medina Bandic, Hans Van~Someren, Sebastian
  Feld, Sergi Abadal, Eduard Alarcon, and Carmen~G. Almudever.
\newblock Mapping quantum algorithms to multi-core quantum computing
  architectures.
\newblock In {\em 2023 {IEEE} International Symposium on Circuits and Systems
  ({ISCAS})}, pages 1--5.
\newblock {ISSN}: 2158-1525.

\bibitem{jnane2022multicore}
Hamza Jnane, Brennan Undseth, Zhenyu Cai, Simon Benjamin, and B{\'a}lint
  Koczor.
\newblock Multicore quantum computing.
\newblock {\em Physical Review Applied}, 18(4):044064, 2022.

\bibitem{wei2022towards}
Shi-Hai Wei, Bo~Jing, Xue-Ying Zhang, Jin-Yu Liao, Chen-Zhi Yuan, Bo-Yu Fan,
  Chen Lyu, Dian-Li Zhou, You Wang, Guang-Wei Deng, Hai-Zhi Song, Daniel Oblak,
  Guang-Can Guo, and Qiang Zhou.
\newblock Towards real-world quantum networks: a review.
\newblock {\em Laser \& Photonics Reviews}, 16(3):2100219, 2022.

\bibitem{azuma2021tools}
Koji Azuma, Stefan B{\"a}uml, Tim Coopmans, David Elkouss, and Boxi Li.
\newblock Tools for quantum network design.
\newblock {\em AVS Quantum Science}, 3(1), 2021.

\bibitem{liu1973scheduling}
Chung~Laung Liu and James Layland.
\newblock Scheduling algorithms for multiprogramming in a hard-real-time
  environment.
\newblock {\em Journal of the ACM (JACM)}, 20(1):46--61, 1973.

\bibitem{hambarde2014survey}
Prasanna Hambarde, Rachit Varma, and Shivani Jha.
\newblock The survey of real time operating system: Rtos.
\newblock In {\em 2014 International Conference on Electronic Systems, Signal
  Processing and Computing Technologies}, pages 34--39. IEEE, 2014.

\bibitem{burns2017survey}
Alan Burns and Robert Davis.
\newblock A survey of research into mixed criticality systems.
\newblock {\em ACM Computing Surveys}, 50(6):1--37, 2017.

\bibitem{hillery1999quantum}
Mark Hillery, Vladim{\'\i}r Bu{\v{z}}ek, and Andr{\'e} Berthiaume.
\newblock Quantum secret sharing.
\newblock {\em Physical Review A}, 59(3):1829, 1999.

\bibitem{network-scheduling}
Thomas Beauchamp, Hana Jirovsk\'{a}, Scarlett Gauthier, and Stephanie Wehner.
\newblock A modular quantum network architecture for integrating network
  scheduling with local program execution.
\newblock In preparation. Private communication, 2025.

\bibitem{skrzypczyk2021architecture}
Matthew Skrzypczyk and Stephanie Wehner.
\newblock An architecture for meeting quality-of-service requirements in
  multi-user quantum networks.
\newblock {\em arXiv preprint arXiv:2111.13124}, 2021.

\bibitem{leichtle2021verifying}
Dominik Leichtle, Luka Music, Elham Kashefi, and Harold Ollivier.
\newblock Verifying {B}{Q}{P} computations on noisy devices with minimal
  overhead.
\newblock {\em PRX Quantum}, 2(4):040302, 2021.

\bibitem{ruf2021quantum}
Maximilian Ruf, Noel Wan, Hyeongrak Choi, Dirk Englund, and Ronald Hanson.
\newblock Quantum networks based on color centers in diamond.
\newblock {\em Journal of Applied Physics}, 130(7), 2021.

\bibitem{krutyanskiy2023telecom}
Victor Krutyanskiy, Marco Canteri, Martin Meraner, James Bate, Vojtech
  Krcmarsky, Josef Schupp, Nicolas Sangouard, and Ben Lanyon.
\newblock Telecom-wavelength quantum repeater node based on a trapped-ion
  processor.
\newblock {\em Physical Review Letters}, 130(21):213601, 2023.

\bibitem{vardoyan2022quantum}
Gayane Vardoyan, Matthew Skrzypczyk, and Stephanie Wehner.
\newblock On the quantum performance evaluation of two distributed quantum
  architectures.
\newblock {\em ACM SIGMETRICS Performance Evaluation Review}, 49(3):30--31,
  2022.

\bibitem{drmota2023robust}
Peter Drmota, Dougal Main, David Nadlinger, Bethan Nichol, Marius~Alfons Weber,
  Ellis Ainley, Ayush Agrawal, Raghavendra Srinivas, Gabriel Araneda, Chris
  Ballance, and David Lucas.
\newblock Robust quantum memory in a trapped-ion quantum network node.
\newblock {\em Physical Review Letters}, 130(9):090803, 2023.

\bibitem{vidick2023introduction}
Thomas Vidick and Stephanie Wehner.
\newblock {\em Introduction to Quantum Cryptography}.
\newblock Cambridge University Press, 2023.

\bibitem{coopmans2021netsquid}
Tim Coopmans, Robert Knegjens, Axel Dahlberg, David Maier, Loek Nijsten, Julio
  de~Oliveira~Filho, Martijn Papendrecht, Julian Rabbie, Filip Roz{\k{e}}pdek,
  Matthew Skrzypczyk, Leon Wubben, Walter de~Jong, Damian Podareanu, Ariana
  Torres-Knoop, David Elkouss, and Stephanie Wehner.
\newblock Netsquid, a network simulator for quantum information using discrete
  events.
\newblock {\em Communications Physics}, 4(1):164, 2021.

\bibitem{diadamo2021qunetsim}
Stephen DiAdamo, Janis N{\"o}tzel, Benjamin Zanger, and Mehmet~Mert Be{\c{s}}e.
\newblock Qu{N}et{S}im: A software framework for quantum networks.
\newblock {\em IEEE Transactions on Quantum Engineering}, 2:1--12, 2021.

\bibitem{fang2023quantum}
Kun Fang, Jingtian Zhao, Xiufan Li, Yifei Li, and Runyao Duan.
\newblock Quantum {NETwork}: from theory to practice.
\newblock 66(8):180509, 2023.

\bibitem{satoh2022quisp}
Ryosuke Satoh, Michal Hajdu{\v{s}}ek, Naphan Benchasattabuse, Shota Nagayama,
  Kentaro Teramoto, Takaaki Matsuo, Sara~Ayman Metwalli, Poramet Pathumsoot,
  Takahiko Satoh, Shigeya Suzuki, and Rodney Van~Meter.
\newblock Qu{I}{S}{P}: a quantum internet simulation package.
\newblock In {\em 2022 IEEE International Conference on Quantum Computing and
  Engineering (QCE)}, pages 353--364. IEEE, 2022.

\bibitem{squidasmrepo}
QuTech.
\newblock Squid{A}{S}{M}.
\newblock https://github.com/QuTech-Delft/squidasm, 2022.

\bibitem{polychronopoulos1991hierarchical}
Constantine Polychronopoulos.
\newblock The hierarchical task graph and its use in auto-scheduling.
\newblock In {\em Proceedings of the 5th International Conference on
  Supercomputing}, pages 252--263, 1991.

\bibitem{girkar1994hierarchical}
Milind Girkar and Constantine Polychronopoulos.
\newblock The hierarchical task graph as a universal intermediate
  representation.
\newblock {\em International Journal of Parallel Programming}, 22(5):519--551,
  1994.

\bibitem{silberschatz2020operating}
Abraham Silberschatz, Peter~B Galvin, and Greg Gagne.
\newblock {\em Operating System Concepts}.
\newblock Wiley, 10th edition, 2020.

\bibitem{evaluation-data}
Code and data for running the evalutions.
\newblock 4TU.ResearchData.
  https://doi.org/10.4121/f1e3a0ba-17d5-48f9-a66b-2c45520f229c, 2025.

\bibitem{greenberger1989going}
Daniel Greenberger, Michael Horne, and Anton Zeilinger.
\newblock Going beyond bell's theorem.
\newblock In {\em Bell's theorem, quantum theory and conceptions of the
  universe}, pages 69--72. Springer, 1989.

\bibitem{bradley2019ten}
Conor Bradley, Joe Randall, Mohamed Abobeih, Remon Berrevoets, Maarten Degen,
  Michiel Bakker, Matthew Markham, Daniel Twitchen, and Tim Taminiau.
\newblock A ten-qubit solid-state spin register with quantum memory up to one
  minute.
\newblock {\em Physical Review X}, 9(3):031045, 2019.

\bibitem{hermans2022qubit}
Sophie Hermans, Matteo Pompili, Hans Beukers, Simon Baier, Johannes Borregaard,
  and Ronald Hanson.
\newblock Qubit teleportation between non-neighbouring nodes in a quantum
  network.
\newblock {\em Nature}, 605(7911):663--668, 2022.

\bibitem{topcuoglu2002performance}
Haluk Topcuoglu, Salim Hariri, and Min-You Wu.
\newblock Performance-effective and low-complexity task scheduling for
  heterogeneous computing.
\newblock {\em IEEE transactions on parallel and distributed systems},
  13(3):260--274, 2002.

\bibitem{baruah2011scheduling}
Sanjoy Baruah, Vincenzo Bonifaci, Gianlorenzo d'Angelo, Haohan Li, Alberto
  Marchetti-Spaccamela, Nicole Megow, and Leen Stougie.
\newblock Scheduling real-time mixed-criticality jobs.
\newblock {\em IEEE Transactions on Computers}, 61(8):1140--1152, 2011.

\bibitem{andersson2006multiprocessor}
Bj{\"o}rn Andersson and Eduardo Tovar.
\newblock Multiprocessor scheduling with few preemptions.
\newblock In {\em 12th IEEE International Conference on Embedded and Real-Time
  Computing Systems and Applications (RTCSA'06)}, pages 322--334. IEEE, 2006.

\bibitem{jensen1993timeliness}
Douglas Jensen.
\newblock A timeliness model for asychronous decentralized computer systems.
\newblock In {\em Proceedings ISAD 93: International Symposium on Autonomous
  Decentralized Systems}, pages 173--182. IEEE, 1993.

\bibitem{li2004utility}
Peng Li.
\newblock {\em Utility accrual real-time scheduling: Models and algorithms}.
\newblock PhD thesis, Virginia Polytechnic Institute and State University,
  2004.

\bibitem{lattner2004llvm}
Chris Lattner and Vikram Adve.
\newblock L{L}{V}{M}: A compilation framework for lifelong program analysis \&
  transformation.
\newblock In {\em International symposium on code generation and optimization,
  2004. CGO 2004.}, pages 75--86. IEEE, 2004.

\bibitem{lattner2021mlir}
Chris Lattner, Mehdi Amini, Uday Bondhugula, Albert Cohen, Andy Davis, Jacques
  Pienaar, River Riddle, Tatiana Shpeisman, Nicolas Vasilache, and Oleksandr
  Zinenko.
\newblock M{L}{I}{R}: Scaling compiler infrastructure for domain specific
  computation.
\newblock In {\em 2021 IEEE/ACM International Symposium on Code Generation and
  Optimization (CGO)}, pages 2--14. IEEE, 2021.

\bibitem{avis2023requirements}
Guus Avis, Francisco Ferreira~da Silva, Tim Coopmans, Axel Dahlberg, Hana
  Jirovsk{\'a}, David Maier, Julian Rabbie, Ariana Torres-Knoop, and Stephanie
  Wehner.
\newblock Requirements for a processing-node quantum repeater on a real-world
  fiber grid.
\newblock {\em npj Quantum Information}, 9(1):100, 2023.

\end{thebibliography}

\appendices

\onecolumn

\section*{Appendix}
Appendix containing additional information for the paper \textbf{Qoala: an Application Execution Environment for Quantum Internet Nodes}.
This Appendix is structured as follows.
\begin{itemize}
\item \cref{app:program_structure} provides details of the \textbf{Qoala program format}.
\item \cref{app:runtime_environment} provides details of the \textbf{Qoala runtime environment}.
\item In \cref{app:scheduling_execution}, our \textbf{scheduler implementation} (\cref{sec:implementation}) is explained more in-depth.
\item \cref{app:simulator} gives an overview of our \textbf{simulator implementation}.
\item In \cref{app:evaluation} we elaborate on our \textbf{evaluations}.
\end{itemize}

\section{Program structure}
\label{app:program_structure}
This section provides details about the structure and contents of Qoala programs as described in \cref{sec:program_structure}.

\subsection{Program representation and components}
A Qoala program is represented in human-readable text format.
This allows one to directly write Qoala programs, although our vision is that programmers write their code in a higher-level language, and that a compiler translates this into a Qoala program.

In the main text, some parts of example programs were omitted for brevity.
In \cref{fig:app:example_full_program} we show an example of a full Qoala program.

A Qoala program encompasses both classical and quantum code.
These different code segments are put into different sections in the program.
The host section contains QoalaHost code which is to be run on the CPS.
The NetQASM section contains local routines (containing NetQASM instructions) which are meant to be run on the QPS.
The request section contains specifications of requests for remote entanglement generation, to be handled by the QPS.
Furthermore, there is a meta section which defines global information about the program.
Each of these sections is explained in more detail below.

In all of the sections in a Qoala program, values may be replaced by a \textbf{template}.
A template represents a value that is not defined for the program, but is filled in at program instantiation. For example, a QKD program might have a request object in its request section containing the entry \texttt{num\_pairs: {N}}, where \texttt{{N}} is a template. This construction allows one to instantiate the same program with different values for \texttt{N}, and it is hence not needed to define separate programs for each different number of pairs to generate in the QKD program.

\subsection{Program Meta}
Program metadata contains:

\begin{itemize}
    \item \textbf{Name}: The name of this program.
    \item \textbf{Parameters}: Global arguments to this program. These arguments may be used as templates (see above) in the program. Examples may be the name of a remote node, or the number of EPR pairs to generate.
    \item \textbf{Classical Sockets}: A mapping from IDs to remote node names. The IDs are local identifiers that can be used by Host code to distinguish different classical sockets.
    \item \textbf{EPR Sockets}: A mapping from IDs to remote node names. The IDs are local identifiers that can be used by Host code to distinguish different EPR sockets.
\end{itemize}

\begin{figure*}[htbp]
    \centering
    \begin{minipage}{\textwidth}
\begin{lstlisting}[language=qoala, caption={}]
META_START
    name: bob
    parameters: alice_id
    csockets: 0 -> alice
    epr_sockets: 0 -> alice
META_END

(*@\textcolor{purple}{\textasciicircum b1}@*) {type = QC}:
    run_request(vec<>) : req
  
(*@\textcolor{purple}{\textasciicircum b2}@*) {type = QL}:
    vec<m0> = run_routine(vec<>) : post_epr

(*@\textcolor{purple}{\textasciicircum b3}@*) {type = CL}:
    return_result(m0)

SUBROUTINE post_epr
    params: 
    returns: m0
    uses: 0
    keeps:
    request:
  NETQASM_START
    set Q0 0
    meas Q0 M0
    store M0 @output[0]
  NETQASM_END

REQUEST req
  callback_type: wait_all
  callback: 
  return_vars:
  remote_id: {alice_id}
  epr_socket_id: 0
  num_pairs: 1
  virt_ids: all 0
  timeout: 1000
  fidelity: 1.0
  typ: create_keep
  role: receive
\end{lstlisting}
    \end{minipage}
    \caption{Example Qoala program which creates an EPR pair with remote program Alice, measures the local qubit, and returns the classical outcome value.
    \textit{Meta section.} The meta section defines the name of this program, the global arguments (input values, in this case: the node ID of the Alice program),
    the classical sockets used (mapping local socket ID to the name of the remote node, and the EPR sockets used (also mapping local socket ID to remote node names)).
    \textit{Host section.} This example host section consists of three blocks (\texttt{b1, b2, b3}). \texttt{b1} calls request routine \texttt{req} (no result values).
    \texttt{b2} calls local routine \texttt{post\_epr}, resulting in a classical vector with one value (\texttt{m0}).
    \texttt{b3} returns \texttt{m0} as the result of this program.
    \textit{Local routines section.} Consists of a single local routine called \texttt{post\_epr}.
    It requires the virtual qubit (see \cref{app:runtime_environment}) with ID 0 to be allocated, and acts on this qubit.
    Upon finishing the local routine, this qubit is not in use anymore (the \texttt{keeps} entry is empty).
    The NetQASM code represents measuring the qubit, and then storing the result (in register \texttt{M0}) to the \texttt{@output} array (see \cref{app:runtime_environment}), which is in shared memory and can be accessed by host code by the name \texttt{m0}.
    \textit{Request routines section.} Consists of a single request routine called \texttt{req}.
    It represents a request to the network stack for generating a single entangled pair (\texttt{num\_pairs} is 1), which is kept in memory (\texttt{typ: create\_keep}; not measured immediately).
    This program acts as a `receiver' for entanglement generation (\texttt{role} attribute), which breaks symmetry in the entanglement generation process (the remote Alice program will have \texttt{role: sender}). Symmetry breaking is needed for the network stack to organize the entanglement generation.
    No callbacks are used, and all qubits (in this case: one) are stored in virtual qubit 0.
    }
    \label{fig:app:example_full_program}
\end{figure*}

\subsection{Host section}
The host section contains code the be executed by the CPS.
It consists of both local processing (like calculation and conditional logic), and
communication (sending and receiving classical messages to and from other nodes in the network).

The language in which host code is represented is called QoalaHost.
This is a low-level instruction set with well-defined semantics and types,
and is meant to be executed by a virtual machine or interpreter.
One can also imagine QoalaHost code to be translated (either ahead-of-time or at just-in-time) to native CPS code, such as x86 or ARM. However, for the sake of simplicity and of implementation independence, we treat here only the QoalaHost language and its semantics itself.

The QoalaHost (QH) language was designed to resemble intermediate representations as found in LLVM~\cite{lattner2004llvm} and MLIR~\cite{lattner2021mlir},
such that integration with future compilers is accessible.
Specifically, one may imagine a compiler that uses MLIR for its intermediate representation (IR).
When this compiler then produces the host code of the program, the translation of its own IR to QoalaHost code should be straightforward.

\textbf{Blocks.} 
The Host section consists of a list of blocks.
A block consists of a block metadata and a list of QH instructions.

The block metadata contains the following entries:
\begin{itemize}
\item \textbf{Name}: The name of this block. Host code can refer to this name in QH branch instructions.
\item \textbf{Type}: one of CL, CC, QL or QC (see below).
\item \textbf{Deadlines}: Deadlines relative to other blocks.
The deadlines are specified in terms of EHI arguments. Upon program instantiation, concrete values are filled in based on the actual EHI value.
\item \textbf{Time hints}: Duration estimate of executing the block.
The estimates are specified in terms of EHI arguments. Upon program instantiation, concrete values are filled in based on the actual EHI value.
\end{itemize}

\subsection{Block types}
Blocks are categorized into the following four types:
\begin{itemize}
\item \textbf{CL}: Classical Local. The block contains only instructions that are classical, local and only involve the CPS
\item \textbf{CC}: Classical Communication. CPS-only instructions, but starts with a `receive message` instruction.
\item \textbf{QL}: Quantum Local. The block contains calls to local routines.
\item \textbf{QC}: Quantum Communication. The block contains calls to request routines.
\end{itemize}

\textbf{QoalaHost Language.}
The QH Language describes a fixed set of QH instructions as well as QH Variable types.
Host code is represented as blocks containing QH instructions.
These instructions may be directly interpreted by a processor or OS.

All basic values are 32-bit signed integers (i32) or floating point values (f32).
A variable in Host code can either be
\begin{itemize}
\item singleton variable, holding one basic value. Has a single name. E.g. \texttt{x}
\item vector, holding an arbitrary number of basic values. Has a single name. E.g. \texttt{x<>}
\end{itemize}

The QH Language allows for expressing multiple variables in a single expression, called a \textit{tuple}.
A tuple holds a fixed number of basic values. E.g. \texttt{tuple<x, y, z>}.

\textbf{Local Memory.}
Host code is assumed to have access to a local memory space that is logically organized as a mapping of \textit{names} to \textit{values}.
For example, the local memory may at some point during execution contain the following items:

\begin{lstlisting}
"var_x" -> 3
"my_vec" -> <1, 2, 5>
\end{lstlisting}

\textbf{Shared Memory.}
The QH Language does not allow direct access to shared memory.
Only variables from the local memory can be used.
When calling and getting results from Local Routines (LRs) and Request Routines (RRs), values are automatically
moved from local memory to shared memory. 
Shared memory is discussed in more detail in \cref{app:shared_memory}.

\subsubsection{Block format}

A block has the following format:
\begin{qoalacode}
(*@\textcolor{purple}{\textasciicircum \#name}@*) {type = #type}:
    <list of QH instructions>
\end{qoalacode}

Example:
\begin{qoalacode}
(*@\textcolor{purple}{\textasciicircum b0}@*) {type = CL}:
    x = assign() : 3
    return_result(x)
\end{qoalacode}

\begin{figure*}
    \centering
    \includegraphics[width=\textwidth]{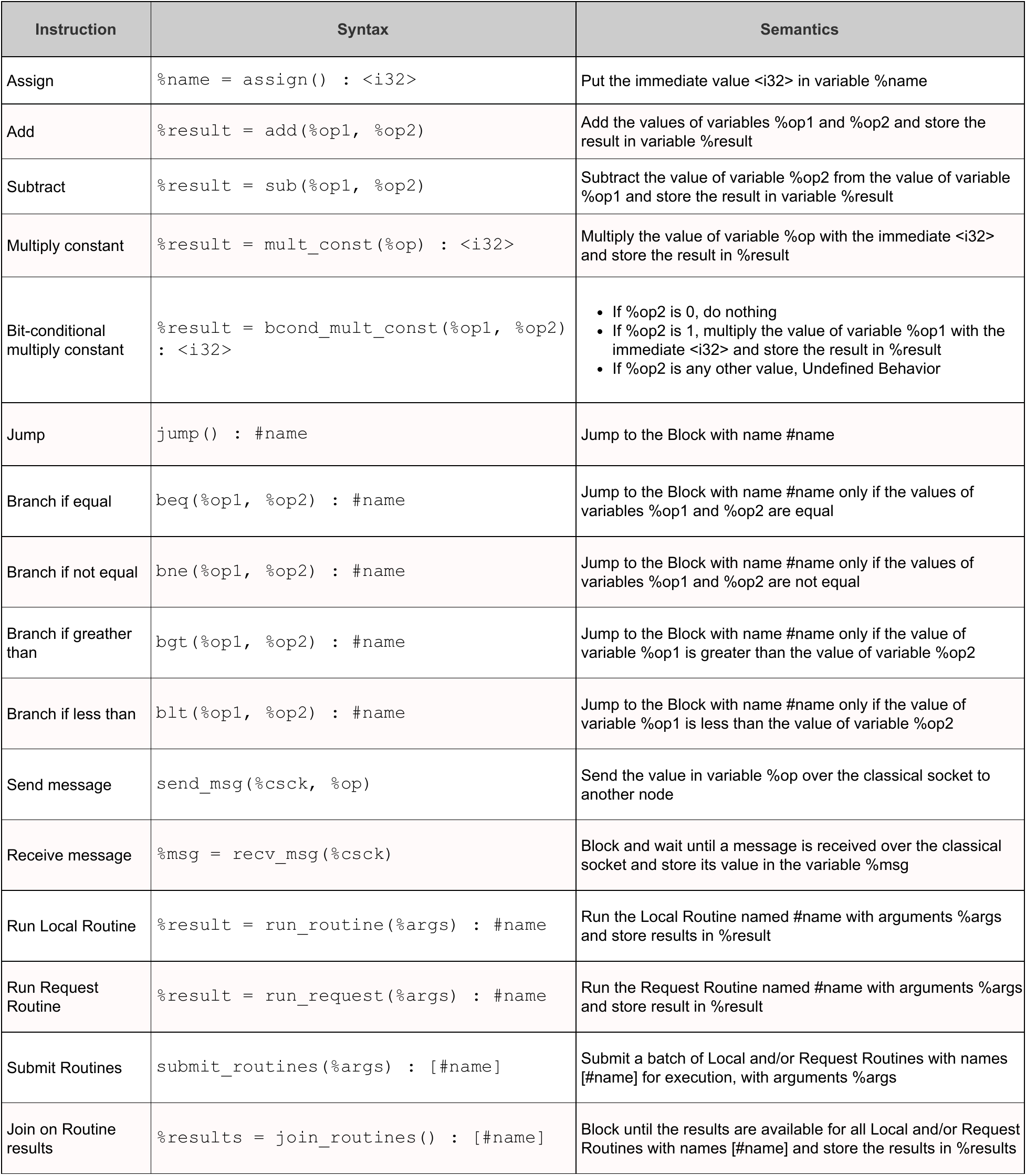}
    \caption{Overview of all host code (QoalaHost) instructions, their syntax and their semantics.}
    \label{fig:app:qh_table}
\end{figure*}

\subsubsection{QH instructions}
A full list of QoalaHost instructions is given in \cref{fig:app:qh_table}.

\subsection{NetQASM section}
\label{app:netqasm}
The NetQASM section consists of a list of local routines that are to be executed on the QPS.
A local routine is only executed when it is called by host code using the \texttt{run\_routine} instruction. A local routine may be run multiple times, again depending on the host code.

The instructions of a local routine are represented using the NetQASM 2.0 format.
This is an updated format compared to NetQASM 1.0 as presented in~\cite{dahlberg2022netqasm}.

\begin{figure*}
    \centering
    \includegraphics[width=\textwidth]{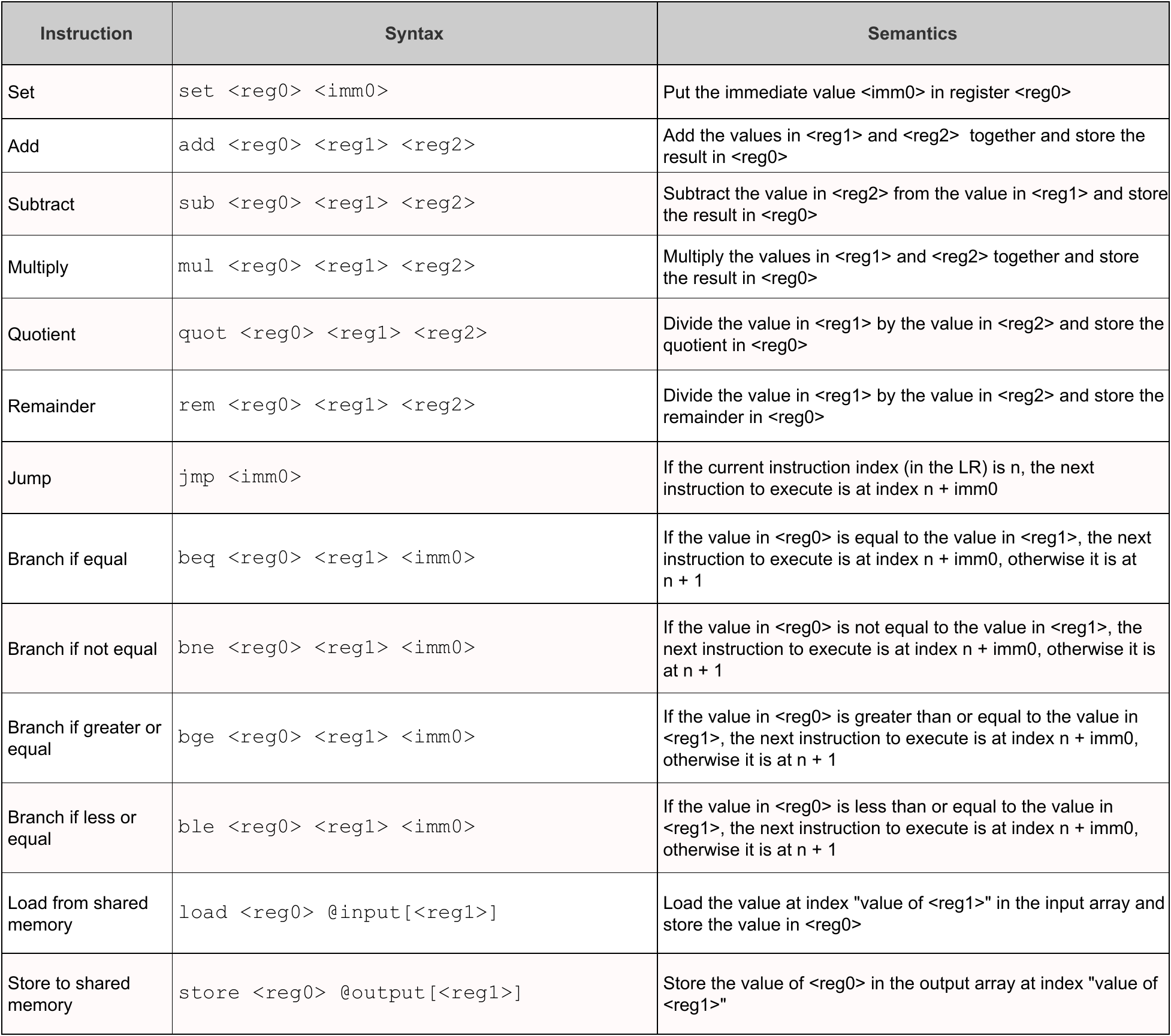}
    \caption{Overview of all NetQASM classical instructions, their syntax and their semantics.
    Quantum instructions depend on the particular flavour~\cite{dahlberg2022netqasm} that is being used.
    Semantics of quotient and remainder for non-positive integers are the same as in the C language standard.
    Note that \texttt{jmp 1} is a no-op.}
    \label{fig:app:netqasm_table}
\end{figure*}

\textbf{NetQASM values.}
All values are 32-bit signed integers. Floating-point values are not supported. Angles for qubit rotations must be expressed as discrete values.
Booleans are represented as follows: \texttt{true} is the 32-bit 0 value, \texttt{false} is the 32-bit 1 value. Any other 32-bit value is not a valid boolean.
The reason for keeping the different types limited is to keep the QPS implementation simple.

\textbf{NetQASM Local Memory}
The QPS is expected to have a local memory (only accessible by the QPS itself) consisting of 64 32-bit registers:
\begin{itemize}
\item 16 \textbf{R} registers: \texttt{R0} to \texttt{R15}
\item 16 \textbf{C} registers: \texttt{C0} to \texttt{C15}
\item 16 \textbf{M} registers: \texttt{M0} to \texttt{M15}
\item 16 \textbf{Q} registers: \texttt{Q0} to \texttt{Q15}
\end{itemize}

The four groups of registers are not inherently different. A compiler producing NetQASM code may use a certain group only for certain values, but this is not mandatory.

\textbf{Shared Memory}
See \cref{app:shared_memory} for more information about Shared Memory and arrays.
The QPS is expected to have access to Shared Memory (accessible by both the CPS and QPS).
Two shared memory Arrays are available:
\begin{itemize}
\item an \texttt{@input} array, containing the LR input variables
\item an \texttt{@output} array, with space to write the LR results to
\end{itemize}

The length of the \texttt{@input} array is equal to the number of LR parameters.
The length of the \texttt{@output} array is equal to the number of LR return variables.

\begin{itemize}
\item The \texttt{@input} and \texttt{@output} arrays are the only arrays accessible from within the LR.
\item The QPS can \textbf{only read} from the \texttt{@input} array (see \texttt{load} instruction below).
\item The QPS can \textbf{only write} to the \texttt{@output} array (see \texttt{store} instruction below).
\end{itemize}

\textbf{NetQASM Instruction}
Each instruction consists of the instruction type followed by a list of operands.
The text form of an instruction is:

\begin{lstlisting}
instr_name  op0 op1 ... opn
\end{lstlisting}

where the number of operands can be 0 or more (no limit).

A list of all NetQASM 2.0 instructions can be found in \cref{fig:app:netqasm_table}.

These instructions can be classified as:
\begin{itemize}
\item shared memory access: \texttt{load} for reading LR inputs, \texttt{store} for writing LR results
\item classical logic and control-flow: like \texttt{set} , \texttt{add}, or \texttt{jmp}
\item quantum operations: gates from a specific flavour~\cite{dahlberg2022netqasm}
\end{itemize}

NetQASM instructions representing quantum operations are either \textit{core instructions} or \textit{flavour-specific} instructions.
Core instructions are quantum hardware independent and are expected to be compatible with any QPS implementation. On top of the core instructions, flavour-specific instructions may be added and supported by a specific QPS implementation. For example, a QPS that controls an NV-centre may support NetQASM instructions of the NV flavour, which contain gate operations only available on this particular quantum hardware. Which NetQASM instructions are supported by the QPS is exposed to higher layers (including a compiler) as part of the EHI (see \cref{app:ehi}). Using this information, a compiler may produce optimized NetQASM code using the flavour-specific NetQASM instructions.

Note that NetQASM 2.0 \textbf{does not} contain (in contrast to NetQASM 1.0~\cite{dahlberg2022netqasm}):
\begin{itemize}
\item Allocation instruction (\texttt{qalloc} in NetQASM 1.0): The memory manager allocates virtual qubits based on the LR header information. Note that qubit allocation is different from \textit{qubit initialization} (\texttt{init} instruction).
\item Instructions for EPR generation: This is handled by request routines.
\item Waiting instructions: Waiting is handled by the scheduler choosing which tasks to execute when.
\end{itemize}

\textbf{Local Routine}
A Local Routine (LR) represents a block of local program operations that are executed on the QPS. An LR is:
\begin{itemize}
\item local: there is no interaction whatsoever with external nodes or controllers
\item atomic: execution of an LR cannot be pre-empted; when the QPS start executing an LR, it will not do anything else until the LR has finished (unless an abort happens)
\end{itemize}

An LR consists of a \textit{header} and a \textit{body}. The header contains metadata such as the resource usage of the LR, and its input/output interface. The body contains the actual instructions in the form of NetQASM code.

\textbf{Arguments and Returns.}
An LR may have zero or more \textit{arguments}: values that are provided to the LR only at runtime.
They can be seen as inputs or parameters to the LR.
These values appear in the \texttt{@input} array in shared memory, and are put there by the CPS.

An LR may also have zero or more \texttt{returns}: values that are provided by the LR only at runtime.
They can be seen as outputs or results of the LR.
These values must be written to the \texttt{@output} array in shared memory, and can then be used by the CPS.

Arguments and returns are always 32-bit signed integers. There is no limit to the number of arguments and returns an LR may have.

\textbf{Local routine header.}
A Local routine (LR) header contains the following entries:
\begin{itemize}
\item \textbf{Name}: The name of this LR. Host code refers to this name in a \texttt{run\_routine} QoalaHost instruction.
\item \textbf{Uses}: A list of virtual qubits IDs. These refer to all virtual qubits that are used by this LR. At runtime, the memory manager makes sure that these virtual qubits are allocated before execution of the LR starts. (They may already have been allocated earlier; alternatively the memory manager allocates them just before the LR starts.)
\item \textbf{Keeps}: A list of virtual qubit IDs. These refer to all virtual qubits that should \textit{remain allocated} after finishing the LR. (They may e.g. be used in subsequent LRs.)
\item \textbf{Args}: A list of names for the arguments of the LR. They are in the same order as how their values are accessible from the \texttt{@input} Array.
\item \textbf{Returns}: A list of names for the returns of the LR. They are in the same order as how their values are put into the \texttt{@output} Array.
\end{itemize}

\textbf{Quantum memory usage annotations.}
The LR header indicates which virtual qubits are used and freed by the LR. This makes it possible for the scheduler to decide which \texttt{LocalRoutine} task it may schedule when. For more information, see section \cref{app:scheduling_execution} on scheduling.
The following listing provides an example:

\begin{qoalacode}
SUBROUTINE subrt1
    uses: 0, 1
    keeps: 0
    returns: m0
    <rest omitted>
  NETQASM_START
    set Q0 0
    set Q1 1
    init Q0
    init Q1
    cnot Q0 Q1
    meas Q1 M1
    store M1 @output[0]
  NETQASM_END
\end{qoalacode}
This local routine initializes virtual qubits 0 and 1 and then applies a CNOT gate on them.
It measures qubit 1 and stores the output in the \texttt{@output} array which can then be accesses by host code using the name \texttt{m0}.
Using the metadata, a scheduler knows the following information even before executing this LR: virtual qubits 0 and 1 need to be free before this LR can run, and after running the LR, qubit 1 is free (again) but qubit 0 remains occupied.

It is the responsibility of the compiler to make sure that the use and free values correspond to the actual NetQASM code.

\subsection{Request section}
The callback (which is an LR) can have zero or more arguments (just like standard LRs). The runtime values of these arguments are provided by the QPS directly.
A Request Routine (RR) may have zero or more returns: outputs or results of the entire RR. The only allowed results at this moment are measurement outcomes in case of Measure Directly requests.
RR callbacks can have (just like standard LRs) zero or more returns.

\textbf{Request routine header}
A Request routine (RR) header contains the following entries:
\begin{itemize}
\item \textbf{Name}: The name of this RR. Host code refers to this name in a \texttt{run\_request}.
\item \textbf{Returns}: A list of names for the returns of the RR. Since the returns can only be measurement outcomes, these names are either (1) the name of a single QoalaHost vector variable which will hold all outcomes, or (2) a list of names for each individual outcome stored in its own QoalaHost int variable.
\item \textbf{Callback type}: Either \texttt{sequential} or \texttt{wait\_all}. Sequential means that the callback of this RR is executed for each generated pair, before the next pair is generated. Wait-all means that the callback is only executed once, namely when all pairs have been generated.
\item \textbf{Callback}: The name of the LR that acts as the callback for this RR. Can be empty (no callback is used).
\end{itemize}

\textbf{Request Parameters}
\begin{itemize}
\item \textbf{Remote ID}: The node ID of the remote node with which to generate entanglement.
\item \textbf{EPR Socket ID}: The ID of the EPR Socket to use.
\item \textbf{Number of pairs}: The number of entangled pair to generate.
\item \textbf{Virtual IDs}: A specification of the virtual IDs to assign to the entangled qubits. This may be in one of three formats:
\begin{itemize}
  \item \texttt{all <N>}: all qubits get virtual ID \texttt{<N>}. This might be used when a sequential callback is used that measures the qubit immediately after generating; thereby freeing up virtual ID \texttt{<N>} immediately for the next pair
  \item \texttt{increment <N>}: the first generated qubit gets ID \texttt{<N>}, the next \texttt{<N> + 1}, etc.
  \item \texttt{custom <N1, N2, ...>}: a custom list of IDs that should have the same length as the number of pairs
\end{itemize}
\item \textbf{Fidelity}: The desired fidelity \texttt{F} of the generated pairs.
If this request routine is for multiple pairs and the callback type is \texttt{wait\_all}, this value is used to specify that all pairs, after they have all been created, should have fidelity at least \texttt{F}. (How this is realized, which may involve multiple retries, is up to the network stack implementation in the QPS.)
\item \textbf{Type}: Create and Keep (\texttt{create\_keep}), Measure Directly (\texttt{measure\_directly}), or Remote State Preparation (\texttt{rsp}) \cite{dahlberg2019link}.
\item \textbf{Role}: \texttt{create} or \texttt{receive}. These roles are used to break symmetry between two nodes participating in entanglement generation (they should always have different roles). The `create' node is the initiating one.

\end{itemize}

\clearpage
\section{Runtime environment}
\label{app:runtime_environment}
\begin{figure*}[ht]
    \centering
    \includegraphics[width=\textwidth]{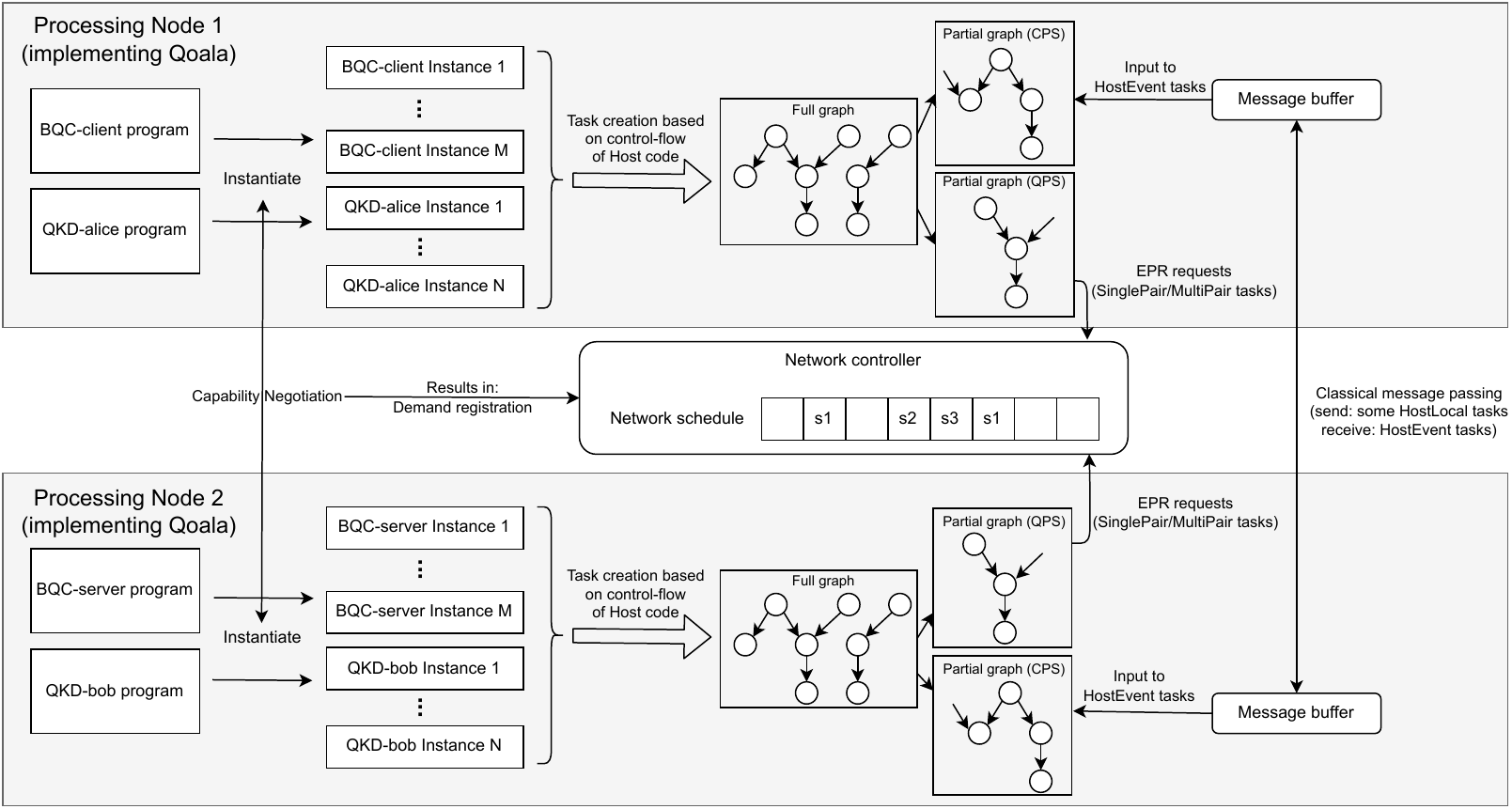}
    \caption{Detailed Qoala runtime overview.}
    \label{fig:app:runtime_detailed}
\end{figure*}

In this section we provide more information about the runtime environment described in \cref{sec:runtime_environment}.
\cref{fig:app:runtime_detailed} provides an overview of the runtime architecture.

\subsection{Program instantiation}
A program instance is a Qoala program with additional runtime- and context-specific information that is supplied when preparing execution of the program.
A program instance represents a single execution of a Qoala program.

The additional information consists of:
concrete values for the global arguments of the program,
the Exposed Hardware Info (EHI),
an explicit Unit Module (see below), and
results from capability negotiation.

Based on the above additional information, a program instance can be created which has the following properties:
\begin{itemize}
\item \textbf{Program ID}: A unique ID for distinguishing multiple program instances that all need to be scheduled and run.
\item \textbf{Program}: The static Qoala program (without runtime information).
\item \textbf{Program Inputs}: The values for the program's global arguments.
\item \textbf{Unit Module}: The virtual quantum memory space that this program instance may use at runtime.
\item \textbf{Timing Information}: Deadlines for individual tasks. Computed using both the program's timing hints and information from the EHI.
\end{itemize}

\begin{figure}[ht]
    \centering
    \includegraphics[width=0.5\columnwidth]{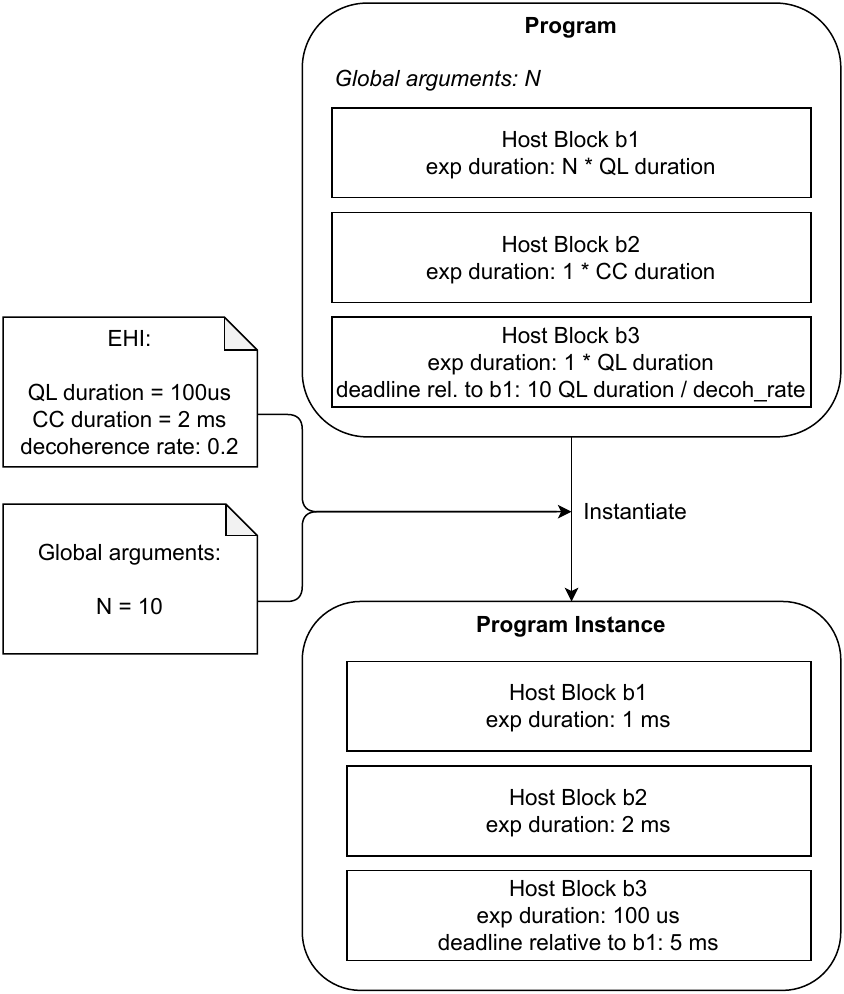}
    \caption{Schematic example of program instantiation.
    A program containing global arguments ($N$) is instantiated using a concrete value for the arguments ($N = 10$) and the EHI (containing values for the expect duration of a QL block, the expected duration of a CC block, and the qubit noise parameter expressed as the \textit{decoherence rate}). This results in a program instance for which the expected durations have concrete values.
    }
    \label{fig:app:instantiation}
\end{figure}

\cref{fig:app:instantiation} provides a schematic example of program instantiation.

\subsection{Program versus program instance}
A program is typically the output of a compiler.
For example, a compiler might produce a BQC-server program, including global arguments for the remote ID of the client (i.e. the client ID is \textit{not} hardcoded into it).
A program instance represents a single execution of a Qoala program with concrete values for its global arguments.
For instance, the client ID now has the explicit value of 3, since the remote client happens to have node ID 3.
Often many program instances may be created for a single program.
For example, if 1000 runs of the BQC program are desired, 1000 program instances are created based on the single Qoala program.

\paragraph{Batches}
A program may be submitted for execution in a batch.

A batch $B$ consists of a program $P$, the number of execution $N$ and inputs for each execution. Based on this, $N$ program instances are created.

\subsection{Shared memory}
\label{app:shared_memory}
The CPS and QPS need to exchange information in order to execute local routines and request routines. They do so using shared memory.
The CPS writes routine arguments and reads results.
The QPS reads routine arguments and writes results.

Conflicts in writing and reading are avoided by the runtime itself (it is not assumed the hardware itself enforces read-only or write-only regions of memory).
This is achieved by strict read/write rules in Qoala: certain regions can only be written to by the CPS (QPS) while only be read from the QPS (CPS).
No region can be written to by both CPS and QPS.
Note that this design leaves open how the shared memory can be implemented: either as real physical shared memory, or as a message passing protocol.

\begin{figure}[ht]
    \centering
    \includegraphics[width=0.6\columnwidth]{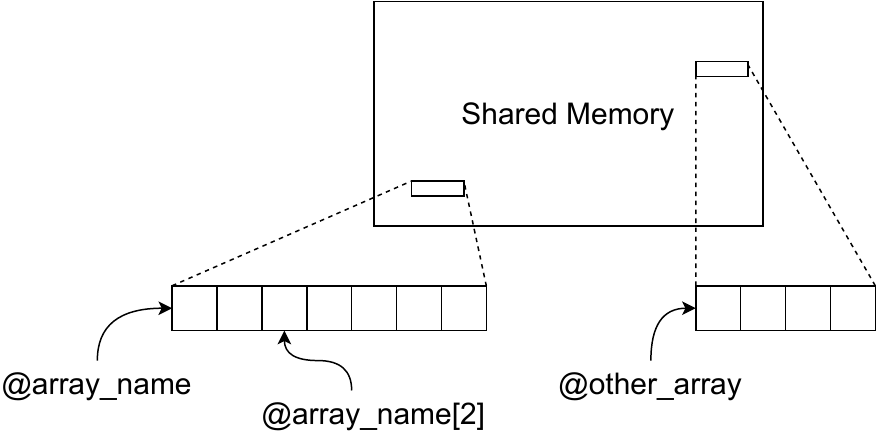}
    \caption{Schematic overview of shared memory, which is organized as \textit{arrays}.
    Arrays are allocated by the CPS with a certain size (the number of \textit{array elements}).
    Each array element holds a single classical value.
    Arrays are identified using the \texttt[@<name>] syntax.
    Particular array elements may be accessed using the \texttt{[index]} syntax.
    }
    \label{fig:app:arrays}
\end{figure}

\paragraph{Arrays}
The shared memory is logically divided into \textit{array elements} that can be allocated only by the CPS (\cref{fig:app:arrays}).
Each element can hold a single 32-bit signed integer.
The CPS can allocate shared memory space by specifying a \textit{size}, resulting in an allocated array.
An array is an ordered list of array elements. 
One can think of an array being a region in Shared Memory consisting of a consecutive list of elements.

Shared Memory is similar to the heap in classical OSes. Allocating an array is similar to \texttt{malloc} in C. Each program instance has its own view in the global shared memory, just like in classical OS, each program instance (or `process') has its own virtual memory space.

Elements that have been allocated but never written to have an undefined value.

An array may be named; it is written as \texttt{@arrayname}. An element in an array at index \texttt{i} is written as \texttt{@arrayname[i]}. This notation is used in NetQASM(\cref{app:netqasm}).

Arrays are used to share data between the CPS and the QPS.
They are used for executing both LRs and RRs.

\begin{figure}[ht]
    \centering
    \includegraphics[width=0.5\columnwidth]{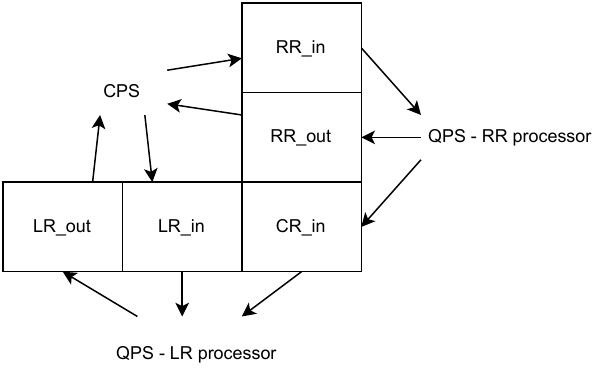}
    \caption{Shared memory regions.
    The CPS writes local routine arguments to the \texttt{LR\_in} section and request routine arguments to the \texttt{RR\_in} section.
    The CPS reads local routine results from the \texttt{LR\_out} section and request routine results from the \texttt{RR\_out} section.
    The QPS reads local routine arguments from \texttt{LR\_in} and write results to \texttt{LR\_out}.
    The QPS reads request routine arguments from \texttt{RR\_in} and write results to \texttt{RR\_out}.
    Callbacks for request routines use the separate \texttt{CR\_in} section to use request routine results as arguments of the callback local routine.
    }
    \label{fig:app:shared_memory}
\end{figure}

The shared memory is logically divided into 5 \textit{regions} (\cref{fig:app:shared_memory}).
Each of the regions contains array elements, and in each region, arrays can be allocated.
The regions are only a logical division, where each arrays in a certain region are only used to hold data for a specific use-case:

\begin{itemize}
\item \texttt{LR\_in}: Argument values for LRs. CPS writes, QPS reads.
\item \texttt{LR\_out}: Result values for LRs. CPS reads, QPS writes.
\item \texttt{RR\_in}: Argument values for RRs. CPS writes, QPS reads.
\item \texttt{RR\_ou}t: Result values for RRs. CPS reads, QPS writes.
\item \texttt{CR\_in}: Argument values for callback LRs. QPS reads, QPS writes.
\end{itemize}

\paragraph{Arrays for local routines}
Before an local routine (LR) can be executed, two arrays must be allocated by the CPS:
\begin{itemize}
\item An array in the \texttt{LR\_in} region. Its size needs to match the number of arguments for the LR.
\item An array in the \texttt{LR\_out} region. Its size needs to match the number of results of the LR.
\end{itemize}

The array in the \texttt{LR\_in} region can be accessed by the NetQASM code in the LR body using the name \texttt{@input}.
The array in the \texttt{LR\_out} region can be accessed by the NetQASM code in the LR body using the name \texttt{@output}.

Note that each program instance allocates (at runtime) its own arrays. Each individual LR in each individual program instance has access to two arrays called \texttt{@input}  and \texttt{@output}, but in practice there can hence be multiple "input" and "output" arrays, each occupying a different part of the global Shared Memory.

\paragraph{Arrays for request routines}
Before a request routine (RR) can be executed, multiple arrays must be allocated by the CPS:
\begin{itemize}
\item An array in the \texttt{RR\_in} region. 
\item An array in the \texttt{RR\_out} region. Its size needs to match the number of names in the "Results" entry in the RR header.
\item An array in the \texttt{CR\_in} region. Its size needs to match the number of arguments for the callback LR of the RR.
\end{itemize}

The results of the RR are written to the array in the \mbox{\texttt{RR\_out}} region. Arguments to the callback LR are written to the array in the \texttt{CR\_in} region.

\subsection{Quantum memory}
\label{app:quantum_memory}
The QPS is assumed to have access to a quantum random access memory (QRAM) consisting of \textit{qubits}.
Each qubit is a single location in the QRAM and can hold a single 2-dimensional quantum value, like $\ket{0}$ or $\ket{+}$.

We distinguish between (1) the \textit{physical quantum memory space (PQMS)} consisting of \textit{physical qubits}
and (2) a \textit{virtual quantum memory space (VQMS)} for each program instance (\cref{sec:runtime_environment}).

The topology (qubit connectivity) and noise characteristics of the PQMS are exposed as part of the EHI.
Each program instance has access to its own VQMS, which is represented as a Unit Module~\cite{dahlberg2022netqasm}.
The VQMS for each program instance is created when instantiating the program.
This can be seen as virtual memory allocation for the program.
At runtime, the VQMS of each running program instance is mapped to the PQMS.

\paragraph{Unit Modules}
A Unit Module (UM) describes the topology of a VQMS as well as its noise characteristics.
That is, a UM contains:
\begin{itemize}
\item \textbf{Qubit Info}: a list of all qubits available in the VQMS, with for each qubit the following information:
its virtual ID,
whether it is a communication qubit or not, and
its decoherence rate per second.

\item \textbf{Gate Info}: a list of all quantum gates and quantum local operations available for the qubits in the VQMS, with for each item the following information:
\begin{itemize}
  \item Which NetQASM instruction it is represented by (may be in a particular NetQASM flavor).
  \item On which sets of qubits the gate or operation can be applied.
  \item Its duration.
  \item The decoherence rate per second on each of the qubits it acts on.
\end{itemize}
\end{itemize}

A UM can be seen as a subset of the full EHI of a node, specifically containing a subset of all qubits available in the node.

\begin{figure}[ht]
    \centering
    \includegraphics[scale=0.4]{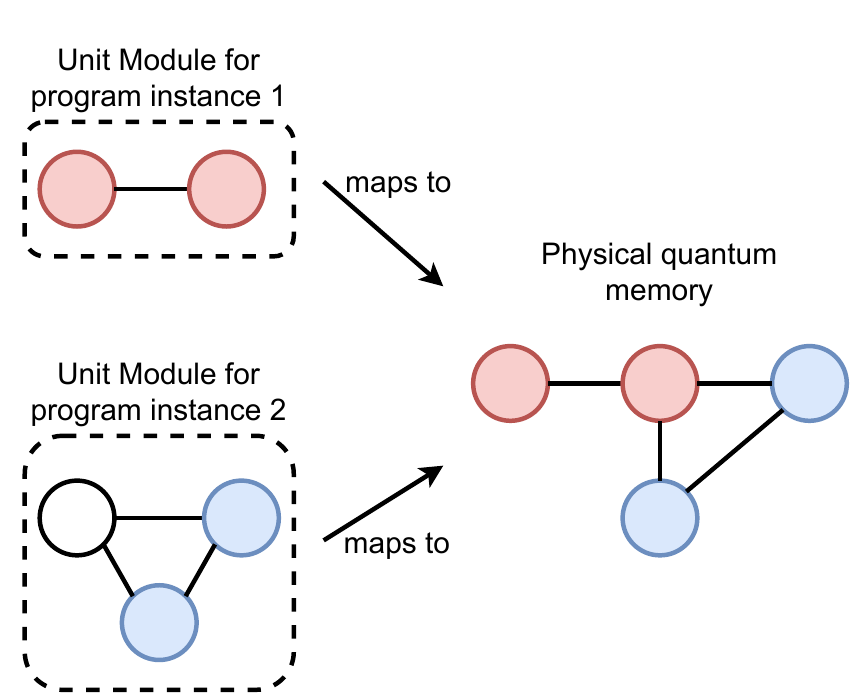}
    \caption{Example of a physical quantum memory available in a node (four qubits) and two allocated unit modules. The colors of the qubits represent the physical locations they map to. Note that the top-left qubit of Unit Module 2 is not currently mapped and that it also cannot be mapped. Therefore, tasks that require program instance 2 to use a third qubit cannot be executed at this time.}
    \label{fig:app:unit_module}
\end{figure}

Qubits in the Unit Module are called \textit{virtual qubits}. They are identified by their \textit{virtual IDs} and are mapped to physical qubits (\cref{fig:app:unit_module}).

\paragraph{Memory manager}
Quantum memory allocation and freeing is handled by a memory manager, which lives in the QPS.
The memory manager keeps track of the unit modules of all program instances, and maps virtual qubits to physical qubits. 

Before starting a local routine or request routine, the memory manager allocates the corresponding qubits.
For example, if a local routine for program instance $P$ defines in its metadata (see \cref{app:program_structure}) that it uses virtual qubits 0 and 1, the memory manager allocates virtual qubits 0 and 1 (if not already allocated).
This involves finding currently unused physical qubits and mapping new virtual qubit to these free physical qubits.

\subsection{Exposed hardware interface}
\label{app:ehi}
The Qoala execution environment exposes certain information related to the hardware and software capabilities.
This information includes noise characteristics of quantum memory and of entanglement generation, as well as estimates of classical latencies.

All information that is exposed falls under the Exposed Hardware Interface (EHI).
The EHI can be divided into \textit{node info} and \textit{network info}.

\paragraph{EHI Node Info}
The EHI node info consists of:

\begin{itemize}
\item \textbf{Qubit Info}: a list of all qubits available at the node, with for each qubit the following information: 
(1) its ID,
(2) whether it is a communication qubit or not, and
(3) its decoherence rate per second.
\item \textbf{Gate Info}: a list of all quantum gates and quantum local operations available at the node, with for each item the following information:
(1) which NetQASM instruction it is represented by (may be in a particular NetQASM flavor,
(2) on which sets of qubits the gate or operation can be applied,
(3) its duration, and
(4) the decoherence rate per second on each of the qubits it acts on.

\item \textbf{NetQASM flavor}: a list of all supported NetQASM instructions. All NetQASM instructions mentioned in Gate Info must be in this list

\item \textbf{Classical latencies}:
Covers
(1) duration of executing a single QH Instruction, and
(2) duration of executing a classical NetQASM instruction (Note that the duration of quantum operations is covered by the Gate Info).
\end{itemize}

\paragraph{EHI Network Info}
The EHI network info consists of \textbf{Link Info} for each link in the network, with
(1) the expected duration of generating an entangled pair on this link, and
(2) the expected fidelity of generating an entangled pair on this link.

\subsection{Sockets}
Connections with remote nodes are modeled as \textit{sockets}.
Each program instance running on a node has access to classical sockets an EPR sockets.
Classical sockets represent an endpoint for connections over which classical messages can be sent.
A program instance can have classical sockets with any other nodes in the network.

An EPR socket represents an endpoint of a quantum connection.
Through the EPR socket, a program can ask for entanglement with a remote node.

\clearpage
\section{Scheduling and execution}
\label{app:scheduling_execution}
This section provides more details about tasks, task creation and scheduling (\cref{sec:architecture}) as well as about our scheduler implementation (\cref{sec:implementation}).

\subsection{Tasks}
\label{app:scheduling_tasks}

\paragraph{Task creation}

Tasks are created based on the blocks in a program.
Specifically, a block $B$ in the program is mapped to a set $T(B)$ of tasks.
Since a block may be executed multiple times, multiple instances of $T(B)$ can be created at runtime.

\begin{figure*}
    \centering
    \includegraphics[width=\textwidth]{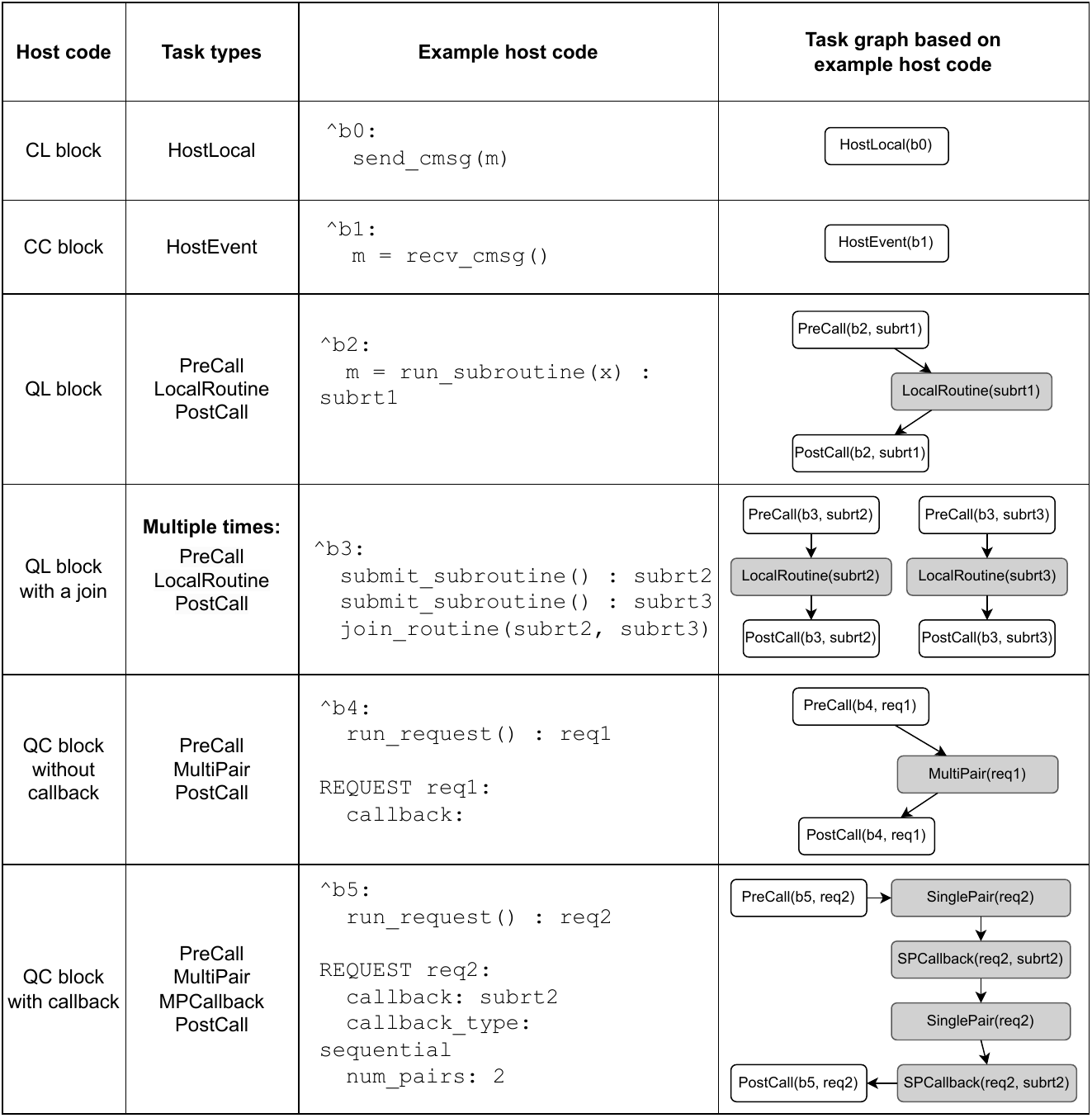}
    \caption{Overview of different host blocks with corresponding tasks. In the rightmost column, tasks with a dark background are QPS tasks, the others are CPS tasks. This example shows that tasks contain data about the program segment they correspond to, such as \texttt{LocalRoutine} tasks having the name of the routine they are executing.}
    \label{fig:app:task_creation}
\end{figure*}

CL and CC blocks are mapped to CPS tasks only.
QL and QC blocks are mapped to a sequence of CPS- and QPS tasks.
\begin{itemize}
    \item CL block. A single \texttt{HostLocal} task is created.
    \item CC block. A single \texttt{HostEvent} task is created.
    \item QL block. If there is a single \texttt{run\_routine} call, a \texttt{LocalRoutine} task is created for the QPS, as well as a \texttt{PreCall} tasks and a \texttt{PostCall} task for the CPS.
    Two precedence constraints are added: the \texttt{PreCall} task precedes the \texttt{LocalRoutine} task, and the \texttt{LocalRoutine} task preceded the \texttt{PostCall} task.
    If there is a \texttt{join\_routines} on multiple local routine, multiple \texttt{PreCall}-\texttt{LocalRoutine}-\texttt{PostCall} task sets are created, without any dependencies between the task sets.
    \item QC block. If the request that is called from this block is for a single pair,
    a \texttt{SinglePair} task is created. If the request is for more than 1 pair, a \texttt{MultiPair} task is created. In both cases, an additional \texttt{PreCall} and a \texttt{PostCall} task are created with precedence constraints like for QL blocks.
    If there is a \texttt{join\_routines} on multiple request routines, multiple \texttt{PreCall}-Pair-\texttt{PostCall} task sets are created, without any dependencies between the task sets.
\end{itemize}
\cref{fig:app:task_creation} shows an overview of blocks and corresponding tasks and their precedence constraints.

\begin{figure}[ht]
    \centering
    \includegraphics[width=0.7\columnwidth]{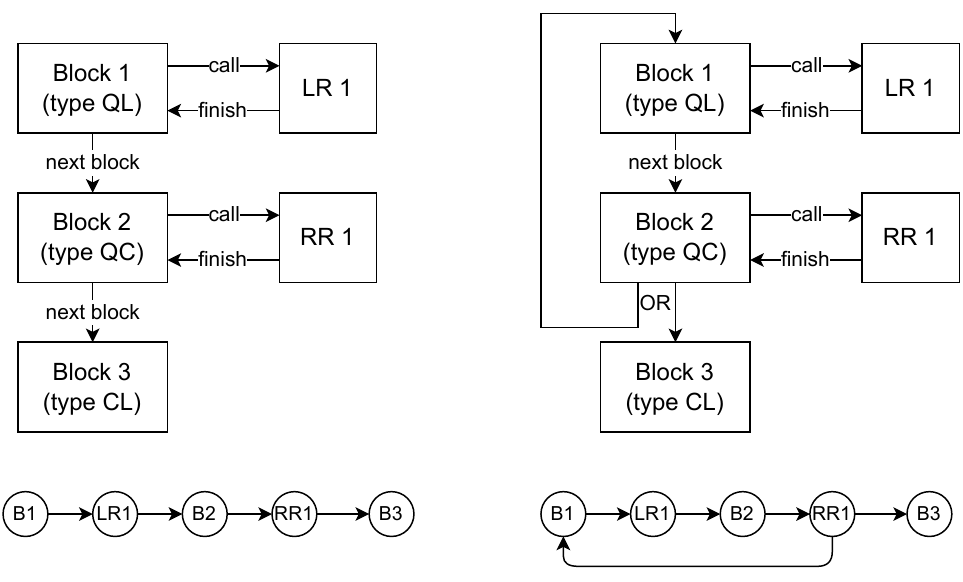}
    \caption{Schematic overview of the difference between predictable and non-predictable programs.
    The control-flow of the predictable program (left) is linear: first block 1 is executed (calling local routine (LR) 1), then block 2 (calling request routine (RR) 1), and finally block 3.
    Therefore, the number of tasks is fixed and known before execution.
    The non-predictable program is similar but after executing block 2, control-flow may go back to block 1 (again), depending on a runtime value (e.g. the result of RR 1).
    Hence, the number of times that blocks 1 and 2 are executed is not known beforehand, and therefore the number of tasks is also not known.
    }
    \label{fig:app:linear_non_linear}
\end{figure}

\paragraph{Predictable vs unpredictable programs}
Tasks are created based on the contents of a program instance, and their precedence relations are defined by the control-flow of the blocks in the program's host code.
Because of jump and branch instructions in the host code, a block may be executed zero, one or multiple times.
Furthermore, the exact number of executions of a block may not be known ahead of time.
For example, a program might loop through a sequence of blocks by using a conditional branch instruction at the end of the last block of the sequence.
The condition could depend on a runtime value (such as the result of a quantum measurement).
We say that control-flow is \textit{predictable} if it can be completely known before runtime. 
\textit{Unpredictable} control-flow, on the other hand, depends on values available only at runtime. 
For predictable programs, all its tasks can be created before runtime.
For unpredictable programs, (some of) its tasks must be created on-the-fly during program execution.
\cref{fig:app:linear_non_linear} illustrates the difference between predictable and unpredictable programs.

\paragraph{Task execution}
Tasks are executed by the CPS or the QPS, and the specific operations involved depend on the type of the task.

\textbf{\texttt{HostLocal} task execution.} A \texttt{HostLocal} task $t_{hl} = (P, B)$ for program instance $P$ and block $B$ is handled by executing each of the instructions in $B$. When the task finishes, the name of the next block to execute is recorded. If $B$ ends with a branch instruction, this is the target block; otherwise it is the next block in the program (if this was the last block, the next block is nil).

\textbf{\texttt{HostEvent} task execution}. A \texttt{HostEvent} task represents a block $B$ of type $CC$, which must start with exactly one \texttt{recv\_cmsg} instruction. Handling the task involves reading a message from the message buffer and assigning it to the result variable of the receive instruction. Then, the remaining instructions in $B$ are executed just like in a \texttt{HostLocal} task.

\textbf{\texttt{PreCall} task execution.} A \texttt{PreCall} task corresponds to a LR call instruction in Host code. The CPS allocates space in the shared memory for arguments and results. It then writes argument values to the shared memory.

\textbf{\texttt{PostCall} task execution.} A \texttt{PostCall} task corresponds to a LR call instruction in Host code. The CPS reads the results from the shared memory and copies them to the corresponding variables in the host local memory.

\textbf{\texttt{LocalRoutine} task execution.} A \texttt{LocalRoutine} task is executed by the QPS. It involves the following steps. First, based on information in the uses/keeps metadata, virtual quantum memory is allocated. Then all NetQASM instructions are executed, which may involve loading values from shared memory (reading arguments) and storing values to shared memory (populating results). Finally, quantum memory is freed.

\textbf{\texttt{SinglePair} task execution.} A \texttt{SinglePair} task is executed by the QPS. First, arguments are read from shared memory. Then, an EPR request (see \cref{app:entanglement_distribution}) is sent to the network controller.

\textbf{\texttt{MultiPair} task execution.} A \texttt{MultiPair} task is executed by the QPS. First, arguments are read from shared memory. Then, multiple EPR requests are sent to the network controller. Whether these requests are all sent at once or consecutively and waiting for intermediate responses is up to the implementer; the choice may depend on efficiency and resource considerations.

\textbf{\texttt{SinglePairCallback} and \texttt{MultiPairCallback} task execution.} First read results (from a \texttt{SinglePair} or \texttt{MultiPair} task) from shared memory. Then execute the callback routine just like a \texttt{LocalRoutine} task.

\paragraph{Deadlines}
Deadlines can be specified for blocks relative to other blocks using the syntax:

\begin{qoalacode}[caption=Pseudocode for the Algorithm, label=lst:pseudocode]
(*@\textcolor{purple}{\textasciicircum block\_0}@*):
    ...

(*@\textcolor{purple}{\textasciicircum block\_1}@*) { deadlines = [b0: 3ms] }:  // relative deadline of 3 ms compared to block_0
    ...
\end{qoalacode}

A relative deadline to some block $B$ is always with respect to the last task in $T(B)$, for the last task set instance (in case of multiple execution of this task set).
The deadline value may be an explicit value (like $3 ms$) or it can be in terms of EHI values, such as for example $0.1 * CC$ where $CC$ is the expected classical node-node latency provided by the EHI.

\paragraph{Precedence constraints}
By default, blocks are executed in the order they are given in the program.
Blocks ending with a jump or branch instruction define precedence constraints at runtime for unpredictable programs.

Scheduling happens at runtime and involves choosing which task to execute next.
In Qoala, there are three schedulers per node: the \textit{CPS scheduler} controls task execution on the CPS,
the \textit{QPS scheduler} controls task execution on the QPS, and the \textit{node scheduler} controls the CPS- and QPS schedulers.
The CPS- and QPS schedulers are both processor schedulers.

\subsection{Scheduling}
In this and the following sections we describe the scheduler from our implementation (\cref{sec:implementation}).

\begin{figure}[ht]
    \centering
    \includegraphics[width=0.6\columnwidth]{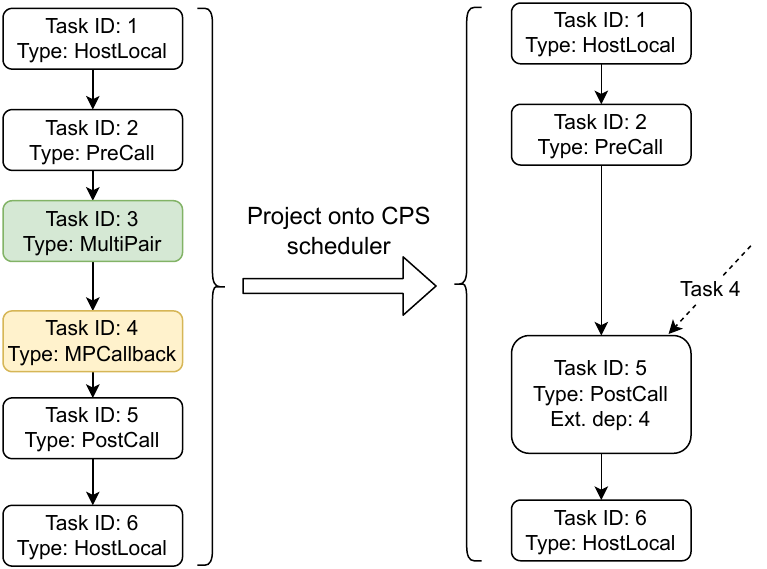}
    \caption{
        Example of a mapping from a full task graph (containing both CPS and QPS tasks) to a partial graph (containing only CPS tasks).
        Task 5 depends on task 4, which is external from the perspective of the CPS scheduler (indicated using the \texttt{external-dependencies} attribute).
        Note that Task 3 is not needed at all in the partial graph; only the dependency on task 4.
    }
    \label{fig:app:task_graph_partial}
\end{figure}

Each scheduler maintains their own task graph,
which is a directed acyclic graph (DAG) in which the nodes represent tasks and edges represent precedence constraints.
The node scheduler task graph contains all tasks (CPS or QPS) that are to be executed.
Each processor scheduler task graph is a partial copy of the node scheduler task graph containing only the tasks that can be executed by its own processor.
Edges in the node scheduler graph between heterogenous tasks (i.e. between CPS and QPS tasks) are represented in the partial processor graphs by the \textit{external dependencies} node attribute. See \cref{fig:app:task_graph_partial} for an example.
When a processor scheduler finishes a task, it is removed from the task graph and a signal is sent to the node scheduler.
The node scheduler updates its own task graph accordingly, and may then add new tasks to the task graph of the processor scheduler.
Note that although the processor task graphs are accessible by both the owning processor scheduler and the node scheduler, there are no read/write conflicts
since tasks can only be added by the node scheduler, and tasks can only be removed by the processor scheduler.

\paragraph{Task graph}
A task graph consists of 
\begin{itemize}
\item \textit{tasks} to be scheduled (the nodes),
\item \textit{precedence constraints} between the tasks (precedence edges),
\item \textit{external precedence constraints} for tasks in the case of processor task graphs (annotated on the nodes),
\item \textit{relative deadlines} between tasks (deadline edges),
\item \textit{trigger annotations} for some tasks (like incoming messages or network schedule timestamps)
\end{itemize}

Upon program instantiation, all created tasks are added to the node scheduler task graph, and the relevant tasks are added to the processor schedulers.
The number of tasks that are created (and can hence be added to the task graphs) depends on the predictability of the program.
During runtime, the node scheduler may create new tasks based on the control-flow of the program.

\paragraph{Task graph splitting}
\label{app:task_graph_splitting}
The node scheduler creates a heterogenous task graph consisting of both CPS and QPS tasks.
This graph needs to be split into a partial CPS and a partial QPS graph.
This is done using the following algorithm.

We consider creating the partial graph for the CPS, and hence the QPS is `the other processor'.
For the partial graph of the QPS the procedure is exactly the same but with reversed roles.

For a heterogenous task graph $G$ containing tasks $T$ (all tasks for both CPS and QPS), precedence constraints $P$ ($(t_1, t_2) \in P$ means that $t_1$ must precede $t_2$), compute the partial CPS graph $G_{CPS}$ as follows:

\begin{itemize}
    \item Split $T$ into a set $T_{CPS}$ consisting of all tasks that run on the CPS, and $T_{QPS}$ consisting of all other tasks in $T$. $G_{CPS}$ will consist of only tasks in $T_{CPS}$.
    \item Let $P_{CPS} \subset P$ consist of all precedence constraints $(t_1, t_2)$ where $t_1 \in T_{CPS}$ and $t_2 \in T_{CPS}$.
    These constraints will remain the same in $G_{CPS}$ since they are between tasks in $T'$.
    \item Compute the `immediate cross-predecessors' set $I$ of all tasks $t_{cp} \in T_{QPS}$ such that there exists a task $t \in T_{CPS}$ and $(t_{cp}, t) \in P$.
    In other words, $I$ contains all tasks running on the QPS that are immediate predecessors of CPS tasks.
    \item For each $t_i \in I$, compute the `closest CPS ancestor' task $t_{anc} \in T_{CPS}$, which is a CPS task that has a direct precedence constrain with the closest ancestor of $t_i$.
    Add $(t_{anc}, t_i)$ to the precedence constraints of $G_{CPS}$.
\end{itemize}

\paragraph{Scheduler communication}
Here we describe how the schedulers communicate in our implementation (\cref{sec:implementation}).

The three schedulers need to exchange information in order to work together.
All schedulers can broadcast a \textit{signal} with short information such as `task N completed' or `memory freed'. Each scheduler receives these signals.
Furthermore, the following read and write access is given:
\begin{itemize}
    \item The CPS scheduler can read from the completed task ID list of the QPS and vice versa. This makes it possible for the CPS (QPS) scheduler to directly update their remote dependencies without having to wait for a signal from the node scheduler, leading to overall improvement in efficiency
    \item The node scheduler can add new tasks to the partial graphs of the CPS and QPS. Note that the node scheduler will only add tasks to the partial graph of a processor scheduler when this scheduler is in a waiting state; that is, after the processor scheduler has sent a `waiting` signal and before the node scheduler has sent a `task added` signal (only after this signal will the processor scheduler continue). In this way, there are no read/write conflicts in the partial graphs of processor schedulers.
    \item The CPS (QPS) scheduler can only remove tasks from its own partial graph, not add any.
\end{itemize}

\subsection{Scheduler algorithms}

\paragraph{Node scheduler algorithm}
Below we describe the high-level steps involved in the node scheduler algorithm implementation of \cref{sec:implementation}.
\label{app:node_scheduler_algorithm}
\begin{enumerate}
    \item Split the current task graph into a partial CPS graph and a partial QPS graph. For the algorithm, see `Task graph splitting' above.
    \item Add the CPS (QPS) tasks to the partial graph of the CPS (QPS) scheduler
    \item Wait for a `task finished` signal from either CPS or QPS scheduler
    \item Remove the corresponding task from the task graph.
    \item If the finished task was a \texttt{HostLocal} task for some program instance P, and if the CPS partial graph is empty, check which block the program instance should jump to. This information is given by the task itself (and stored in the completed task list of the CPS scheduler), after evaluating the last instruction (a jump or branch instruction) in the BB that the task represented. For this new BB, create corresponding tasks for both the CPS and QPS. Task creation is discussed in \cref{app:scheduling_tasks}.
    \item If the task graph is empty, idle until new programs are instantiated.
    \item Go back to step 1.
\end{enumerate}

Note that the role of the node scheduler is much smaller when only predictable programs are run.
When predictable programs are instantiated, all of their tasks are created at once, resulting in a large task graph in the node scheduler, which never gets new tasks created at runtime.
In this scenario, after steps 1 and 2 the CPS and QPS schedulers possess a partial graph which will never get any new tasks.
Both processor schedulers will work on their tasks until they are both empty, after which all program instances have finished. Meanwhile, the node scheduler just loops through steps 1, 2, 3 and 6, not doing anything.

\paragraph{CPS scheduler algorithm}
Below we describe the high-level steps involved in the CPS scheduler algorithm implementation of \cref{sec:implementation}.
\begin{enumerate}
    \item Check which new tasks were completed by the QPS by reading from the shared task memory. Remove external dependency edges that correspond to QPS tasks that have completed.
    \item Find all tasks in the partial graph that are ready to execute. These are tasks that fulfill all following requirements:
        \begin{itemize}
            \item The task has no incoming precedence constraints (there are no unfinished tasks in the task graph that must precede this task)
            \item The task has no external precedence constraints (there are no unfinished QPS tasks that must precede this task)
            \item If the task is a \texttt{HostEvent} task, there must be at least one message in the CPS' message buffer
            \item If the task has a specific start time, the current time should be at least the start time
        \end{itemize}
    \item If there is no task ready to execute, send a `waiting` signal and wait until a signal is received that indicates one of the following events:
        \begin{itemize}
            \item The node scheduler has added one or more tasks to the partial graph
            \item The QPS scheduler has completed a task
            \item The start time has arrived of one of the tasks that were previously not ready only because their start time had not yet passed
            \item One or more new messages have been put into the message buffer
        \end{itemize}
        After one of these signals is received, go back to step 1.
    \item If there is at least one task ready to execute, choose which one to execute now. This depends on the scheduling policy that is being used. The policy may or may not use information about the deadlines of the available tasks. Scheduling policies that were implemented for our evaluation are described in \cref{app:evaluation}.
    \item If the task failed, go back to step 1
    \item If the task completed, remove it from the partial graph, add its ID to the completed task ID list, and broadcast a signal that the task was finished. If the task was a \texttt{HostLocal} task, then also store (in the completed task list) an entry containing the name of the next block to execute. (In this way, the node scheduler knows which task(s) to create and add to the full task graph. See \cref{app:scheduling_tasks} for more details.)
    Update the deadlines of all other tasks in the task graph.
    Then go back to step 1.
\end{enumerate}

\paragraph{QPS scheduler algorithm}
Below we describe the high-level steps involved in the QPS scheduler algorithm implementation of \cref{sec:implementation}.
\begin{enumerate}
    \item Check which new tasks were completed by the CPS by reading from the shared task memory. Remove external dependency edges that correspond to CPS tasks that have completed.
    \item Find all tasks in the partial graph that are ready to execute. These are tasks that fulfill all following requirements:
        \begin{itemize}
            \item The task has no incoming precedence constraints (there are no unfinished tasks in the task graph that must precede this task)
            \item The task has no external precedence constraints (there are no unfinished CPS tasks that must precede this task)
            \item If the task is a \texttt{SinglePair} or \texttt{MultiPair} task, the current time should be the beginning of a network time slot that corresponds to this task. (For example, if the task is for creating EPR pairs for program instance 1 on this node (called `Alice') and program instance 2 on node `Bob', then the current time should be the start of a $(Alice, 1, Bob, 2)$ time slot).
            \item If the task has a specific start time, the current time should be at least the start time
        \end{itemize}
    \item If there is no task ready to execute, wait for a signal that indicates one of the following events:
        \begin{itemize}
            \item The node scheduler has added one or more tasks to the partial graph
            \item The CPS scheduler has completed a task
            \item The start time has arrived of one of the tasks that were previously not ready only because their start time had not yet passed
            \item The start of a time slot has arrived which corresponds to one of the tasks that were previously only blocked on the arrival of this time slot
        \end{itemize}
        After one of these signals is received, go back to step 1.
    \item If there is at least one task ready to execute, choose which one to execute now. This depends on the scheduling policy that is being used. The policy may or may not use information about the deadlines of the available tasks.
    \item If the task failed, go back to step 1
    \item If the task completed, remove it from the partial graph, add its ID to the completed task ID list, and broadcast a signal that the task was finished.
    Update the deadlines of all other tasks in the task graph.
    Then go back to step 1.
\end{enumerate}

\paragraph{Task graph updates}
The node scheduler may add tasks to the current task graph of the CPS or QPS.
When a processor scheduler has finished a task, it is removed from the task graph.
This has the following effects:
\begin{itemize}
    \item Precedence edges from this task are removed, potentially making other tasks available for execution
    \item The time of finishing is recorded; and the deadlines and relative deadlines of all other tasks are updated accordingly
\end{itemize}

\subsection{Other algorithms}
\paragraph{Linear graphs}

When instantiating a program multiple times (for example instantiating a BQC program 1000 times), one has the option to linearize the graphs. Each instantiation has its own graph,
and the full graph of all instances result in many independent tasks.
One can force all instances to be run in sequence, rather than interleaved, resulting in a linear chain of single-instance graphs. This is done using the following algorithm:

\begin{itemize}
    \item For each pair $(i_1, i_2)$ of consecutive instances, add a precedence constraint between the last tasks(s) of $i_1$ and the first task(s) $i_2$.
\end{itemize}

\begin{figure*}
    \newcommand{\networkcontrollerfigheight}{4.5cm}
    \centering
    \subfloat[\centering \label{fig:app:network_controller_requests}]{{\includegraphics[height=\networkcontrollerfigheight, keepaspectratio]{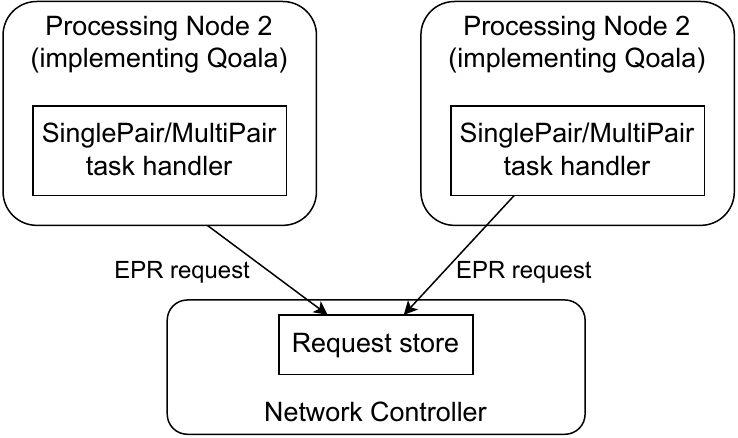}}}%
    \qquad
    \subfloat[\centering \label{fig:app:network_controller_requests_distributed}]{{\includegraphics[height=\networkcontrollerfigheight, keepaspectratio]{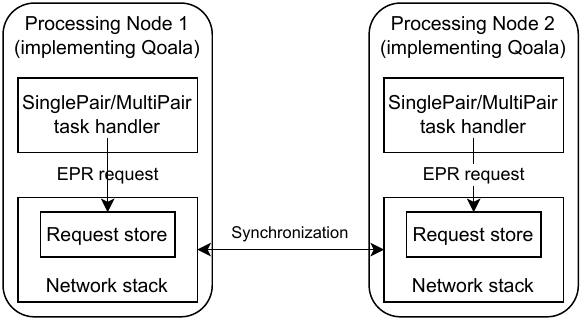}}}%
    \caption{
    Different implementations of network controller and network stack.
    (a) The network controller is centralized and the nodes send requests to this controller
    whenever they are executing \texttt{SinglePair} or \texttt{MultiPair} tasks.
    (b) The network controller is distributed over the nodes. Inside each node there is a network stack which autonomously talks with the network stack of other nodes and synchronizes entanglement generation.
    Execution of \texttt{SinglePair} and \texttt{MultiPair} tasks involves sending a request to the network stack within the node, which then handles pair generation by synchronizing with the network stack in other nodes.
    }%
    \label{fig:app:network_controller_types}
\end{figure*}

\paragraph{Estimating task durations}
The scheduler uses the EHI to estimate the duration of a task.
This duration may then be used by the scheduler to decide which task to execute when.
In our implementation, the scheduler does not make use of these estimates, but we did implement a simple estimator algorithm:

The estimated duration $E$ of a task is computed as follows:
\begin{itemize}
    \item For a \texttt{HostLocal} or \texttt{HostEvent} task representing a program block $B$, $E$ is $N \cdot$ \texttt{host\_latency} where $N$ is the number of HostLanguage operations in $B$ and \texttt{host\_latency} is given in the EHI.
    \item For a \texttt{LocalRoutine} tasks representing a block that call a NetQASM routine $S$, 
        $E$ is the sum of estimated durations of each NetQAM instruction in $S$. The duration of each quantum instruction is obtained from the EHI, and the duration of each classical instruction is given by the \texttt{qnos\_latency} entry in the EHI.
    \item For a \texttt{SinglePair} or \texttt{MultiPair} task based on a block that calls a request $R$ for $N$ EPR pairs, $E$ is $N$ times the duration of a single EPR generation as listed in the EHI.
    \item For \texttt{PreCall} and \texttt{PostCall} tasks, the duration is set to the \texttt{host\_latency} entry in the EHI.
\end{itemize}

\subsection{Entanglement Distribution}
\label{app:entanglement_distribution}
Qoala only defines how program are executed on a node in a quantum network,
and not how and when entanglement is created between nodes.
However, Qoala does assume certain things about how nodes can interact with the entanglement distribution system, however this is implemented.
The assumption about entanglement generation are as follows.

\textbf{Network controller with time slots.}
Conceptually, there is a network controller that oversees entanglement generation and distribution across the whole network.
Qoala does not care whether this controller is implemented as a single entity, or is distributed in some way across multiple (processing) nodes (\cref{fig:app:network_controller_types}).
The network controller maintains a global timeline divided into \textit{time slots}, which can have arbitrary length.
Each time slot may be assigned to a \textit{session}, which is a 4-tuple $(N1, P1, N2, P2)$ where $N1$ ($N2$) is the name of a node in the network and $P1$ ($P2$) is an ID of a program instance running within $N1$ ($N2$).
A session hence represents a pair of running program instances across two nodes, and it is such pairs of program instances that want to create entanglement with each other. 
If a time slot is assigned to some session $(N1, P1, N2, P2)$, only program instances $P1$ and $P2$ (on nodes $N1$ and $N2$) may create entanglement with each other during this time slot.

\begin{figure*}[ht]
    \centering
    \includegraphics[width=0.5\textwidth]{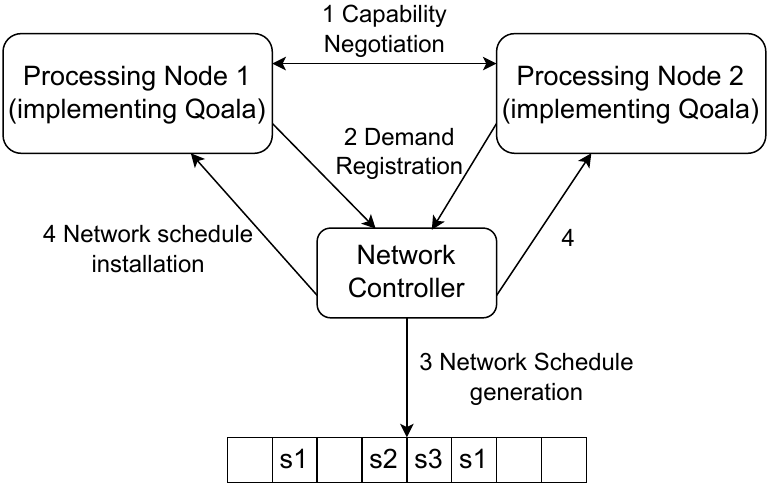}
    \caption{High-level steps of using the network controller.
    1. Nodes discuss among each other constraints about application execution (Capability Negotiation).
    2. The outcome of Capability Negotiation, which contains demands about entanglement generation, is sent to the network controller (Demand Registration).
    3. Based on the demands from the nodes, the network controller constructs a network schedule consisting of time slots. Each time slot is assigned to zero or more \textit{sessions}, which correspond to program instance pairs.
    }
    \label{fig:app:network_controller_setup}
\end{figure*}

Populating the network controller's time slot with sessions is the result of (1) demand registration by nodes in the network, followed by (2) network schedule generation by the network controller itself, which we do not consider here (\cref{fig:app:network_controller_setup}).
In the following, we simply assume that the network controller has a list of time slots assigned to sessions relating to program instances that are being run, and that these time slots are also known by the individual processing nodes.

\textbf{On-demand entanglement requests.} At runtime, nodes implementing Qoala may send requests to the \textit{network stack}.
This network stack then issues \textit{EPR requests} to the network controller.
Upon receiving an EPR request, the network controller stores it and potentially acts on it:
\begin{itemize}
    \item If there is a matching EPR request from the other node, and if the current time slot is assigned to the corresponding session, perform the actual entanglement generation process.
    \item If at least one of the two above conditions does not hold, keep the request until both conditions are satisfied (a matching request from the other node arrives, or the corresponding time slot arrives, or both).
\end{itemize}

An EPR request is a request for a single EPR pair. A \texttt{SinglePair} task is handled by the network stack sending a single EPR request to the network controller.
A \texttt{MultiPair} task is handled by sending multiple EPR requests, possibly interleaved by local QPS processing such as callback routines.

The network stack may fail handling a request. For example, it might timeout trying to produce an EPR pair. In this case, the corresponding task (\texttt{SinglePair} or \texttt{MultiPair}) also fails.
Depending on the scheduler implementation, this task may be executed again at a later time, or the whole program instance may be aborted.

\textbf{Entanglement generation as a black box.} We assume that all nodes can create entanglement with all nodes, orchestrated by the network controller.
Qoala does not assume anything about the existence of repeater nodes or entanglement routing algorithms.
Rather, a node sending a request for entanglement (in a suitable time slot) will either get this entanglement (created in some way, irrelevant to Qoala) or not (creation failed for some reason, again irrelevant to Qoala).
The network stack and controller may be implemented in various ways, such as illustrated in \cref{fig:app:network_controller_types}.

\clearpage
\section{Simulator implementation}
\label{app:simulator}

\subsection{Package overview}
The simulator has been implemented as a Python package and is available at~\cite{qoala2023simulator}.
It is divided into three subpackages: (1) \texttt{lang}, defining the format of Qoala programs and of the EHI, (2) \texttt{runtime} defining common types for the runtime system, and (3) \texttt{sim} containing Netsquid objects that implement the Qoala runtime.

The division into subpackages is made in such a way that only \texttt{sim} depends on (imports from) Netsquid; the other two subpackages are implementation independent.
The \texttt{lang} can be used standalone in a compiler, without having the compiler to depend on the runtime implementation, whether that is in simulation or on real hardware.

\subsection{Netsquid: Protocols, Components, and Listeners}
The Qoala simulator make heavy use of Netsquid's \textit{Protocol} class,
which can be used to model concurrent software systems. 
Each protocol defines its own run function, and the Netsquid simulator
executes the run functions of each protocol concurrently.
(Netsquid uses only a single thread, but protocols are run interleaved, i.e. NetSquid provides provides an asynchronous runtime).

The simulator implements a hierarchy of protocols. 
Each node in the network is a protocol, containing subprotocols for a \textit{Host, Qnos, and Netstack}. The Host represents the CPS, and the Qnos and Netstack together represent the QPS, where Qnos handle local quantum processing, and the Netstack handles requests to the central network controller. 

The protocol objects implement the runtime logic of the subsystems.
The Netsquid \textit{Component} holds static information about the subsystem, and contains \textit{Ports} for communicating with other components. Protocols use these ports to send messages to other protocols. 

Listener objects are a feature of the Qoala simulator that are protocols with the sole purpose to wait for any incoming messages on a port and then notifying the corresponding protocol of them.

\subsection{Interfaces and configuration}
The Qoala simulator allows for a lot of configuration.
The Low-level Hardware Interface (LHI) defines a format for defining physical quantum instructions, durations, and noise models.
Default values are provided for NV-centers and trapped ions, but custom hardware models can easily be added.
The LHI allows for representing real-life validated hardware, but also for simulating hardware that does not (yet) exist.

The LHI allows for the configurations of
\begin{itemize}
    \item Allowed gate types, gate durations, and gate noise models
    \item Qubit decoherence model and qubit topology in a node
    \item Topology of the network
    \item Entanglement fidelity and generation duration between pairs of nodes
    \item Classical communication latency between nodes
    \item Internal communication latency between scheduler components
    \item Duration of CPS instructions and of classical QPS instructions
\end{itemize}

The Native-To-Flavour (NTF) interface is used to define the translation from LHI physical quantum instructions to a NetQASM flavour.
The QPS is expected to provide an implementation of the NTF, such that it can translate instruction in Local Routines (which are from a particular NetQASM flavour) into the corresponding hardware instructions.

The Exposed Hardware Interface (EHI) is described in \cref{sec:architecture}.
The simulator provides automated tools for translating a combination of an LHI instance and a NTF into an EHI.

\subsection{Simulator Architecture}
\begin{figure*}[hp]
    \centering
    \includegraphics[width=\textwidth]{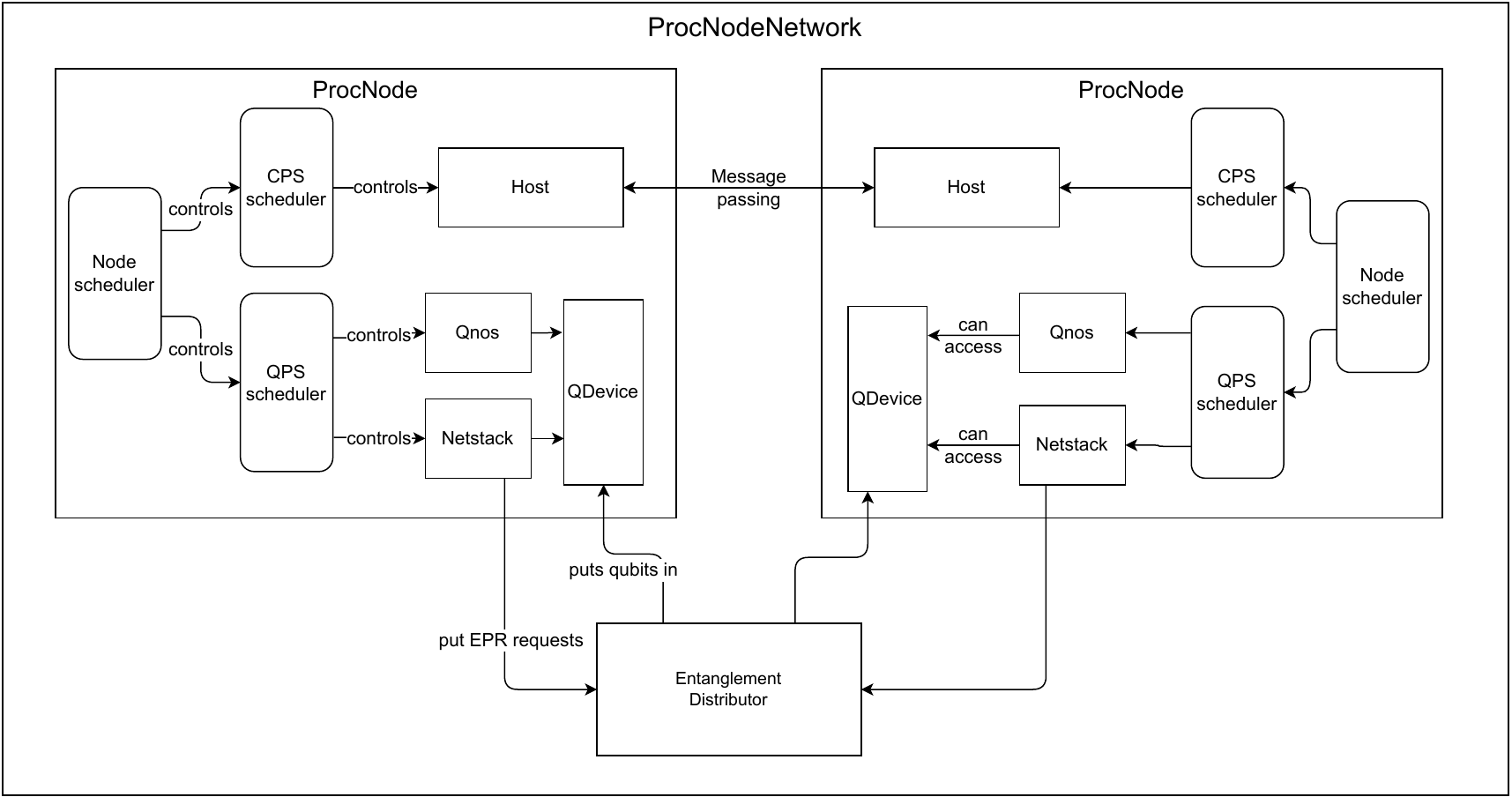}
    \caption{High-level overview of the simulator architecture. Each box represents a component.
    A single network component can contain multiple nodes (two are shown, but there can be more).
    Each node contains a node scheduler, CPS scheduler and QPS scheduler, implementing the algorithms as described in \cref{app:scheduling_execution}.
    A single entanglement distributor object creates entangled pairs and puts the qubits directly in the quantum memory of the nodes, in the QDevice.
    }
    \label{fig:app:simulator}
\end{figure*}

\begin{figure*}[hp]
    \centering
    \includegraphics[width=\textwidth]{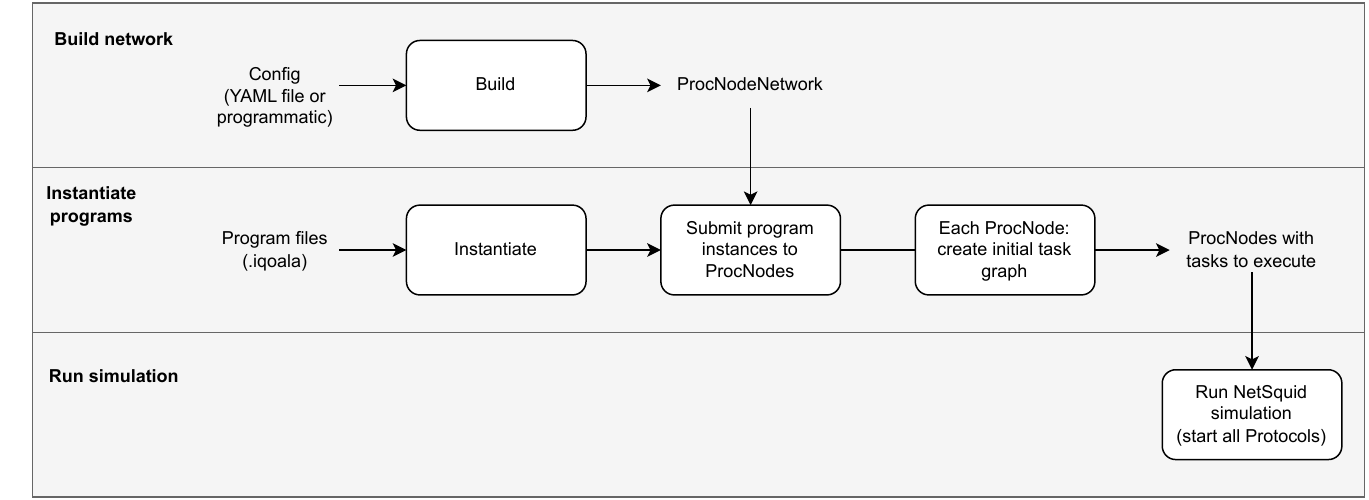}
    \caption{High-level overview of sequence of steps performed in order to simulate a network running programs on nodes implemented Qoala.
    }
    \label{fig:app:simulator_sequence}
\end{figure*}

The simulator defines various component and protocol classes representing the concepts defined in \cref{sec:architecture}. These classes can be instantiated in a custom way, and can hence be seen as building blocks. The simulator provides a default way of using these blocks (namely, in the way explained in the Qoala architecture), but it is possible to use these blocks in another way to investigate different architectures.
In \cref{fig:app:simulator} a schematic overview of the most important classes and their roles is given, and in \cref{fig:app:simulator_sequence} an overview is provided of the general sequence of actions involved when simulating the execution of one or more applications on a quantum network.

In our simulator, the network controller (\cref{app:entanglement_distribution})
is implemented as a single centralized entity called EntDist (\cref{fig:app:simulator}).

\clearpage
\section{Evaluation details}
\label{app:evaluation}

\subsection{Simulator setup}
All simulations have been done with the simulation package found at~\cite{qoala2023simulator}
and were run on a machine using 80 Intel Xeon Gold cores at 3.9 GHz and 192 GB of RAM.

\textbf{Code availability.}
All code used for the evaluations is available in the \texttt{evaluation/} folder~\cite{qoala2023simulator}.
For each evaluation done (each subsection in \cref{sec:evaluation}, it includes the Qoala program source code, the scripts for running the simulations, the scripts for producing the plots, and a \texttt{README} that explains how to use the code.
In the source code, the term \textit{time bin} is used for what we here call \textit{time slot}.

\subsection{Hardware parameters}
In this section we describe the characteristics of the hardware types that have been used in our evaluations.
For the evaluation in \cref{sec:demonstrating_architecture_effectiveness} we simulated all three types below;
for the other evaluations we only considered the generic hardware type.

\subsubsection{Generic hardware}
The allowed gate set is expressed as a particular NetQASM flavour~\cite{dahlberg2022netqasm}.

\begin{itemize}
  \item Allowed single-qubit gates (vanilla NetQASM flavour~\cite{dahlberg2022netqasm}):
  \texttt{init}, \texttt{rot\_x}, \texttt{rot\_y}, \texttt{rot\_z}, \texttt{x}, \texttt{y}, \texttt{z}, \texttt{h}, \texttt{meas}.
  \item Allowed two-qubit gates (vanilla NetQASM flavour): \texttt{cnot}, \texttt{cphase}.
\end{itemize}

Qubit decoherence times are expressed as T1 (amplitude damping) and T2 (dephasing time), which is commonly done in quantum computing.
Unless stated otherwise in the evaluation details below, the default noise and duration parameters used for the generic hardware are:
\begin{itemize}
  \item Single-qubit duration: $5 \cdot 10^3$ ns.
  \item Two-qubit duration: $200 \cdot 10^3$ ns.
  \item Qubit T1 time: $10^9$ ns.
  \item Qubit T2 time: $10^8$ ns.
\end{itemize}

\subsubsection{NV hardware}
Values from~\cite{avis2023requirements} and private communication.

\begin{itemize}
  \item Allowed single-qubit gates on communication qubit (NV NetQASM flavour): \texttt{init}, \texttt{rot\_x}, \texttt{rot\_y}, \texttt{meas}.
  \item Allowed single-qubit gates on memory qubit (NV NetQASM flavour): \texttt{init}, \texttt{rot\_x}, \texttt{rot\_y}, \texttt{rot\_z}, \texttt{meas}.
  \item Allowed two-qubit gates between communication qubit and memory qubit (NV NetQASM flavour): \texttt{crot\_x}, \texttt{crot\_y}.
\end{itemize}

Unless stated otherwise in the evaluation details below, the default noise and duration parameters used for the NV hardware are:
\begin{itemize}
  \item Single-qubit duration on communication qubit: 300 ns.
  \item Single-qubit duration on memory qubit: $1.2$ ms.
  \item Two-qubit duration: 1 ms.
  \item Communication qubit T1 time: 3600 ms
  \item Communication qubit T2 time: 500 ms
  \item Memory qubit T1 time: 35000 ms
  \item Memory qubit T2 time: 1 ms
\end{itemize}

\subsubsection{Trapped-ion hardware}
Values from~\cite{avis2023requirements} and private communication.

\begin{itemize}
  \item Allowed single-qubit gates (trapped-ion NetQASM flavour): \texttt{init}, \texttt{rot\_z}, \texttt{meas}.
  \item Allowed all-qubit gates (trapped-ion NetQASM flavour): \texttt{init\_all}, \texttt{meas\_all}, \texttt{rot\_x\_all}, \texttt{rot\_y\_all}, \texttt{rot\_z\_all}, \texttt{bichromatic}.
\end{itemize}

The effect of applying a bichromatic gate is expressed as
\[
   U_{XX}(\theta) = \exp(-i \frac{\theta}{2} \sum_{i<j} \sigma_X^{(i)} \sigma_X^{(j)})
\]

for some angle $\theta$.

Unless stated otherwise in the evaluation details below, the default noise and duration parameters used for the trapped-ion hardware are:
\begin{itemize}
  \item Single-qubit duration on communication qubit: 26.6 $\mu s$.
  \item All-qubit duration: 85 ms.
  \item Qubit T1 time: $\infty$.
  \item Qubit T2 time: 85 ms.
\end{itemize}

\subsubsection{NetQASM gate sequence for CNOT on trapped- ion hardware}
We list the sequence of netqasm instructions to effectively apply a CNOT gates on two qubits, which is non-trivial.

Assuming 2 qubits are in use, CNOT gate between qubit 0 and qubit 1 on trapped ion:
\begin{qoalacode}
  NETQASM:
    // cnot between q0 and q1
    rot_x_all 8 4
    rot_z Q0 8 4
    rot_x_all 24 4
    bichromatic 8 4
    rot_x_all 24 4
    rot_x_all 8 4
    rot_z Q0 24 4
    rot_x_all 24 4
\end{qoalacode}

\begin{table*}[]
\begin{tabular}{|l|l|l|l|l|l|}
\hline
\textbf{Application} & \textbf{Number of nodes} & \textbf{\begin{tabular}[c]{@{}l@{}}Number of EPR\\ pairs per instance\end{tabular}} & \textbf{\begin{tabular}[c]{@{}l@{}}Max number of\\ qubits per node\end{tabular}} & \textbf{\begin{tabular}[c]{@{}l@{}}Number of \\ instances\end{tabular}} & \textbf{\begin{tabular}[c]{@{}l@{}}Simulation\\ duration (s)\end{tabular}} \\ \hline
A1. QKD              & 2   & 1000   & 1     & 1000     & 2166   \\ \hline
A2. BQC              & 2   & 2      & 2     & 1000     & 227    \\ \hline
A3. Teleportation    & 2   & 1      & 2     & 1000     & 24     \\ \hline
A4. Ping-pong        & 2   & 2      & 2     & 1000     & 35     \\ \hline
A5. GHZ              & 3   & 4      & 2     & 1000     & 41     \\ \hline
\end{tabular}
\caption{Overview of application used in the evaluation described in \cref{sec:demonstrating_architecture_effectiveness}.
Each application was simulated three times, once for each hardware type (generic, NV, trapped-ion).
Each simulation was for 1000 instances of the application.
The simulation duration is an average over the three simulations per application.
}
\label{tab:app:applications}
\end{table*}

\begin{figure*}
    \centering

    \subfloat[\centering \label{fig:app:qkd_circuit}]{\includegraphics[scale=0.9]{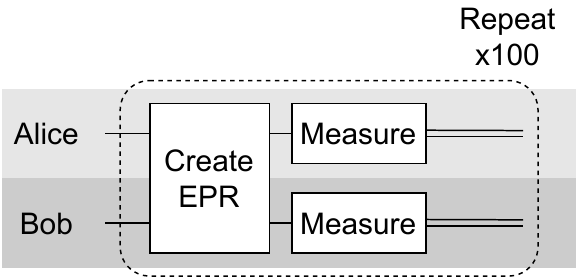}}
    \vspace{1cm}
    \subfloat[\centering \label{fig:app:teleport_circuit}]{\includegraphics[scale=0.8]{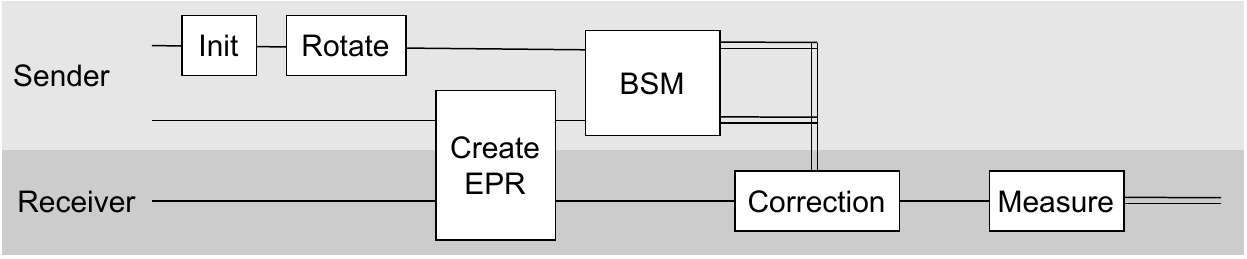}}
    \vspace{1cm}
    \subfloat[\centering \label{fig:app:pingpong_circuit}]{\includegraphics[scale=0.6]{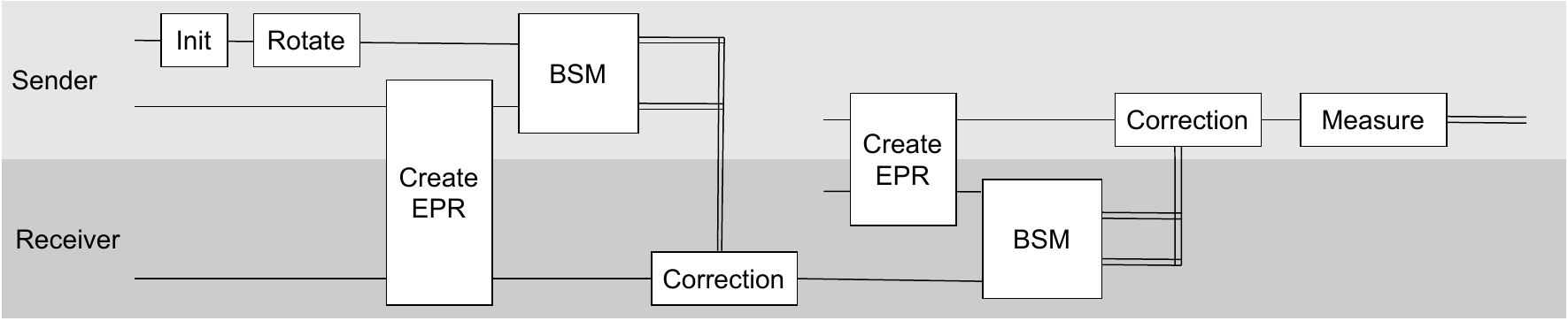}}
    \caption{
    Circuit for applications A1 (QKD), A3 (Teleport) and A4 (ping-pong) from \cref{sec:evaluation}.
    Single lines represent qubits. Double lines represent classical values.
    (a) QKD (A1). Two nodes repeatedly generate entangled pairs which are immediately measured.
    (b) Teleport (A3). A sender node (having 2 qubits) teleport a state to a receiver node. The sender applies local quantum operations (initialization, qubit rotation gates).
    The sender and receiver create an entangled pair. The sender performs local quantum gates and measurements resulting in classical outcomes.
    The sender sends the classical outcomes to the receiver. Based on the outcomes the receiver applies local quantum gates and measurement.
    (c) Ping-pong (A4). The sender teleports a state to the receiver and the receiver immediately teleports it back to the sender. In total, 2 entangled pairs are created.
    }
    \label{fig:app:circuits_1}
\end{figure*}

\begin{figure*}
    \centering

    \subfloat[\centering \label{fig:app:vbqc_circuit}]{\includegraphics[scale=0.45]{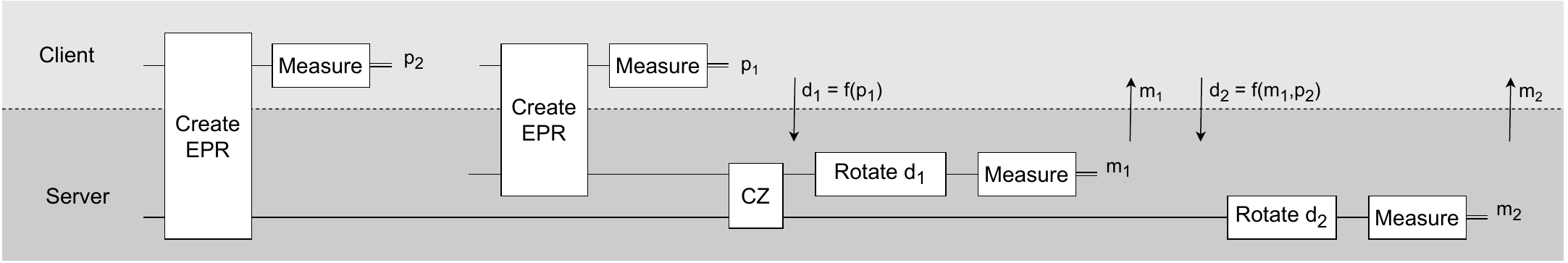}}
    \vspace{1cm}
    \subfloat[\centering \label{fig:app:ghz_circuit}]{\includegraphics[scale=0.45]{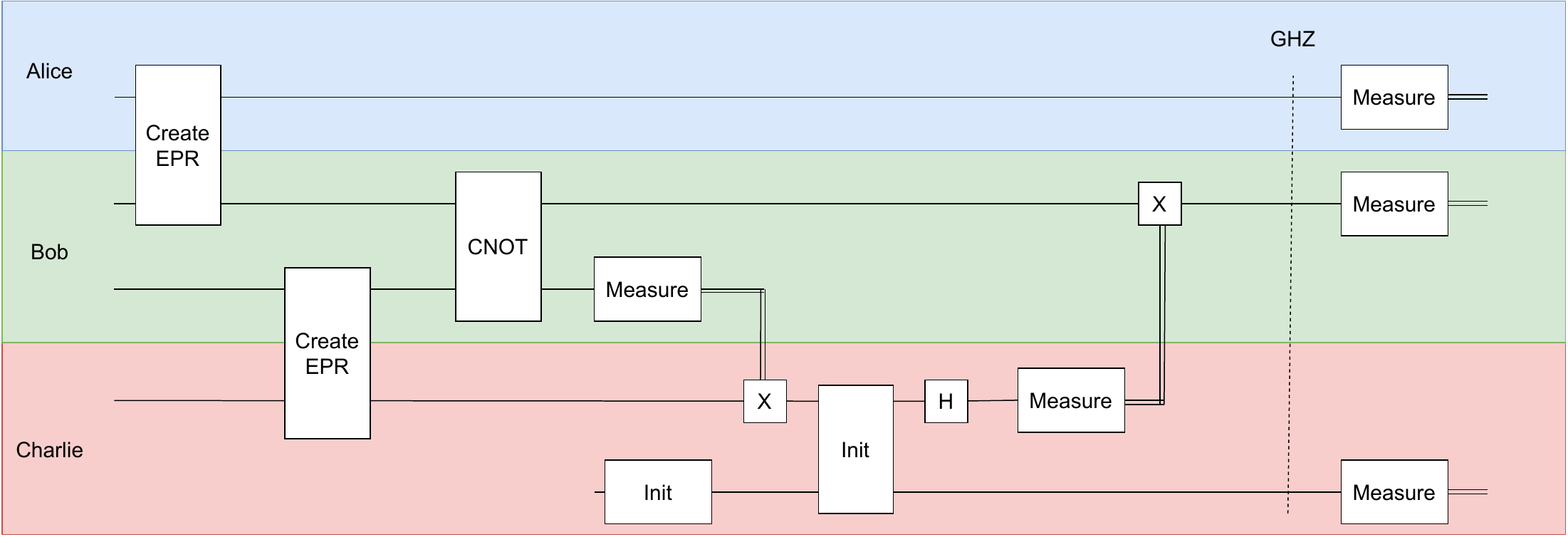}}
    \vspace{1cm}

    \caption{
    Circuit for applications A2 (BQC) and A5 (GHZ) from \cref{sec:evaluation}.
    Single lines represent qubits. Double lines represent classical values.
    (a) BQC (A2). A client node remotely prepares two qubits on a server node by creating an entangled pair and locally measuring its qubits, resulting in classical outcomes $p_1$ and $p_2$.
    The server applies a local two-qubit gate (CZ or cphase) on its qubits.
    The client sends a classical value $d_1$ which it calculates based on $p_1$ and other application input values.
    The server applies local gates based on $d_1$ and measures, resulting in classical value $m_1$ which it sends to the client.
    The client sends a classical value $d_2$ which it calculates based on $p_2$, $m_1$ and other application input values.
    The server applies local gates based on $d_2$ and measures, resulting in classical value $m_2$ which it sends to the client.
    The client uses the values $m_2$ to calculate the final result (not in the Figure).
    (b) Three nodes (Alice, Bob, and Charlie) create pair-wise entangled pairs. Bob applies local gates and measures one of his qubits, sending the outcomes to Charlie.
    Based on this outcome, Charlie performs local operations and sends a measurement outcome back to Bob.
    At the time of the vertical dashed line, the three nodes share a 3-qubit GHZ state. They all measure and check their correlations.
    }
    \label{fig:app:circuits_2}
\end{figure*}

\subsection{Details for VI.A (Demonstrating the architecture's effectiveness)}
\label{app:details_6_1}
A free network schedule was used, meaning that there were no specific time slots, and entanglement generation was allowed at any time.
Such a free network schedule was justified since we considered only whether the application ran successfully

All applications were run with both (a) hardware-validated parameters (see above); all executions were successful and (b) no-noise versions of these parameters (same durations of all operations but no decoherence nor gate noise); these were used to check if the expected outcomes were obtained.

\Cref{tab:app:applications} provides an overview of the applications.

\textbf{Quantum Key Distribution (QKD)}. 
Two programs (on two nodes): Alice and Bob. $N$ EPR are pairs are generated. Each generated pair is immediately measured by both programs.
This results in both programs having $N$ classical outcome bits.
See \cref{fig:app:qkd_circuit} for the circuit.
Success per instance is determined by checking that Alice and Bob got the same $N$ outcomes bits.
For the evaluation, we ran 1000 instances, each creating 1000 EPR pairs.

\textbf{Teleportation}.
Two programs (on two nodes): Sender and Receiver. The Sender teleports a \textit{state} (which is state is an argument to the Sender program) to the Receiver.
The Receiver measures in a \textit{basis} (which basis is an argument to the Receiver program) and obtains a single classical outcome bit which is the result of the application.
See \cref{fig:app:teleport_circuit} for the circuit.
Success per instance is determined by checking that the Receiver got the expected outcome bit (which depends on the combination of state and basis).
For the evaluation, we ran 1000 instances.

For each of the Sender program instances, the \textit{state} argument was chosen evenly from the following:
$\ket{0}$,
$\ket{1}$,
$\ket{+} = 1/\sqrt{2} (\ket{0} + \ket{1})$,
$\ket{-} = 1/\sqrt{2} (\ket{0} - \ket{1})$,
$\ket{+i} = 1/\sqrt{2} (\ket{0} + i \ket{1})$,
$\ket{-i} = 1/\sqrt{2} (\ket{0} - i \ket{1})$.

For each corresponding Receiver program instance, the \textit{basis} argument was chosen such that the expected outcome bit is always 1.
Hence, a single application instance succeeded if the Receiver outcome was 1.

\textbf{Ping-pong}. Teleportation from Sender to Receiver and immediately back to Sender.
Same as teleportation application, but the Receiver does not measure; the Sender receives the state back by teleportation and measures.
See \cref{fig:app:teleport_circuit} for the circuit.
State and basis per instance were chosen similarly as for the teleportation application.
Success now depends on the Sender measurement outcome being 1.
For the evaluation, we ran 1000 instances.

\textbf{Blind Quantum Computation (BQC)}.
Two programs (on two nodes): Client and Server.
Two EPR pairs are generated, after which 2 rounds happen. In each round, the client sends a classical message to the server, after which the server performs a measurement on one of its qubits, sending the measurement outcome back.
The same BQC application was used as in~\cite{dahlberg2022netqasm}, using values $\alpha = \pi/2$ and $\beta = -\pi/2$.
See \cref{fig:app:vbqc_circuit} for the circuit.
Success per instance is determined by checking that the Client received the expected classical bit $m_2$.
For the evaluation, we ran 1000 instances.

\textbf{GHZ}. 
Three programs (on three nodes): Alice, Bob, and Charlie.
Alice creates an EPR pair with Bob, and Bob creates an EPR pair with Charlie.
Then, local gates and classical messages are sent between the nodes, resulting in a 3-qubit state (one qubit per node) that is an entangled \textit{GHZ state}~\cite{greenberger1989going}.
At the end, each program measures its own qubit.
See \cref{fig:app:ghz_circuit} for the circuit.
Success per instance is determined by checking that the all three programs got the same measurement outcome.
For the evaluation, we ran 1000 instances.

\subsection{Details for VI.B (Demonstrating Qoala's multitasking potential and Network schedule impact)}

\subsubsection{Multitasking of teleportation and of BQC}
\textbf{Sequential vs Interleaved execution.}
Sequential: All tasks for all instances were created and added to the task graph at the beginning, but additional precedence constraints were added between the last task for each instance and the first task of the next instance. This resulting in the sequential execution of the 10 instances.
Interleaved: All tasks for all instances were created and added to the task graph at the beginning, and no additional precedence constraints were added. We used an FCFS scheduler to pick tasks; since there were no precedence constraints between tasks of different instances, the execution of instances was interleaved.

\textbf{Teleportation multitasking scenario.}
One sender node and one receiver node.
The teleportation application (A3 in \cref{sec:evaluation}, see also \cref{fig:app:teleport_circuit}) was instantiated 100 times.

Classical node-node communication latency: $10^7$ ns.
Sender node: 2 qubits.
Receiver node: sweep over range $[1, \dots, 6]$.
For each number of qubits $Q \in [1, \dots, 6]$, we ran a simulation using both a sequential and an interleaved scheduling approach.

For the self-preemption case, the teleportation application was only instantiated 5 times.

\textbf{BQC multitasking scenario.}
10 client nodes and one server node.
The BQC application (A2 in \cref{sec:evaluation}, see also \cref{fig:app:vbqc_circuit}) was instantiated 10 times for each client, for a total for 100 program instances.

Classical node-node communication latency: $10^5$ ns.
Client node: 2 qubits.
Server node: sweep over $\{2, 5, 10\}$.

For each number of qubits $Q \in \{2, 5, 10\}$, we ran a simulation using both a sequential and an interleaved scheduling approach.

\textbf{QKD-BQC multitasking scenario.}
One client node and one server node.
Client and server execute both (a) 50 instances of QKD (A1 in \cref{sec:evaluation}, see also \cref{fig:app:qkd_circuit}) and (b) 50 instances of BQC (A2 in \cref{sec:evaluation}, see also \cref{fig:app:vbqc_circuit}).

We compared two network schedules.
Sequential network schedule with repeating pattern $P_{seq}$. $P_{seq}$ consists of time slots $QKD_1$, $QKD_2$, $\dots$, $QKD_{50}$, $BQC_1$, $BQC_2$, $\dots$, $BQC_{50}$.
Alternating network schedule with repeating pattern $P_{alt}$. $P_{alt}$ consists of time slots $QKD_1$, $BQC_1$, $QKD_2$, $BQC_2$, $\dots$, $QKD_{50}$, $BQC_{50}$.

\subsection{Details for VI.C (Improvement over NetQASM architecture)}

\textbf{Scenario.}
Two nodes: client and server.
The client and server execute a remote measurement-based quantum computing (MBQC) application.
The server initializes local qubits and applies two-qubit gates on them, resulting in a cluster state of three qubits.
Then, three rounds of communication happen.
In each round, the client sends a classical message containing a measurement basis to the server, the server measures one of its qubits, and finally sends the measurement outcome to the client.
After three rounds, the application ends; the last message from the server is the result of the application.
This result has an expected value, which is used to determine if a single application instance succeeded or not.
The success probability is calculated as the fraction of instances that resulted in the expected value.

We consider a program implementation $P$ for the server.
The steps of $P$ are as follows.
\begin{enumerate}
  \item (Quantum) Initialize all three qubits and apply gates until the desired cluster state is realized.
  \item (Classical) Wait for a message $\theta_0$ from the client, representing the first measurement basis.
  \item (Quantum) Measure the first qubit in basis $B(\theta_0)$, resulting in classical bit $m_0$.
  \item (Classical) Send $m_0$ to the client.
  \item (Classical) Wait for a message $\theta_1$ from the client, representing the second measurement basis.
  \item (Quantum) Measure the second qubit in basis $B(\theta_1)$, resulting in classical bit $m_1$.
  \item (Classical) Send $m_1$ to the client.
  \item (Classical) Wait for a message $\theta_2$ from the client, representing the third measurement basis.
  \item (Quantum) Measure the third qubit in basis $B(\theta_2)$, resulting in classical bit $m_2$.
  \item (Classical) Send $m_2$ to the client.
\end{enumerate}

In our evaluation, we considered a program $P_{netqasm}$ written in the NetQASM runtime format~\cite{dahlberg2022netqasm}, which would be written in Python.
Specifically, $P_{netqasm}$ contains the above steps in Python code, in the same order.
The quantum steps are converted on-the-fly into NetQASM subroutines.
This means that, in the NetQASM runtime, we have the following execution:

$P_{netqasm}$ execution:
\begin{enumerate}
  \item NetQASM subroutine for initializing the three qubits.
  \item Classical Python code for waiting for $\theta_0$.
  \item NetQASM subroutine for measuring the first qubit.
  \item Classical Python code for sending for $m_0$.
  \item Classical Python code for waiting for $\theta_0$.
  \item NetQASM subroutine for measuring the second qubit.
  \item Classical Python code for sending for $m_1$.
  \item Classical Python code for waiting for $\theta_2$.
  \item NetQASM subroutine for measuring the third qubit.
  \item Classical Python code for sending for $m_2$.
\end{enumerate}

We note that since the NetQASM runtime does not allow for compilation across classical and quantum segments of the code,
there is no way to change the order of the steps.
In our evaluation, we represented $P_{netqasm}$ as a Qoala program $Q_{netqasm}$ with the exact same contents, but with classical code represented as host code, and the NetQASM subroutines as Qoala local routines.

$P$ can be optimized by noting that some qubit operations can be delayed until a later time, decreasing the duration the some qubits have to stay in memory.
This mitigates decoherence and it is expected that overall such an optimized program $P_{opt}$ leads to a higher success probability.

The steps of $P_{opt}$ are:
\begin{enumerate}
  \item Wait for a message $\theta_0$ from the client, representing the first measurement basis.
  \item Initialize the first 2 qubits and apply gates until a partial cluster state is realized.
  \item Measure the first qubit in basis $B(\theta_0)$, resulting in classical bit $m_0$.
  \item Send $m_0$ to the client.
  \item Wait for a message $\theta_1$ from the client, representing the second measurement basis.
  \item Initialize the third qubit and apply gates until the remaining partial cluster state is realized.
  \item Measure the second qubit in basis $B(\theta_1)$, resulting in classical bit $m_1$.
  \item Send $m_1$ to the client.
  \item Wait for a message $\theta_2$ from the client, representing the third measurement basis.
  \item Measure the third qubit in basis $B(\theta_2)$, resulting in classical bit $m_2$.
  \item Send $m_2$ to the client.
\end{enumerate}
where we note that the end-to-end behavior of $P_{netqasm}$ and $P_{opt}$ are the same, and hence $P_{opt}$ is a valid optimized version of $P$.
In our evaluation, we represented $P_{opt}$ as a Qoala program $Q_{opt}$ with the exact same steps.

For the client, we used a single program implementation $Q_{client}$, optimized for Qoala.

We compared running (NETQASM): $Q_{client}$ on the client node and $Q_{netqasm}$ on the server node with (QOALA): $Q_{client}$ on the client node and $Q_{opt}$ on the server node. In both cases we instantiated the application 1000 times.
We obtained success probabilities $66\%$ for NETQASM and $82\%$ for QOALA.

\subsection{Details for VI.D (Tradeoffs between classical and quantum performance metrics)}
\textbf{Scenario.}
Two nodes: Alice and Bob.
Bob executes an interactive quantum program where classical input is given by Alice.
Bob also executes a `busy' program consisting only of CPS tasks.

\textbf{Interactive program.}
The interactive program does the following steps:
(1) prepare a local qubit in a state (state given as program instance argument) by initialization and qubit rotation,
(2) send an `acknowledge' message to Alice,
(3) wait for a message from Alice,
(4) measure the local qubit in a basis (basis given as program instance argument)
(5) return the measurement result (classical bit).
For each combination of state and basis, an expected measurement result value is computed.
The success probability of the interactive program is given by the fraction of program instances that produces the expected value.
The interactive program was instantiated 1000 times.

\textbf{Busy program.}
The busy program consists only of a block which waits for some duration (input argument).
This waiting time mimics the CPS being busy with some local classical computation.

\textbf{Fixed parameters.}
Qubit coherence times: $T_1 = 10^{10}$ ns, $T_2 = 10^8$ ns.
Classical node-to-node communication latency: $10^7$ ns.
Rate of arrival of busy programs instances: once every $10^6$ ns.

\subsection{Details for VI.E (Success probabilities with quantum multitasking)}
\textbf{Local program.}
A program which prepares a single qubit to the $\ket{-} = 1/\sqrt{2} (\ket{0} + \ket{1})$ state, then waits for duration $d$, and then measures the qubit in the $X$-basis. The expected outcome bit is hence 1.

\textbf{Scenario.}
Two nodes: Alice and Bob.
Alice and Bob execute the teleportation application (A3 in \cref{sec:evaluation}, see also \cref{fig:app:teleport_circuit}) $T$ times.
Bob concurrently executes the local program described above $L$ times.
For each combination of $T \in [1, 15]$ and $L \in [1, 15]$, we ran a simulation 200 times, where in each simulation, all instances and all their tasks are created at the same time and added to the task graphs of the node.
The success per teleportation instance is calculated as in \cref{app:details_6_1}.
Success probability is calculated as the fraction of successful instances.
The success probability of the local program is calculated as the fraction of all local program instances (across all 200 runs) gave the expected classical result 1.

\textbf{Fixed parameters.}
Qubit coherence times: $T_1 = 10^{10}$ ns, $T_2 = 10^7$ ns.
Classical node-node communication latency: $0.1$ ms.
Network schedule: repeating pattern $P$ of time slots, where $P = \langle 0, \dots, n \rangle$ where $n$ is the number of teleportation instances and where each $i$ is associated with a single teleportation instance.
Number of qubits at Bob: $10$.
Number of qubits at Alice: $20$.

\subsection{Details for VI.F (Performance sensitivity)}
\textbf{Scenario.}
One server node runs 10 BQC applications (A2 in \cref{sec:evaluation}, see also \cref{fig:app:vbqc_circuit}) concurrently with 10 client nodes (one BQC application per client node).
One BQC instance is run for each client.
This scenario was repeated 100 times for each of the three evaluations (impact of node-node-latencies, impact of internal latencies, impact of time slot length) in order to obtain statistics.
The success per BQC instance is calculated as in \cref{app:details_6_1}.
Success probability is calculated as the fraction of successful instances.

\textbf{Fixed parameters.}
Number of client nodes: 10.
Number of qubits per client node: 1.
Number of qubits for server node: 20.
Qubit coherence time: $T_2 = 1\cdot 10^7$ ns.
Network schedule: repeating pattern of 10 slots, each assigned to one client-server pair.

\textbf{Impact of node-to-node classical communication latencies.}
Network time slot length: $1\cdot 10^5$ ns.
Internal scheduler communication latency: $0$ ns.

Values used for node-to-node classical communication latencies: $10^5$, $10^6$, $10^7$ ns (i.e. $0.01$, $0.1$, $1$ times the $T_2$ coherence time).

\textbf{Impact of internal scheduler latencies.}
Network time slot length: $1\cdot 10^5$ ns.
Classical communication latency: $10^5$ ns.

Values used for internal scheduler communication latencies: $10^3$, $10^5$, $10^7$ ns, where we obtained success probabilities $0.89(2)$, $0.89(2)$, $0.83(2)$, respectively.

\textbf{Impact of network schedule time slot length.}
Classical communication latency: $10^5$ ns.
Internal scheduler communication latency: $0$ ns.

Values used for time slot length: $10^5$, $10^6$, $10^7$ ns, (i.e. $0.01$, $0.1$, $1$ times the $T_2$ coherence time).

\EOD

\end{document}